\documentclass{article}
\usepackage{PRIMEarxiv}
\usepackage{amssymb}
\usepackage{graphicx}
\usepackage{amssymb}
\usepackage{amsthm}
\usepackage{amsmath}
\usepackage{stmaryrd}
\usepackage{color}
\usepackage{float}
\usepackage{booktabs}
\usepackage{tabularx}
\usepackage{placeins}
\usepackage{multirow}
\usepackage[colorlinks,linkcolor=blue]{hyperref}      
\usepackage[misc]{ifsym}
\usepackage[numbers]{natbib}  
\graphicspath{{media/}}     

\title{On a general multi-layered hyperelastic plate theory of growth}

\author{
Ping Du\textsuperscript{1}, Zhanfeng Li\textsuperscript{1}, Xiaoyi Chen\textsuperscript{3}, Jiong Wang\textsuperscript{1,2,*} \\
\textsuperscript{1} School of Civil Engineering and Transportation, South China University of Technology, China \\
\textsuperscript{2} State Key Laboratory of Subtropical Building Science, South China University of Technology, China \\
\textsuperscript{3} Division of Science and Technology, BNU-HKBU United International College, China \\
\Letter: ctjwang@scut.edu.cn ( Jiong Wang)} 

\begin{document}
\maketitle

\begin{abstract}
In this paper, we propose a multi-layered hyperelastic plate theory of growth within the framework of nonlinear elasticity. First, the 3D governing system for a general multi-layered hyperelastic plate is established, which incorporates the growth effect, and the material and geometrical parameters of the different layers. Then, a series expansion-truncation approach is adopted to eliminate the thickness variables in the 3D governing system. An elaborate calculation scheme is applied to derive the iteration relations of the coefficient functions in the series expansions. Through some further manipulations, a 2D vector plate equation system with the associated boundary conditions is established, which only contains the unknowns in the bottom layer of the plate. To show the efficiency of the current plate theory, three typical examples regarding the growth-induced deformations and instabilities of multi-layered plate samples are studied. Some analytical and numerical solutions to the plate equation are obtained, which can provide accurate predictions on the growth behaviors of the plate samples. Furthermore, the problem of `shape-programming' of multi-layered hyperelastic plates through differential growth is studied. The explicit formulas of shape-programming for some typical multi-layered plates are derived, which involve the fundamental quantities of the 3D target shapes. By using these formulas, the shape evolutions of the plates during the growing processes can be controlled accurately. The results obtained in the current work are helpful for the design of intelligent soft devices with multi-layered plate structures.
\end{abstract}


\section{Introduction}
\label{sec:1}

Soft biological tissues with multi-layered plate or shell structures are commonly observed in nature. For example, the diverse structures of plant leaves \cite{koch2008}, the reflectors in the eyes of squid \cite{math2009} and the structures in the horn sheath of a cattle \cite{li2010}. From the perspectives of biomimetic mechanics, these examples in nature raise the possibility of design and application of soft intelligent devices with multi-layered plate or shell forms. In the engineering fields, soft materials have attracted extensive attentions due to their fantastic characteristics, including multifunctionality, biocompatibility, low cost and responsiveness to stimuli. Recently, multi-layered soft devices have been designed and applied in the areas of drug delivery \cite{fern2012}, soft robotics \cite{li2017,li2021}, nano-scale semiconductor tubes \cite{stoy2012,egun2016}, etc. The soft biological tissues in nature and the soft devices in engineering applications usually exhibit the changes of body mass or volume (e.g., the growth or atrophy of soft biological tissues, the swelling or shrinkage of hydrogels). For convenience, these changes are referred to as `growth' of the soft material samples in the current work. During the growing processes, the growth fields in soft material samples are usually inhomogeneous or incompatible, which is called differential growth and can result in various shape changes of the samples \cite{ambr2011,li2012,liu2015}.

Within the framework of nonlinear elasticity, soft materials can be viewed as certain kinds of hyperelastic materials. A number of theoretical models have been reported to study the growth-induced deformations of soft materials \cite{luba2002,ben2005,gori2017}. Usually, to capture the growth effect, the total deformation gradient is decomposed into the multiplication of the elastic strain tensor and the growth tensor \cite{kond1987,rodr1994}. As the elastic deformations of soft materials are generally isochoric, the constraint equation of elastic incompressibility should also be adopted. It is known that the responses of soft material samples are closely related to the residual stresses \cite{skal1996,tabe2001,hump2003}. Especially in the multi-layered plate or shell samples, the different growth parameters in the neighboring layers lead to the incompatibility of the growth fields, which can induce residual stresses in the samples. To release the misfit energy, the samples will undergo various bending or torsional deformations. In the current work, we mainly focus on the growth-induced deformations of multi-layered hyperelastic plates.

To model the mechanical behaviors of plate samples, one usually needs to adopt suitable plate theories. Some classical plate theories (e.g., the Kirchhoff-Love theory, the F\"{o}ppl-von K$\acute{\rm a}$rm$\acute{\rm a}$n (FvK) theory and the Mindlin-Reissner theory) were established within the small strain range and by making a priori hypotheses on the in-plane or transverse displacement components, which are not applicable for the large deformations of soft material samples. In \citet{dai2014}, a finite-strain plate theory was proposed for compressible hyperelastic materials, which can achieve the term-wise consistency with the variational formulation of the 3D governing system. This plate theory has been developed for the dynamic cases \cite{song2016} and for incompressible hyperelastic materials \cite{wang2016}. The numerical implementation of this plate theory has also been realized in \citet{fu2021}. In \citet{wang2018}, a consistent finite-strain plate theory of growth was further proposed, which takes the growth effect and elastic incompressibility into account and has no restrictions on the displacement components. Thus, it is suitable for modeling the growth-induced deformations in soft material samples. Applications of this plate theory can be found in \citet{wang2019,kada2021,meht2021,li2022,wang2022}.

Regarding the mechanical behaviors of multi-layered hyperelastic plates induced by growth, swelling or prestrain, a number of modeling works have also been reported in the literature. For example, \citet{tsai2004} derived the analytical solutions for the swelling-induced deformations of dual zone hyperelastic samples. By extending the $\Gamma$-convergence approach, \citet{schm2007} and \citet{delg2021} established the plate theories with the prestrain effect for heterogeneous multi-layered samples. \citet{derv2010} developed a FvK-type plate theory with the growth effect, which was applied to investigate the wrinkling properties of a growing film-substrate system. \citet{armo2011} studied the helical configurations and sharp morphological transitions of strips cut from a bilayer hyperelastic plate. \citet{budd2014} and \citet{ben2017} studied the convolutions of mammalian brains by using the growth models of multi-layered soft material samples. \citet{luca2014} and \citet{nard2017} analyzed the swelling-induced deformations and wrinkling properties in bilayer gel beams. \citet{pezz2016} investigated the deformations of bilayer plates with arbitrary shapes under the isotropic in-plane growth. \citet{van2017} solved the inverse problem of growth-induced shape-programming for anisotropically growing elastic bilayer structures. The obtained results can be used to achieve any target configurations. \citet{acke2022} investigated the mechanics of growing epithelia based on a bilayer FvK plate theory. Following the approach of \citet{wang2018}, \citet{du2020,du2022} proposed a finite-strain plate theory to study the growth-induced plane-strain deformations and instabilities of multi-layered hyperelastic plates. Some analytical solutions of the plate equations were obtained, which can provide accurate predictions on the response of the plate samples. From the analytical results, the influences of the geometrical and material parameters can also be revealed.

Despite the existence of the above modeling works on the growth behaviors of multi-layered hyperelastic plates, the research state has not attained a satisfactory level. Most of the models focus on some special types of samples (e.g., the bilayer samples) or deformation styles (e.g., the plane-strain deformations). Currently, there still lacks a plate theory that can be applied for general multi-layered plate samples (with general material properties and geometrical shapes) and arbitrary growth-induced deformations. This is just the motivation for carrying out the current work.

In the current work, we aim to propose a general multi-layered hyperelastic plate theory of growth within the framework of nonlinear elasticity. Compared with the previous modeling works, this new plate theory is applicable for the plate samples with general geometrical shapes, number of layers, material properties and external loading conditions, which can also incorporate arbitrary growth fields in the different layers. To establish the plate theory, we start from the 3D governing system and eliminate the thickness variables through a series expansion-truncation approach \cite{dai2014,wang2018}. An elaborate calculation scheme is then applied to derive the iteration relations of the coefficient functions in the series expansions \cite{du2020,du2022}. Through some further manipulations, a 2D vector plate equation system with the associated boundary conditions can be established. To show the efficiency of the new plate theory, three typical examples regarding the growth-induced deformations and instabilities of multi-layered plate samples will be studied. Some analytical and numerical solutions to the plate equation can be obtained. It will be shown that these analytical results can provide accurate predictions on the mechanical behaviors of the plate samples. Furthermore, the problem of `shape-programming' of multi-layered hyperelastic plates through differential growth will be solved, from which the explicit formulas of shape-programming for some typical multi-layered plate samples can be derived. By using these formulas, the shape evolutions of the plate samples during the growing processes can be controlled accurately.

This paper is organized as follows. In Section 2, we first formulate the 3D governing system for a general multi-layered hyperelastic plate, then the 2D vector plate equation system is established through a series expansion-truncation approach. In Section 3, three typical examples with different sample shapes, growth conditions and boundary restrictions are studied, from which some analytical and numerical solutions to the plate equation system will be obtained. In Section 4, we further study the problem of `shape-programming' of multi-layered hyperelastic plates through differential growth. Finally, some conclusions are drawn in Section 5.

\section{The plate equation system for a multi-layered hyperelastic plate}
\label{sec:2}

In this section, we first formulate the 3D governing system for a multi-layered hyperelastic plate, where the growth effect will be taken into account. After that, through a series expansion-truncation approach and by adopting an elaborate calculation scheme, the 2D vector plate equation system will be derived.


\subsection{The 3D governing system}
\label{sec:2.1}

\begin{figure}[h]
\begin{minipage}{0.49\textwidth}
\centering \includegraphics[width=0.95\textwidth]{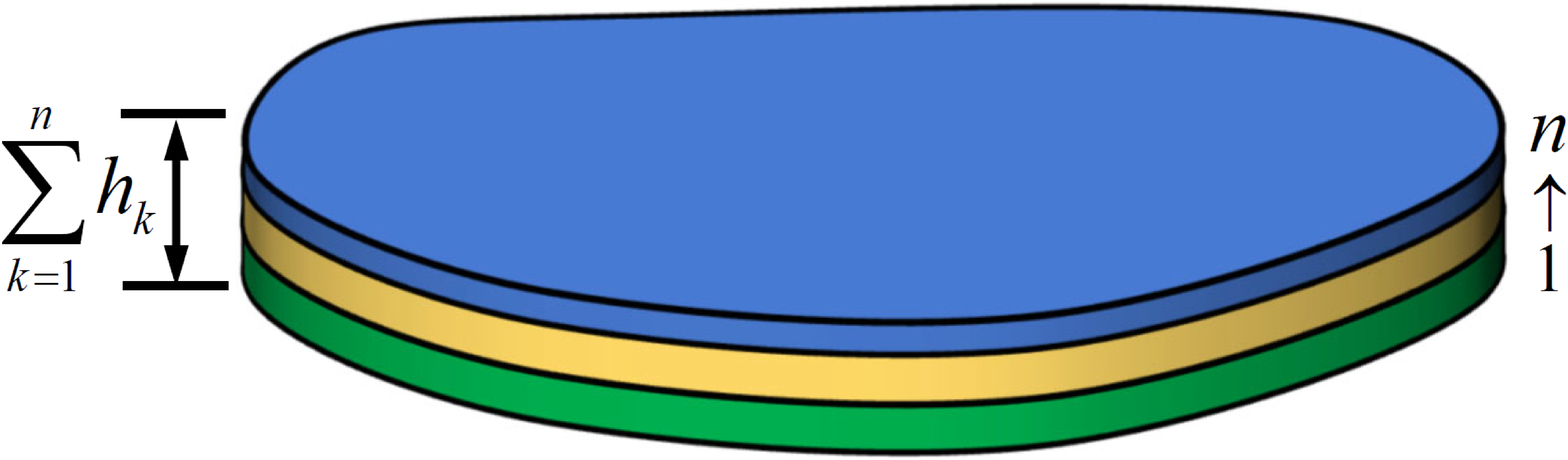}\\(a)\vspace{0.2cm}
\end{minipage}
\begin{minipage}{0.49\textwidth}
\centering \includegraphics[width=0.95\textwidth]{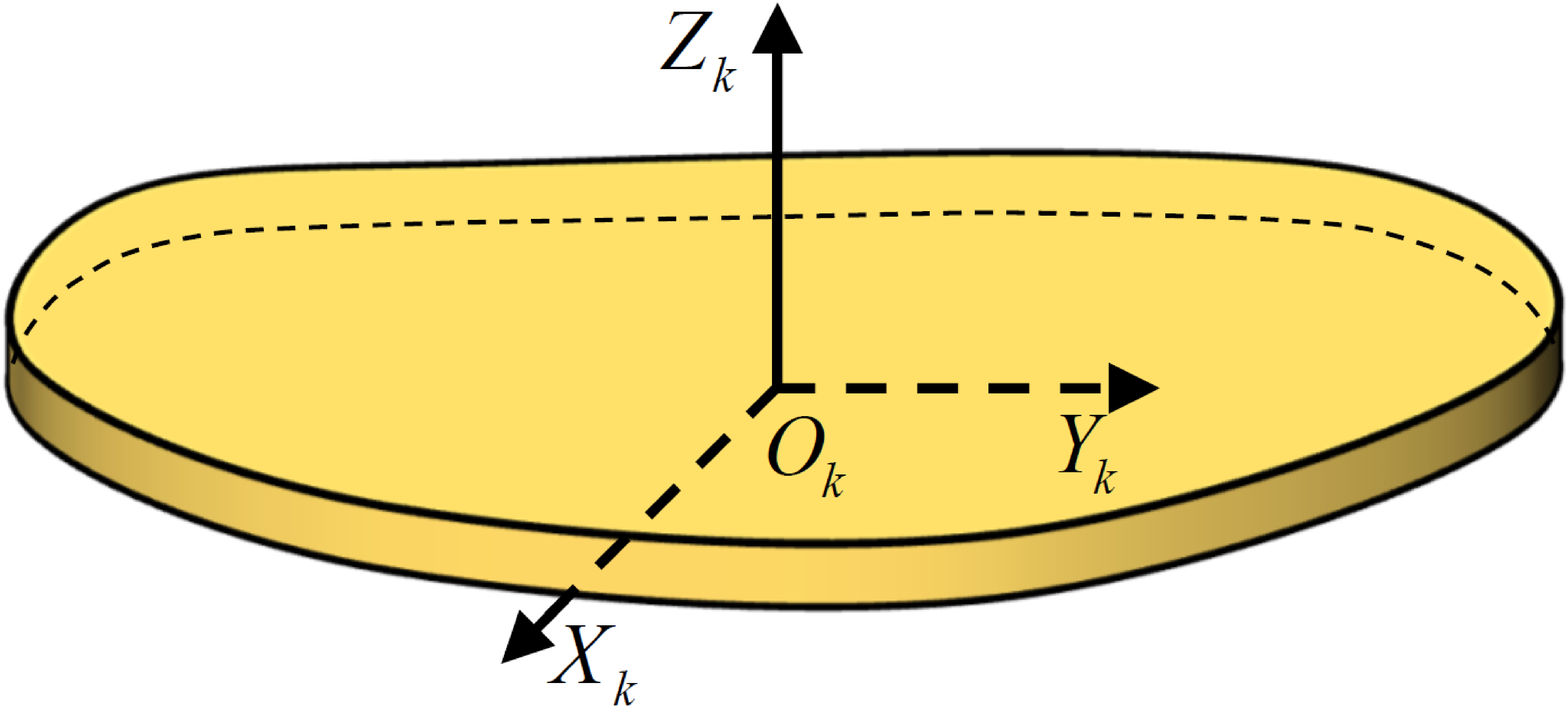}\\(b)\vspace{0.2cm}
\end{minipage}\\
\caption{(a) Illustrations of the reference configuration of a general multi-layered hyperelastic plate in 3D setting; (b) the local coordinate system of the $k$th layer.}
\label{fig:1}
\end{figure}

We consider a general thin hyperelastic plate as shown in Fig. \ref{fig:1}a, which is composed of $n$ layers with the initial thicknesses $h_1,h_2,\cdots,h_n$. For each layer in the plate, a local coordinate system is established on the bottom surface of the layer (cf. Fig. \ref{fig:1}b). Within this local coordinate system, the reference configuration of the $k$-th layer occupies the region $\kappa_k=\Omega\times[0,h_k]$ ($1\leq k\leq n$), where $\Omega$ is the in-plane area of the plate. The position vector of a material point in $\kappa_k$ is denoted as $\mathbf{X}_k=\mathbf{r}_k+ Z_k\mathbf{k}$, where $\mathbf{k}$ is the unit normal vector of $\Omega$ (i.e., the unit vector directing along the $Z_k$-axis). Under a rectangular cartesian coordinate system, we further denote $\mathbf{r}_k=X_k\mathbf{E}_1 + Y_k\mathbf{E}_2$, where $\mathbf{E}_1$ and $\mathbf{E}_2$ are the unit vectors directing along the $X_k$- and $Y_k$-axes, respectively.

The different layers in the plate can grow independently. Besides that, some external loads are applied on the surface of the plate. Due to the growth effect and the external loads, the configuration of the $k$-th layer in the plate will transform from $\kappa_k$ to the current configuration $\tilde{\kappa}_k$ $(1\leq k\leq n)$. The position vector of a material point in $\tilde{\kappa}_k$ is denoted as $\mathbf{x}_k$. The total deformation gradient tensor in the $k$-th layer is then expressed by
\begin{equation}
\begin{aligned}
\mathbb{F}_k = \frac{\partial \mathbf{x}_k}{\partial \mathbf{X}_k} &= {\nabla_r \mathbf{x}_k} + \frac{\partial \mathbf{x}_k}{\partial Z_k} \otimes \mathbf{k}= \frac{\partial \mathbf{x}_k}{\partial X_k} \otimes \mathbf{E}_1+\frac{\partial \mathbf{x}_k}{\partial Y_k} \otimes \mathbf{E}_2 + \frac{\partial \mathbf{x}_k}{\partial Z_k} \otimes \mathbf{k},
\end{aligned}
\label{eq:1}
\end{equation}
where $\nabla_r$ is the in-plane 2D gradient operator. Following the conventional approach of growth mechanics \cite{kond1987,rodr1994}, the total deformation gradient tensor $\mathbb{F}_k$ is decomposed into
\begin{equation}
\mathbb{F}_k = \mathbb{A}_k\mathbb{G}_k, \ \ \ k=1,\cdots,n,
\label{eq:2}
\end{equation}
where $\mathbb{A}_k$ is the elastic strain tensor and $\mathbb{G}_k$ is the growth tensor. In the current work, we assume that the growth tensor in each layer is uniform along the thickness direction, thus $\mathbb{G}_k$ doesn't depend on the thickness variable $Z_k$. By further assuming the elastic incompressibility of the material (which is satisfied for most soft biological and polymeric materials), we have the following constraint equation
\begin{equation}
R_k \left(\mathbb{F}_k, \mathbb{G}_k \right) = \tilde{R}_{k} \left(\mathbb{F}_k {\mathbb{G}_k}^{-1} \right) = \tilde{R}_{k} \left(\mathbb{A}_k \right) =\mathrm{Det} \left(\mathbb{A}_k \right)-1=0,\ \
\label{eq:3}
\end{equation}
where $\mathrm{Det}(\cdot)$ represents the determinant of a tensor.

We suppose the hyperelastic material in the $k$-th layer of the plate has the strain-energy function $\phi_k(\mathbb{F}_k,\mathbb{G}_k)=J_{G_k}\tilde{\phi}_{k} \left(\mathbb{F}_k{\mathbb{G}_k}^{-1}\right)=J_{G_k}\tilde{\phi}_{k}\left(\mathbb{A}_k\right)$, where $J_{G_k}=\mathrm{Det}(\mathbb{G}_k)$. During the growing process, each layer of the plate keeps in the quasi-static state, which yields the mechanical equilibrium equation
\begin{equation}
\begin{aligned}
\mathrm{Div}\mathbb{S}_k=0,\ \ \ \mathrm{in} \ \kappa_k,\ \ k=1,\cdots,n,
\end{aligned}
\label{eq:4}
\end{equation}
where $\mathbb{S}_k$ is the nominal stress tensor defined by
\begin{equation}
\begin{aligned}
\mathbb{S}_k &= J_{G_k} \left(\frac{\partial \phi_{k}(\mathbb{F}_k,\mathbb{G}_k)}{\partial \mathbb{F}_k}-p_k \frac{\partial R_{k}(\mathbb{F}_k,\mathbb{G}_k)} {\partial \mathbb{F}_k} \right)= J_{G_k} {\mathbb{G}_k}^{-1} \left(\frac{\partial \tilde{\phi}_{k}(\mathbb{A}_k)}{\partial \mathbb{A}_k} - p_k \frac{\partial \tilde{R}_{k}(\mathbb{A}_k)} {\partial \mathbb{A}_k}\right).
\end{aligned}
\label{eq:5}
\end{equation}
In the expression of $\mathbb{S}_k$, $p_k(\mathbf{X})$ is the Lagrangian multiplier related to the constraint of elastic incompressibility, which is an unknown function.

On the top and bottom faces of the plate, we propose the following traction boundary conditions
\begin{equation}
\begin{aligned}
\begin{cases}
\mathbb{S}_1^T\mathbf{k}|_{Z_1=0} = -\mathbf{q}^-(\mathbf{r}_1),\ \ \ \mathrm{on} \ \Omega_1^- ,\\
\mathbb{S}_n^T\mathbf{k}|_{Z_n=h_n} = \mathbf{q}^+(\mathbf{r}_n),\ \ \ \mathrm{on} \ \Omega_n^+.
\end{cases}
\end{aligned}
\label{eq:6}
\end{equation}
where $\mathbf{q}^-$ and $\mathbf{q}^+$ are the applied tractions, $\Omega_1^-$ represents the bottom face of the $1$st layer and $\Omega_n^+$ represents the top face of the $n$-th layer. The lateral surface of each layer is denoted as $\partial\kappa_k$ ($\partial\kappa_k=\partial\Omega\times[0,h_k]$), which is composed of the position boundary region $\partial\kappa_k^{(p)}=\partial\Omega^{(p)}\times[0,h_k]$ and the traction boundary $\partial\kappa_k^{(q)}=\partial\Omega^{(q)}\times[0,h_k]$ (i.e., $\partial\Omega=\partial\Omega^{(p)}\cup\partial\Omega^{(q)}$). On $\partial\kappa_k^{(p)}$ and $\partial\kappa_k^{(q)}$, we propose the following boundary conditions
\begin{equation}
\begin{aligned}
\begin{cases}
\mathbf{x}_k=\mathbf{b}_k(s,Z_k),\ \ \mathrm{on} \ \partial\kappa_k^{(p)},\ \ \ k=1,\cdots,n,\\
\mathbb{S}_k^T\mathbf{N}=\mathbf{q}_k(s,Z_k),\ \ \mathrm{on} \ \partial\kappa_k^{(q)},\ \ \ k=1,\cdots,n.
\end{cases}
\end{aligned}
\label{eq:7}
\end{equation}
where $\mathbf{b}_k$ and $\mathbf{q}_k$ are the prescribed position vectors and applied tractions, and $\mathbf{N}$ is the external normal vector of the lateral surface. On the interfaces between the different layers, we propose the following displacement and stress continuity conditions
\begin{equation}
\begin{aligned}
\mathbf{x}_{k+1}(\mathbf{r}_{k+1},0) - \mathbf{x}_k(\mathbf{r}_k,h_k) = 0,\ \ \ k=1,2,\cdots,n-1,
\end{aligned}
\label{eq:8}
\end{equation}
\begin{equation}
\begin{aligned}
\mathbb{S}_{k+1}^T \mathbf{k}|_{Z_{k+1}=0} - \mathbb{S}_k^T \mathbf{k}|_{Z_k=h_k} = 0,\ \ \ k=1,2,\cdots,n-1.
\end{aligned}
\label{eq:9}
\end{equation}
Eqs. (\ref{eq:3})-(\ref{eq:9}) formulate the 3D governing equation system for the multi-layered hyperelastic plate, which contains the unknowns $\mathbf{x}_{k}$ and $p_k$ $(k=1,\cdots,n)$.


\subsection{Derivation of the 2D vector plate equation}
\label{sec:2.2}

In this subsection, a series expansion-truncation approach will be adopted to derive the 2D vector plate equation from the 3D governing system (\ref{eq:3})-(\ref{eq:9}) (cf. \citet{dai2014,wang2016,wang2018}). For that purpose, the position vector $\mathbf{x}_k(\mathbf{X}_k)$ and the Lagrangian multiplier $p_k(\mathbf{X}_k)$ in each layer are expanded in terms of the thickness variable $Z_k$ as follow
\begin{equation}
\begin{aligned}
&\mathbf{x}_k(\mathbf{X}_k)= \sum_{i=0}^4 \frac{{Z_k}^i}{i!} \mathbf{x}_k^{(i)}(\mathbf{r}_k)+O({Z_k}^5),\ \ \\
&p_k(\mathbf{X}_k)= \sum_{i=0}^3 \frac{{Z_k}^i}{i!} p_k^{(i)}(\mathbf{r}_k)+O({Z_k}^4),\ \ \ k=1,\cdots,n.
\end{aligned}
\label{eq:10}
\end{equation}
In the following derivations, only the $5n$ vector unknowns $\mathbf{x}_k^{(0)}$-$\mathbf{x}_k^{(4)}$ and the $4n$ scalar unknowns $p_k^{(0)}$-$p_k^{(3)}$ $(k=1,2,\cdots,n)$ will be taken into account (i.e., there are totally $19n$ unknowns).

According to the series expansions (\ref{eq:10}), the deformation gradient tensor $\mathbb{F}_k$ and the elastic strain tensor $\mathbb{A}_k$ can also be expanded as
\begin{equation}
\begin{aligned}
&\mathbb{F}_k(\mathbf{X}_k) = \sum_{i=0}^3 \frac{{Z_k}^i}{i!} \mathbb{F}_k^{(i)}+O({Z_k}^4), \ \ \\
&\mathbb{A}_k(\mathbf{X}_k) = \sum_{i=0}^3 \frac{{Z_k}^i}{i!} \mathbb{A}_k^{(i)}+O({Z_k}^4), \ \ k=1,\cdots,n.
\end{aligned}
\label{eq:11}
\end{equation}
As mentioned before, the growth tensor $\mathbb{G}_k$ does not depend on the coordinate $Z$, so we do not need to consider the expansion of $\mathbb{G}_k$. By submitting (\ref{eq:10})$_1$ into (\ref{eq:1}) and comparing with (\ref{eq:11})$_1$, we obtain
\begin{equation}
\mathbb{F}_k^{(i)}=\nabla_r \mathbf{x}_k^{(i)}+\mathbf{x}_k^{(i+1)}\otimes \mathbf{k}, \ \ i=0,1,2,3.
\label{eq:12}
\end{equation}
From (\ref{eq:12}), it can be observed that $\mathbb{F}_k^{(i)}$ and $\mathbf{x}_k^{(i+1)}$ satisfy a linear relation. Further from (\ref{eq:2}) and (\ref{eq:11})$_2$, we derive the following relations
\begin{equation}
\mathbb{A}_k^{(i)} = \mathbb{F}_k^{(i)}\mathbb{G}_k^{-1}, \ \ i=0,1,2,3.
\label{eq:13}
\end{equation}

Similar as the deformation gradient tensor, the nominal stress tensor $\mathbb{S}_k$ can be expanded as
\begin{equation}
\mathbb{S}_k(\mathbf{X}_k, p_k)= \sum_{i=0}^3 \frac{{Z_k}^i}{i!} \mathbb{S}_k^{(i)}+O({Z_k}^4),\ \ \ k=1,\cdots,n.
\label{eq:14}
\end{equation}
On the other hand, by submitting (\ref{eq:11})$_2$ into (\ref{eq:5}), another expression of $\mathbb{S}_k$ can be derived, which is given by
\begin{equation}
\begin{aligned}
\mathbb{S}_k =& \widehat{\mathbb{G}}_k \left(\frac{\partial \tilde{\phi}_k(\mathbb{A}_k)}{\partial \mathbb{A}_k} - p_k \frac{\partial \tilde{R}_k(\mathbb{A}_k )} {\partial \mathbb{A}_k} \right)\\
=& \widehat{\mathbb{G}}_k \bigg\{\mathcal{A}_k^{(0)} (\mathbb{A}_k^{(0)})+\mathcal{A}_k^{(1)}(\mathbb{A} _k^{(0)}) [\mathbb{A}_k -\mathbb{A}_k^{(0)}]\\
&+ \frac{1}{2} \mathcal{A}_k^{(2)}(\mathbb{A}_k^{(0)}) [\mathbb{A}_k -\mathbb{A}_k^{(0)},\mathbb{A}_k -\mathbb{A}_k^{(0)}]\\
&+ \frac{1}{6} \mathcal{A}_k^{(3)}(\mathbb{A}_k^{(0)}) [\mathbb{A}_k -\mathbb{A}_k^{(0)},\mathbb{A}_k -\mathbb{A}_k^{(0)},\mathbb{A}_k -\mathbb{A}_k^{(0)}]\\
&-\Big\{p_k^{(0)}
 +Z_kp_k^{(1)}
+\frac{1}{2}Z_k^2 p_k^{(2)} +\frac{1}{6}Z_k^3 p_k^{(3)} + O(Z_k^4) \Big\}\\
&\times \Big\{\mathcal{R}_k^{(0)} (\mathbb{A}_k^{(0)})+\mathcal{R}_k^{(1)}(\mathbb{A} _k^{(0)}) [\mathbb{A}_k -\mathbb{A}_k^{(0)}]\\
&+ \frac{1}{2} \mathcal{R}_k^{(2)}(\mathbb{A}_k^{(0)}) [\mathbb{A}_k -\mathbb{A}_k^{(0)},\mathbb{A}_k -\mathbb{A}_k^{(0)}]\\
&+ \frac{1}{6} \mathcal{R}_k^{(3)}(\mathbb{A}_k^{(0)}) [\mathbb{A}_k -\mathbb{A}_k^{(0)},\mathbb{A}_k -\mathbb{A}_k^{(0)},\mathbb{A}_k -\mathbb{A}_k^{(0)}] \Big\} \bigg\}
\end{aligned}
\label{eq:15}
\end{equation}
where $\widehat{\mathbb{G}}_k=J_{G_k}\mathbb{G}_k^{-1}$ and
\begin{equation}
\begin{aligned}
\mathcal{A}_k^{(i)}(\mathbb{A}_k^{(0)})={\frac{{\partial}^{i+1}\tilde{\phi}_{k}(\mathbb{A}_k)}{{\partial \mathbb{A}_k}^{i+1}}}\big|_{\mathbb{A}_k=\mathbb{A}_k^{(0)}},\ \ \ \mathcal{R}_k^{(i)}(\mathbb{A}_k^{(0)})={\frac{{\partial}^{i+1}\tilde{R}_{k}(\mathbb{A}_k)}{{\partial \mathbb{A}_k}^{i+1}}}\big|_{\mathbb{A}_k=\mathbb{A}_k^{(0)}}.
\end{aligned}
\label{eq:16}
\end{equation}
Here, $\mathcal{A}_k^{(i)}(\mathbb{A}_k^{(0)})$ $(i=0,1,2,3)$ are the elastic moduli associated with the strain-energy function $\tilde{\phi}_{k}(\mathbb{A}_k)$. By comparing the coefficients of $Z_k^i$ in (\ref{eq:14}) and (\ref{eq:15}), we obtain the following relations
\begin{equation}
\begin{aligned}
\mathbb{S}_k^{(0)} (\mathbf{r}_k) =& \widehat{\mathbb{G}}_k \left(\mathcal{A}_k^{(0)} -p_k^{(0)} \mathcal{R}_k^{(0)} \right),\\
\mathbb{S}_k^{(1)} (\mathbf{r}_k)=& \widehat{\mathbb{G}}_k \left(\mathcal{A}_k^{(1)}[\mathbb{A}_k^{(1)}] - p_k^{(0)} \mathcal{R}_k^{(1)}[\mathbb{A}_k^{(1)}] - p_k^{(1)} \mathcal{R}_k^{(0)} \right),\\
\mathbb{S}_k^{(2)} (\mathbf{r}_k)=& \widehat{\mathbb{G}}_k \bigg\{\mathcal{A}_k^{(1)}[\mathbb{A}_k^{(2)}] + \mathcal{A}_k^{(2)}[\mathbb{A}_k^{(1)},\mathbb{A}_k^{(1)}]\\
&- p_k^{(0)} \left(\mathcal{R}_k^{(1)}[\mathbb{A}_k^{(2)}] + \mathcal{R}_k^{(2)}[\mathbb{A}_k^{(1)},\mathbb{A}_k^{(1)}] \right)\\
&- 2 p_k^{(1)} \mathcal{R}_k^{(1)}[\mathbb{A}_k^{(1)}] - p_k^{(2)} \mathcal{R}_k^{(0)} \bigg\},\\
\mathbb{S}_k^{(3)} (\mathbf{r}_k) =&\widehat{\mathbb{G}}_k \bigg\{ \mathcal{A}_k^{(1)}[\mathbb{A}_k^{(3)}] + 3 \mathcal{A}_k^{(2)}[\mathbb{A}_k^{(2)},\mathbb{A}_k^{(1)}] + \mathcal{A}_k^{(3)}[\mathbb{A}_k^{(1)},\mathbb{A}_k^{(1)},\mathbb{A}_k^{(1)}]\\
&- p_k^{(0)} \left(\mathcal{R}_k^{(1)}[\mathbb{A}_k^{(3)}] + 3 \mathcal{R}_k^{(2)}[\mathbb{A}_k^{(2)},\mathbb{R}_k^{(1)}] + \mathcal{A}_k^{(3)}[\mathbb{A}_k^{(1)},\mathbb{A}_k^{(1)},\mathbb{A}_k^{(1)}] \right) \\
&- 3 p_k^{(1)} \left( \mathcal{R}_k^{(1)}[\mathbb{A}_k^{(2)}] + \mathcal{R}_k^{(2)}[\mathbb{A}_k^{(1)},\mathbb{R}_k^{(1)}] \right)- 3p_k^{(2)} \mathcal{R}_k^{(1)}[\mathbb{A}_k^{(1)}]\\
&- p_k^{(3) } \mathcal{R}_k^{(0)} \bigg\}.
\end{aligned}
\label{eq:17}
\end{equation}

Based on the above preparations, we begin to formulate a closed equation system for the $19n$ unknowns $\mathbf{x}_k^{(0)}$-$\mathbf{x}_k^{(4)}$ and $p_k^{(0)}$-$p_k^{(3)}$ $(k=1,2,\cdots,n)$.

First, by substituting (\ref{eq:11})$_2$ into the constraint equation (\ref{eq:3}), it can be obtained that
\begin{equation}
\begin{aligned}
\tilde{R}_k(\mathbb{A}_k)=&\tilde{R}_k(\mathbb{A}_k^{(0)})+\mathcal{R}_k^{(0)}(\mathbb{A}_k^{(0)})[\mathbb{A}_k-\mathbb{A}_k^{(0)}]\\
&+\frac{1}{2}\mathcal{R}_k^{(1)}(\mathbb{A}_k^{(0)})[\mathbb{A}_k-\mathbb{A}_k^{(0)},\mathbb{A}_k-\mathbb{A}_k^{(0)}]\\
&+\frac{1}{6}\mathcal{R}_k^{(2)}(\mathbb{A}_k^{(0)})[\mathbb{A}_k-\mathbb{A}_k^{(0)},\mathbb{A}_k-\mathbb{A}_k^{(0)},\mathbb{A}_k-\mathbb{A}_k^{(0)}]\\
=&\tilde{R}_k(\mathbb{A}_k^{(0)})+\mathcal{R}_k^{(0)}(\mathbb{A}_k^{(0)}):\left\{Z_k\mathbb{A}_k^{(1)}+\frac{1}{2} Z_k^2\mathbb{A}_k^{(2)}+\frac{1}{6} Z_k^3\mathbb{A}_k^{(3)}\right\}\\
&+\frac{1}{2}\left\{Z_k\mathbb{A}_k^{(1)}+\frac{1}{2} Z_k^2\mathbb{A}_k^{(2)}\right\}:\mathcal{R}_k^{(1)}(\mathbb{A}_k^{(0)})[Z_k\mathbb{A}_k^{(1)}+\frac{1}{2} Z_k^2\mathbb{A}_k^{(2)}]\\
&+\frac{1}{6} Z_k\mathbb{A}_k^{(1)}:\mathcal{R}_k^{(2)}(\mathbb{A}_k^{(0)}) [Z_k\mathbb{A}_k^{(1)}, Z_k\mathbb{A}_k^{(1)}]+O(Z_k^4)=0.
\end{aligned}
\label{eq:18}
\end{equation}
The coefficients of $Z_k^i$ $(i=0,1,2,3)$ in (\ref{eq:18}) should be vanished, which yields that
\begin{equation}
\begin{aligned}
&\tilde{R}_k(\mathbb{A}_k^{(0)})=0,\\
&\mathcal{R}_k^{(0)}:\mathbb{A}_k^{(1)}=0,\\
&\mathcal{R}_k^{(0)}:\mathbb{A}_k^{(2)}+\mathbb{A}_k^{(1)}:\mathbb{R}_k^{(1)}[\mathbb{A}_k^{(1)}]=0,\\
&\mathcal{R}_k^{(0)}:\mathbb{A}_k^{(3)}+3 \mathbb{A}_k^{(2)}:\mathbb{R}_k^{(1)}[\mathbb{A}_k^{(1)}]+\mathbb{A}_k^{(1)}:\mathbb{R}_k^{(2)}[\mathbb{A}_k^{(1)},\mathbb{A}_k^{(1)}]= 0,
\end{aligned}
\label{eq:19}
\end{equation}
where $1\leq k\leq n$ and `$:$' represents the double contraction of the tensors. Eq. (\ref{eq:19}) provides $4n$ equations for the unknowns.

Next, by submitting (\ref{eq:14}) into the mechanical equilibrium equation (\ref{eq:4}) and considering the coefficients of ${Z_k}^i$ $(i=0,1,2)$, we obtain that
\begin{equation}
\begin{aligned}
&\frac{\partial }{\partial X} \left(S_{k,1b}^{(i)}\right) + \frac{\partial }{\partial Y} \left(S_{k,2b}^{(i)}\right) + S_{k,3b}^{(i+1)} = 0,\ \ \ b=1,2,3,
\end{aligned}
\label{eq:20}
\end{equation}
where $1\leq k\leq n$ and $S_{k,ab}^{(i)}$ represents the $ab$-component of $\mathbb{S}_k^{(i)}$. Eq. (\ref{eq:20}) provides another $9n$ equations for the unknowns.

Further substituting (\ref{eq:14}) into the boundary conditions (\ref{eq:6}), the following $6$ equations can be obtained
\begin{equation}
\begin{aligned}
&\left\{\mathbb{S}_1^{(0)}\right\}^T \mathbf{k} =- \mathbf{q}^-(\mathbf{r}_1),\\
&\sum_{i=0}^{3} \left\{\frac{{h_n}^i}{i!} \mathbb{S}_n^{(i)}\right\}^T \mathbf{k} + O(h^4) = \mathbf{q}^+(\mathbf{r}_n).
\end{aligned}
\label{eq:21}
\end{equation}

Finally, from the displacement and stress continuity conditions (\ref{eq:8})-(\ref{eq:9}) between the interfaces of the layers, we can derive the following $6n-6$ equations
\begin{equation}
\begin{aligned}
\mathbf{x}_{k+1}^{(0)} - \sum_{i=0}^{4} \frac{{h_k}^i}{i!} \mathbf{x}_k^{(i)} = \mathbf{0}, \ \ k=1,2,\cdots,n-1,
\end{aligned}
\label{eq:22}
\end{equation}
\begin{equation}
\begin{aligned}
\left\{\mathbb{S}_{k+1}^{(0)}\right\}^T \mathbf{k} - \sum_{i=0}^{3} \left\{\frac{{h_k}^i}{i!} \mathbb{S}_k^{(i)}\right\}^T \mathbf{k} = \mathbf{0}, \ \ k=1,2,\cdots,n-1.
\end{aligned}
\label{eq:23}
\end{equation}

In summary, Eqs. (\ref{eq:19})-(\ref{eq:23}) provide $19n$ equations for the $19n$ unknowns $\mathbf{x}_k^{(0)}$-$\mathbf{x}_k^{(4)}$ and $p_k^{(0)}$-$p_k^{(3)}$ $(k=1,2,\cdots,n)$, which formulate a closed equation system. To derive a 2D vector plate equation system from this closed system, we adopt an elaborate calculation scheme proposed in \citet{du2020,du2022}. The procedure of this calculation scheme is introduced below (which is also shown in Fig. \ref{fig:2}):

\begin{figure}[h]
\centering \includegraphics[width=0.8\textwidth]{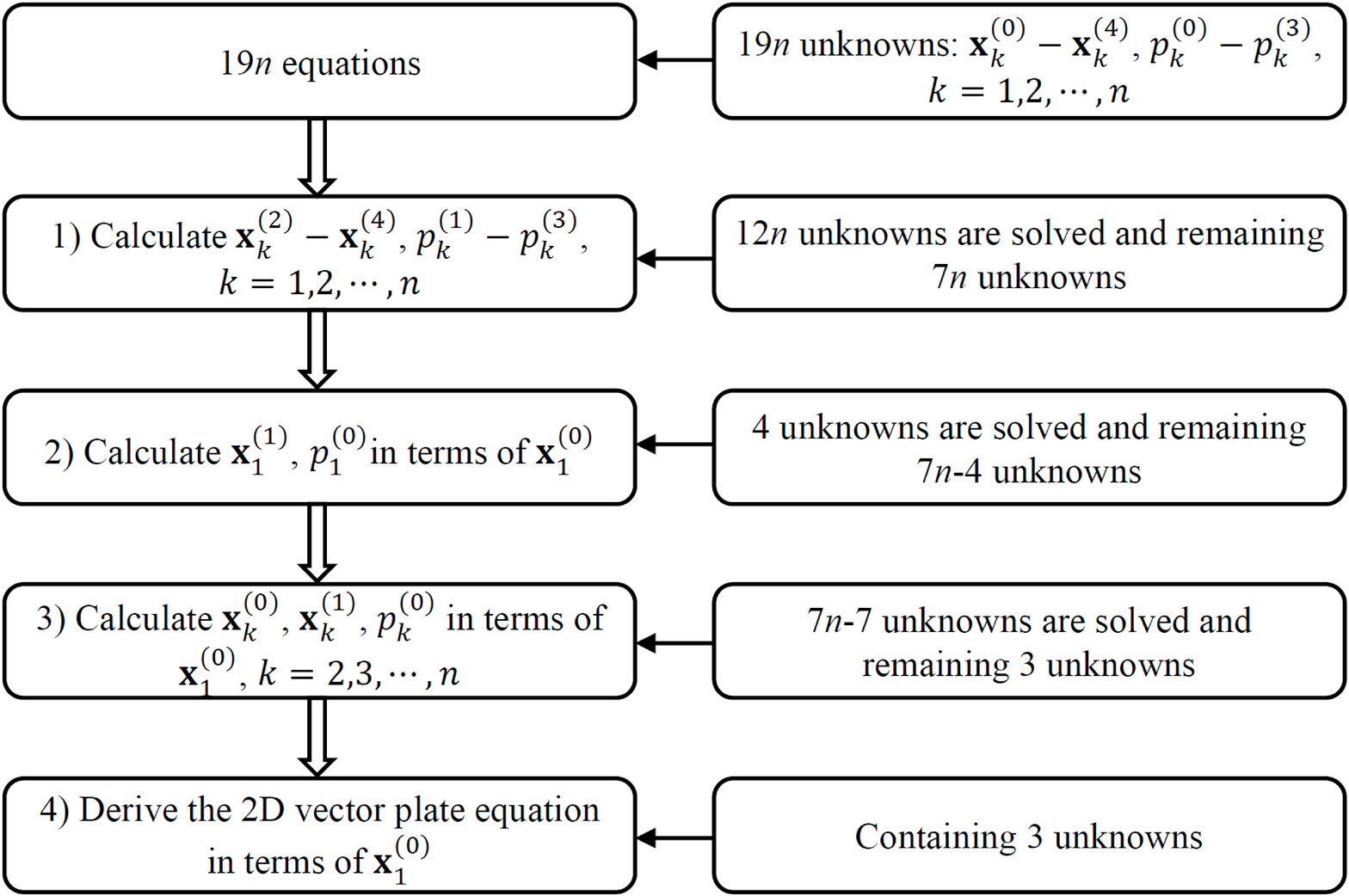}\\
\caption{The flowchart of the calculation scheme to derive the 2D vector plate equation.}
\label{fig:2}
\end{figure}

\begin{itemize}

\item \emph{Step one}. By solving the linear algebraic equations (\ref{eq:19})$_{2-4}$ and (\ref{eq:20}) ($i=0,1,2$), the expressions of $\mathbf{x}_k^{(2)}$-$\mathbf{x}_k^{(4)}$ and $p_k^{(1)}$-$p_k^{(3)}$ $(k=1,\cdots,n)$ can be obtained. In Eq. (\ref{eq:A1}) of \ref{app:a}, the explicit expressions of $\mathbf{x}_k^{(2)}$-$\mathbf{x}_k^{(3)}$ and $p_k^{(1)}$-$p_k^{(2)}$ are presented. The unknowns $\mathbf{x}_k^{(4)}$ and $p_k^{(3)}$ are just adopted to facilitate the derivations, whose explicit expressions are not needed in the current work. After the manipulations in this step, we still have $7n$ equations for the remaining $7n$ unknowns $\mathbf{x}_k^{(0)}$-$\mathbf{x}_k^{(1)}$ and $p_k^{(0)}$ ($k=1,\cdots,n$).

\item \emph{Step two}. Eq. (\ref{eq:21})$_1$ together with Eq. (\ref{eq:19})$_1$ ($k=1$) provide an algebraic equation system for the vector unknown $\mathbf{x}_1^{(1)}$ and the scalar unknown $p_1^{(0)}$. Once this system is solved, we can obtain the expressions of four unknowns in terms of $\mathbf{x}_1^{(0)}$, i.e.,
    \begin{equation}
    \begin{aligned}
    \mathbf{x}_1^{(1)}=\mathbf{f}_1^{(1)}(\mathbf{x}_1^{(0)}),\ \    p_1^{(0)}=&l_1(\mathbf{x}_1^{(0)}).
    \end{aligned}
    \label{eq:24}
    \end{equation}
    It should be noted that for general hyperelastic materials, Eq. (\ref{eq:21})$_1$ provides nonlinear algebraic equations of $\mathbf{x}_1^{(1)}$ and $p_1^{(0)}$. Therefore, it may not be straightforward to derive the explicit expressions of these unknowns. While, if the layers in the plate are made of incompressible neo-Hookean materials, Eq. (\ref{eq:21})$_1$ becomes linear algebraic equations for these unknowns, which can be solved explicitly. The obtained expressions of the unknowns corresponding to incompressible neo-Hookean materials are presented in Eq. (\ref{eq:A2}) of \ref{app:a}. Through the manipulations in this step, we have $7n-4$ equations for the remaining $7n-4$ unknowns $\mathbf{x}_k^{(0)}$ ($k=1,2,\cdots,n$), $\mathbf{x}_k^{(1)}$ ($k=2,3,\cdots,n$) and $p_k^{(0)}$ ($k=2,3,\cdots,n$).

\item \emph{Step three}. Eqs. (\ref{eq:19})$_1$, (\ref{eq:22}) and (\ref{eq:23}) provide another algebraic equation system for the unknowns $\mathbf{x}_k^{(0)}$, $\mathbf{x}_k^{(1)}$ and $p_k^{(0)}$ ($k=2,3,\cdots,n$). By solving this system, one can derive the asymptotic iterative relations of the unknowns between the neighboring layers. The procedure of manipulation is $1\rightarrow 2 \rightarrow \cdots \rightarrow n$ for the layers of the plate. In Eq. (\ref{eq:A3}) of \ref{app:a}, we present the asymptotic expressions of iterative relations for incompressible neo-Hookean materials. By virtue of these iterative relations, the unknowns can be further written into the functions of $\mathbf{x}_1^{(0)}$, i.e.,
    \begin{equation}
    \begin{aligned}
    \mathbf{x}_k^{(0)}=\mathbf{f}_k^{(0)}(\mathbf{x}_1^{(0)}),\ \
    \mathbf{x}_k^{(1)}=\mathbf{f}_k^{(1)}(\mathbf{x}_1^{(0)}),\ \
    p_k^{(0)}=l_k(\mathbf{x}_1^{(0)}),
    \end{aligned}
    \label{eq:25}
    \end{equation}
    where $k=2,3,\cdots,n$. After the manipulations in this step, we only have the unknowns $\mathbf{x}_1^{(0)}$.

\item \emph{Step four}. Finally, the 2D vector plate equation can be derived from the boundary conditions (\ref{eq:21}). In fact, from (\ref{eq:21})$_2$ and by virtue of (\ref{eq:21})$_1$ and (\ref{eq:23}), the following 2D vector plate equation of the plate can be obtain
    \begin{equation}
    \begin{aligned}
    \nabla_r \cdot \bar{\mathbb{S}}(\mathbf{x}_1^{(0)}) = -\tilde{\mathbf{q}},\ \ \ \mathrm{in}\ \ \Omega,
    \end{aligned}
    \label{eq:26}
    \end{equation}
    where $\tilde{\mathbf{q}} = \mathbf{q}^+ +\mathbf{q}^-$ and
    \begin{equation*}
    \begin{aligned}
    \bar{\mathbb{S}}(\mathbf{x}_1^{(0)}) =& \sum_{i=1}^n \int_0^{h_k} \mathbb{S}_k(\mathbf{x}_1^{(0)}) dZ = \sum_{i=1}^n h_k \left(\mathbb{S}_k^{(0)}(\mathbf{x}_1^{(0)}) + \frac{h_k}{2}\mathbb{S}_k^{(1)}(\mathbf{x}_1^{(0)}) + \frac{{h_k}^2}{6} \mathbb{S}_k^{(2)}(\mathbf{x}_1^{(0)})\right).
    \end{aligned}
    \end{equation*}

By submitting the iterative relations (\ref{eq:24}), (\ref{eq:25}) and (\ref{eq:A1}) in \ref{app:a} into Eq. (\ref{eq:26}), the 2D vector plate equation in terms of $\mathbf{x}_1^{(0)}$ can be obtain, which attains the accuracy of $O(h^2)$.

\end{itemize}


\subsection{Edge boundary conditions}
\label{sec:2.3}

To complete the plate equation system, we still need suitable boundary conditions on the edge of the plate area $\partial\Omega$. According to the boundary conditions (\ref{eq:7}) in the 3D governing system, the following edge boundary conditions for the plate equation can be proposed:

\textbf{Case 1. Position boundary conditions}

We suppose the position vector field $\mathbf{b}_k(s,Z_k)$ has been prescribed on the region $\partial\kappa_k^{(p)}=\partial\Omega^{(p)}\times[0,h_k]$ $(k=1,\cdots,n)$, from which we can adopt the following two kinds of edge boundary conditions
\begin{equation}
\begin{aligned}
\mathbf{x}_1^{(0)}(s)=&\mathbf{b}_1(s,0),\ \ \bar{\mathbf{x}}(s) =\bar{\mathbf{b}}(s),\ \ \mathrm{on} \ \partial\Omega^{(p)},
\end{aligned}
\label{eq:27}
\end{equation}
where
\begin{equation*}
\begin{aligned}
\bar{\mathbf{x}}(s)=&\frac{1}{\sum_{k=1}^n h_k} \left[\sum_{k=1}^n \int_0^{h_k} \mathbf{x}_kdZ \right]\\
=&\frac{1}{\sum_{k=1}^n h_k} \left[\sum_{k=1}^n h_k \left[ \mathbf{x}_k^{(0)} + \frac{h_k}{2} \mathbf{x}_k^{(1)} + \frac{h_k^2}{6} \mathbf{x}_k^{(2)} + \frac{h_k^3}{24} \mathbf{x}_k^{(3)} \right] \right],\\
\bar{\mathbf{b}}(s)=&\frac{1}{\sum_{k=1}^n h_k} \left[\sum_{k=1}^n \int_0^{h_k} \mathbf{b}_k(s,Z) dZ \right].
\end{aligned}
\end{equation*}

\textbf{Case 2. Traction boundary conditions}

We suppose the applied traction $\mathbf{q}_k(s,Z_k)$ has been specified on the region $\partial\kappa_k^{(q)}=\partial\Omega^{(q)}\times[0,h_k]$ $(k=1,\cdots,n)$, which yields another two kinds of edge boundary conditions
\begin{equation}
\begin{aligned}
{\mathbb{S}_1^{(0)}}^T \mathbf{N} = \mathbf{q}_1(s,0), \ \ \ \bar{\mathbb{S}}^T \mathbf{N}=\bar{\mathbf{q}},\ \ \mathrm{on} \ \partial\Omega^{(q)},
\end{aligned}
\label{eq:28}
\end{equation}
where
\begin{equation*}
\begin{aligned}
\bar{\mathbb{S}}=&\frac{1}{\sum_{k=1}^n h_k} \left[\sum_{k=1}^n \int_0^{h_k} \mathbb{S}_kdZ \right]\\
=&\frac{1}{\sum_{k=1}^n h_k} \left[\sum_{k=1}^n h_k \left[ \mathbb{S}_k^{(0)} + \frac{h_k}{2} \mathbb{S}_k^{(1)} + \frac{h_k^2}{6} \mathbb{S}_k^{(2)} \right] \right],\\
\bar{\mathbf{q}}=&\frac{1}{\sum_{k=1}^n h_k} \left[\sum_{k=1}^n \int_0^{h_k} \mathbf{q}_kdZ \right].
\end{aligned}
\end{equation*}

\textbf{Case 3. Moment boundary conditions}

Besides the traction boundary conditions (\ref{eq:28}), we can also adopt the following moment boundary condition with the traction $\mathbf{q}_k(s,Z_k)$ applied on $\partial\kappa_k^{(q)}=\partial\Omega^{(q)}\times[0,h_k]$ $(k=1,\cdots,n)$
\begin{equation}
\begin{aligned}
\mathbf{M}_m &= \sum_{k=1}^n \int_0^{h_k} (\mathbf{x}_k-\bar{\mathbf{x}}) \wedge \left\{\mathbb{S}_k\right\}^T \mathbf{N} dZ= \sum_{k=1}^n \int_0^{h_k} (\mathbf{x}_k-\bar{\mathbf{x}}) \wedge \mathbf{q}_k dZ=\mathbf{m}_m,\ \ \ \mathrm{on}\ \partial\Omega^{(q)}.
\end{aligned}
\label{eq:29}
\end{equation}
In Eq. (\ref{eq:29}), the first two components are the bending moments respect to the $X_k$- and $Y_k$-axes, and the third component represents the twisting moment respects to the $Z_k$-axis. Some further discussions on the edge boundary conditions can be found in \citet{dai2014}.

In summary, the 2D vector plate equation (\ref{eq:26}) together with some suitable edge boundary conditions (\ref{eq:27})-(\ref{eq:29}) formulate the plate equation system of the current theory. Some further discussions on the edge boundary conditions can be found in \citet{dai2014}.

\noindent\textbf{Remarks:}
\begin{itemize}

\item By substituting the iterative relations (\ref{eq:24}), (\ref{eq:25}) and (\ref{eq:A1}) of \ref{app:a} into (\ref{eq:26}), the plate equation only involves the components of the unknown vector $\mathbf{x}_1^{(0)}$. In that case, the plate equation will have a very length expression, which is not convenient to be tackled. When solving some concrete problems, the plate equation may be further simplified. For some problems, the plate equation with the asymptotic order $O(h)$ (i.e., only the first two terms of $\bar{\mathbb{S}}$ are substituted in (\ref{eq:26})) can already provide accurate predictions on the growth-induced deformations of the hyperelastic plates \cite{du2020,wang2022}. For some other problems, the magnitudes of the displacement components can be identified in advance, thus the asymptotic analyses may be conducted to simplify the plate equations \cite{wang2018,du2022}. Regarding this issue, some examples are presented in Section \ref{sec:3}.

\item The unknown vector $\mathbf{x}_1^{(0)}$ can be viewed as the parametric equation of the deformed surface $\mathcal{S}$ of the bottom face $\Omega_1^-$ of the plate. We denote $\{E,F,G\}$ and $\{L,M,N\}$ as the fundamental quantities of the first and second fundamental forms of $\mathcal{S}$, respectively. In terms of the position vector $\mathbf{v}=\mathbf{x}_1^{(0)}=\{x_1^{(0)},y_1^{(0)},z_1^{(0)}\}$, we have
    \begin{equation}
    \begin{aligned}
    &E=\mathbf{v}_{,X} \cdot \mathbf{v}_{,X},\ \ F=\mathbf{v}_{,X} \cdot \mathbf{v}_{,Y},\ \ G=\mathbf{v}_{,Y} \cdot \mathbf{v}_{,Y},\\
    &L=\mathbf{v}_{,XX} \cdot \tilde{\mathbf{N}},\ \ M=\mathbf{v}_{,XY} \cdot \tilde{\mathbf{N}},\ \ N=\mathbf{v}_{,YY} \cdot \tilde{\mathbf{N}},\\
    \end{aligned}
    \label{eq:30}
    \end{equation}
    where $\tilde{\mathbf{N}}=\mathbf{v}_N/|\mathbf{v}_N|$ and $\mathbf{v}_N=\mathbf{v}_{,X}\wedge\mathbf{v}_{,Y}$. If the derivative terms of $\mathbf{x}_1^{(0)}$ in the plate equation can be replaced by the fundamental quantities $\{E,F,G\}$ and $\{L,M,N\}$, the lengthy expression of the plate equation can be simplified significantly, which will also facilitate the following work of shape-programming (cf. \citet{li2022} and the examples in Section \ref{sec:4}).

\item Recently, a more rigorous approach has been proposed to establish the edge boundary conditions in finite-strain plate (or rod, shell) theories \cite{yu2020,chen2021}. In this approach, the weak form of the plate (or rod, shell) equation is substituted into the variational formulation of the 3D problem. Then, through some elaborate asymptotic analyses, the traction and moment boundary conditions can be proposed in a consistent manner.

\end{itemize}

\section{Applications of the multi-layered hyperelastic plate theory}
\label{sec:3}

The plate equation system of the multi-layered hyperelastic plate theory has been derived in the previous section. In this section, this plate theory will be applied to studied the growth-induced deformations and instabilities of some typical multi-layered plate samples. To obtain some concrete results, we assume that the layers in the plate samples are made of incompressible neo-Hookean materials, which have the elastic strain-energy functions $\phi_k(\mathbb{F}_k,\mathbb{G}_k)=J_{G_k}\tilde{\phi}_{k}\left(\mathbb{A}_k\right)=J_{G_k}C_k\left[\mathrm{tr}(\mathbb{A}_k{\mathbb{A}_k}^T)-3\right]$ $(k=1,\cdots,n)$. The thickness parameters of the layers in the plate are denoted as $h_k=\beta_kh$ $(k=1,\cdots,n)$, where $h$ is a small parameter representing the characteristic thickness of the layers.


\subsection{Growth-induced plane-strain deformations of multi-layered hyperelastic plates}
\label{sec:3.1}

\begin{figure}[htp]
\centering \includegraphics[width=0.6\textwidth]{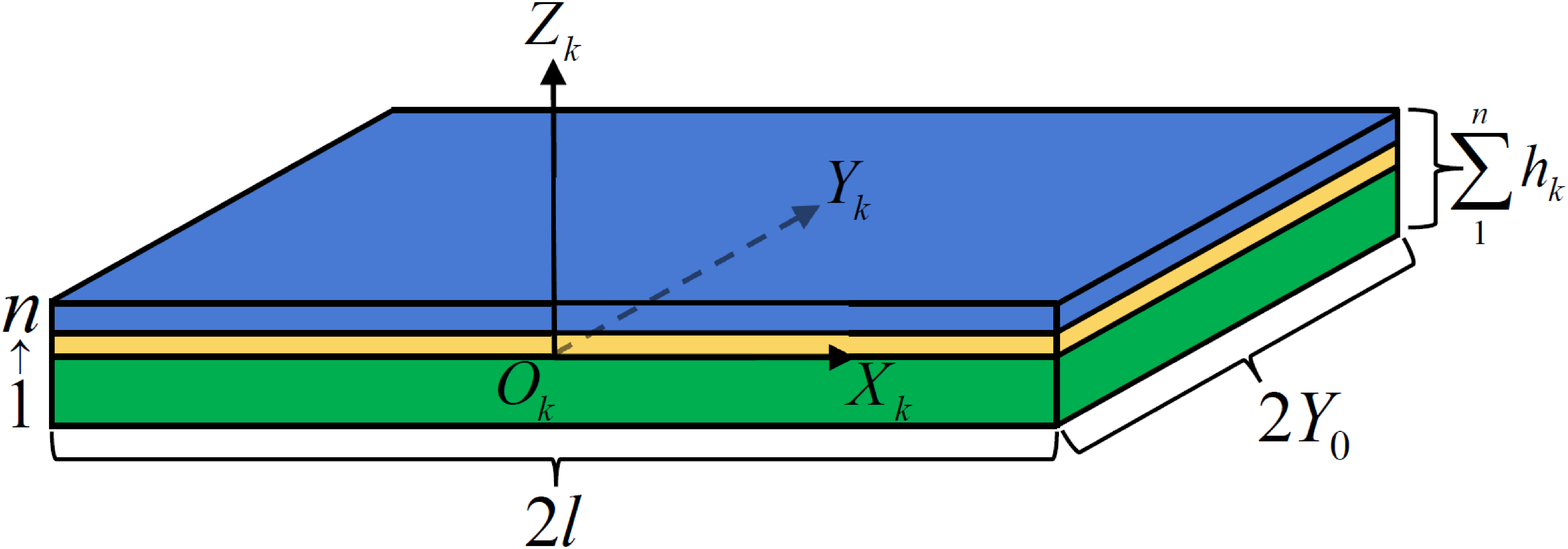}\\(a)\\\vspace{0.5cm}
\centering \includegraphics[width=0.7\textwidth]{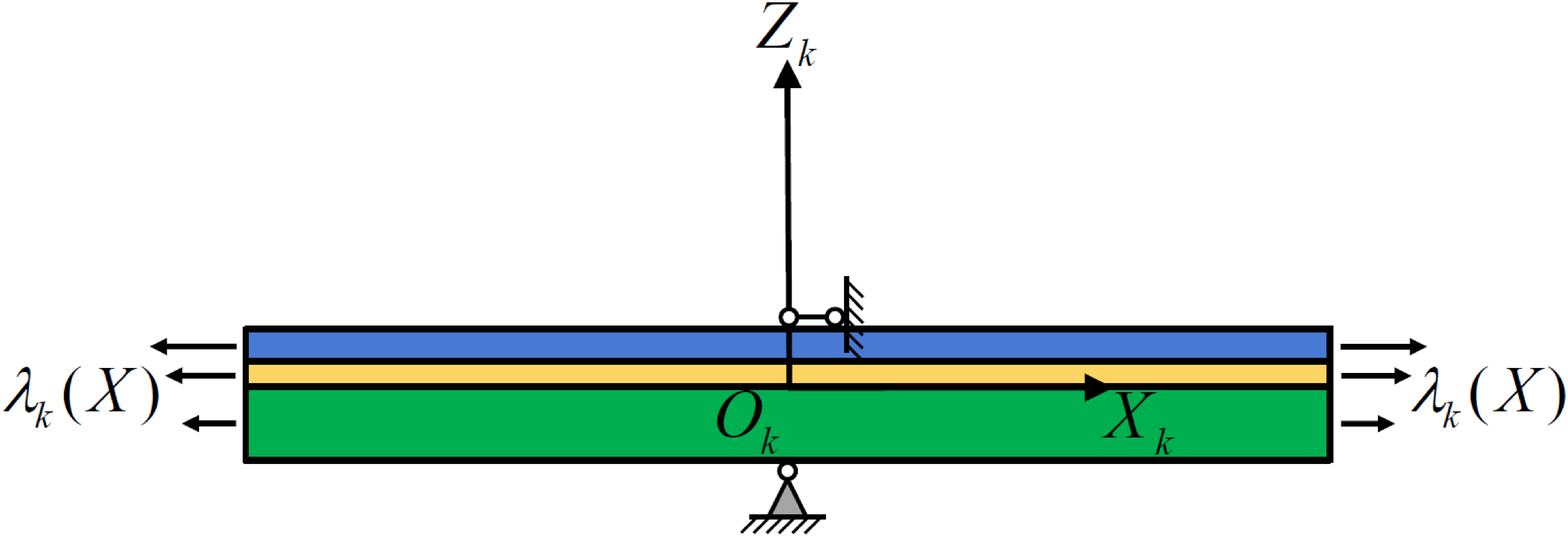}\\(b)\\
\caption{(a) The reference configuration of a multi-layered hyperelastic plate; (b) illustration of the boundary conditions and the growth fields in the plane-strain problems.}
\label{fig:3}
\end{figure}

In the first example, we study the growth-induced plane-strain deformations of multi-layered hyperelastic plates. The reference configuration of a plate sample and the local coordinate system are shown in Fig. \ref{fig:3}a. Without loss of the generality, the half length $l$ of the plate is set to be $1$. We suppose there is no external loads applied on the surface of the plate. The growth in each layer only occurs along the $X_k$-axis, which is represented by the growth function $\lambda_k(X)$ $(k=1,\cdots,n)$ (cf. Fig. \ref{fig:3}b). Along the $Y_k$-axis, the plate undergoes plane-strain deformations. To remove the rigid body motion, the position of the middle point on the bottom face is fixed and the horizontal displacement of the middle point on the top face is restricted. In our previous works \cite{du2020,du2022}, some analytical results of bending deformations for the plane-strain problem of multi-layered hyperelastic plates have been derived, where the growth functions $\lambda_k(X)$ $(k=1,\cdots,n)$ were chosen to be constants. In this example, $\lambda_k(X)$ are arbitrary functions.

To simplify the plate equation system, we only consider the plate equation with the asymptotic order $O(h)$ (i.e., the terms of $O(h^2)$ are neglected). For the plane-strain problem being studied ($y_1^{(0)}=Y$), it can be obtained that the deformed configuration $\mathcal{S}$ of the bottom face $\Omega_1^-$ has the fundamental quantities $F=M=N=0$, $G=1$ and
\begin{equation*}
\begin{aligned}
E = {x_{1,X}^{(0)}}^2 + {z_{1,X}^{(0)}}^2,\ \ L=\frac{-z_{1,X}^{(0)} x_{1,XX}^{(0)} + x_{1,X}^{(0)} z_{1,XX}^{(0)}}{\sqrt{{x_{1,X}^{(0)}}^2 + {z_{1,X}^{(0)}}^2}}.
\end{aligned}
\end{equation*}
To simplify the plate equation, we further denote $\lambda_k(X) = \lambda(X) + h \Delta\lambda_k(X)$, which implies that the differences of the growth functions have $O(h)$. After these manipulations, the plate equation (\ref{eq:26}) for the current plane-strain problem has a relatively simple form, which can be integrated once with respect to $X$ and yields that
\begin{equation}
\begin{aligned}
&-\pmb{\gamma}+\frac{2h\Lambda_n^{(1)}(E^2-\lambda^4)}{E^2 \lambda}\mathbf{v}_{,X}\\
&-\frac{h^2}{E^{\frac{7}{2}}\lambda}\bigg[\frac{2}{\lambda}\big[E^2L\lambda^2\Lambda_n^{(1)}\Lambda_n^{(2)}+E^{\frac{3}{2}}\left(E^2+3\lambda^4\right)g_{\lambda}^{(1)}+L\lambda^6\big(\Lambda_n^{(3)}+\Lambda_n^{(4)}\big)\big]\mathbf{v}_{,X}\\
&-\frac{1}{E^\frac{1}{2}}\left[E^2\Lambda_n^{(1)}\Lambda_n^{(2)}+\lambda^4\left(\Lambda_n^{(3)}+\Lambda_n^{(5)}\right)\right]\mathbf{v}_{N}\bigg]+O(h^3)=\mathbf{0},
\end{aligned}
\label{eq:31}
\end{equation}
where
\begin{equation*}
\begin{aligned}
&g_{\lambda}^{(1)}=\sum_{k=1}^n\beta_kC_k\Delta\lambda_k,\ \ \Lambda_n^{(1)}=\sum_{k=1}^n\beta_kC_k,\ \ \Lambda_n^{(2)}=\sum_{k=1}^n\beta_k,\ \ \Lambda_n^{(3)}=\sum_{k=1}^n{\beta_k}^2C_k,\\
&\Lambda_n^{(4)}=\sum_{k=1}^{n-1}\sum_{p=k+1}^n\beta_k\beta_p(3C_p-C_k),\ \ \Lambda_n^{(5)}=\sum_{k=1}^{n-1}\sum_{p=k+1}^n\beta_k\beta_p(3C_k -C_p).
\end{aligned}
\end{equation*}
Eq. (\ref{eq:31}) contains two independent equations. The integration constant $\pmb{\gamma}$ represents the resultant force on the cross-section, which should be zero due to the traction-free boundary conditions at the two ends \cite{du2020,du2022}. Based on Eq. (\ref{eq:31}), the average traction boundary condition (\ref{eq:28})$_2$ is automatically satisfied. From the moment boundary condition (\ref{eq:29}), we have another equation
\begin{equation}
\begin{aligned}
&2h^2\Lambda_n^{(6)}\left(1-\frac{{\lambda}^4}{E^2}\right)-h^3\left[\frac{2g_{\lambda}^{(2)}}{\Lambda_n^{(2)}\lambda}-\frac{\Lambda_n^{(7)}\lambda L}{3E^{\frac{3}{2}}}-\frac{g_{\lambda}^{(3)}{\lambda}^3}{\Lambda_n^{(2)}E^2}-\frac{\Lambda_n^{(8)}{\lambda}^5L}{3E^{\frac{7}{2}}}\right]=0.
\end{aligned}
\label{eq:32}
\end{equation}
where
\begin{equation*}
\begin{aligned}
g_{\lambda}^{(2)}=&\sum_{k=1}^{n-1}\sum_{p=k+1}^n\left[\beta_k\beta_pC_k(\Delta\lambda_p-\Delta\lambda_k)\left(\beta_p+2\sum_{q=p+1}^n \beta_q\right)\right]\\
&+\sum_{k=2}^n\sum_{p=1}^{k-1}\left[\beta_k\beta_pC_k\left(\beta_p(\Delta\lambda_p-\Delta\lambda_k)+2\sum_{q=p+1}^{k-1} \beta_q(\Delta\lambda_q-\Delta\lambda_k)\right)\right],\\
g_{\lambda}^{(3)}=&2\sum_{k=1}^{n-1}\sum_{p=k+1}^n[\beta_k\beta_p[C_p[4\beta_p\Delta\lambda_p+\beta_k(\Delta\lambda_k+3\Delta\lambda_p)]-C_k[4\beta_k\Delta\lambda_k\\
&+\beta_p(3\Delta\lambda_k+\Delta\lambda_p)]]]\\
&-\sum_{k=1}^{n-2}\sum_{p=k+1}^{n-1}\sum_{q=p+1}^n[4\beta_k\beta_p\beta_q[C_k(3\Delta\lambda_k+\Delta\lambda_p)]-C_q(\Delta\lambda_p+3\Delta\lambda_q)],
\end{aligned}
\end{equation*}
\begin{equation*}
\begin{aligned}
\Lambda_n^{(6)}=&\sum_{k=1}^{n-1}\sum_{p=k+1}^n C_k\beta_k\beta_p-\sum_{k=2}^n\sum_{p=1}^{k-1}C_k\beta_k\beta_p,\\
\Lambda_n^{(7)}=&2\sum_{k=1}^n{\beta_k}^3C_k+\sum_{k=1}^{n-1}\sum_{p=k+1}^n\left[\beta_k\beta_p(\beta_pC_k+2\beta_kC_k+4\beta_kC_p+5\beta_pC_p)\right]\\
&+2\sum_{k=1}^{n-2}\sum_{p=k+1}^{n-1}\sum_{q=p+1}^n\left[\beta_k\beta_p\beta_q(C_k+C_p+4C_q)\right],\\
\Lambda_n^{(8)}=&2\sum_{k=1}^n{\beta_k}^3C_k+\sum_{k=1}^{n-1}\sum_{p=k+1}^n\left[\beta_k\beta_p\left[(20\beta_k+7\beta_p)C_p-(14\beta_k+\beta_p)C_k\right]\right]\\
&-2\sum_{k=1}^{n-2}\sum_{p=k+1}^{n-1}\sum_{q=p+1}^n\left[\beta_k\beta_p\beta_q(C_k+13C_p-20C_q)\right],\\
\end{aligned}
\end{equation*}
Eq. (\ref{eq:32}) is proposed at the two ends $X=\pm1$ of the plate. However, through a simple force analysis, it is known that this equation should also be satisfied in the whole region $-1\leq X\leq1$.

The fundamental quantities $E$ and $L$ can be viewed as two unknowns of the plate equation (\ref{eq:31}). By solving (\ref{eq:31}) together with (\ref{eq:32}) through a regular perturbation method, we obtain
\begin{equation}
\begin{aligned}
&E = {\lambda}^2 + \frac{h}{\Lambda_n^{(1)}}\left[2\lambda g_{\lambda}^{(1)}+ \frac{3\lambda g_{\lambda}^{(4)}(\Lambda_n^{(1)}\Lambda_n^{(2)}+\Lambda_n^{(3)}+\Lambda_n^{(4)})}{\Lambda_n^{(9)}+4\Lambda_n^{(10)}+12\Lambda_n^{(11)}+24\Lambda_n^{(12)}}\right],\\
&L = \frac{6\lambda g_{\lambda}^{(4)}}{\Lambda_n^{(9)}+4\Lambda_n^{(10)}+12\Lambda_n^{(11)}+24\Lambda_n^{(12)}},
\end{aligned}
\label{eq:33}
\end{equation}
where
\begin{equation*}
\begin{aligned}
g_{\lambda}^{(4)}=&\sum_{k=1}^{n-1}\sum_{p=k+1}^n\beta_k\beta_p\left(\beta_k+\beta_p+2\sum_{q=k+1}^{p-1}\beta_q\right)C_kC_p(\Delta\lambda_k- \Delta\lambda_p),\\
\Lambda_n^{(9)}=&\sum_{k=1}^n{\beta_k}^4{C_k}^2,\\
\Lambda_n^{(10)}=&\sum_{k=1}^{n-1}\sum_{p=k+1}^n\left[\left({\beta_k}^3\beta_p+\frac{3}{2}{\beta_k}^2{\beta_p}^2+{\beta_k}{\beta_p}^3\right)C_kC_p \right],\\
\end{aligned}
\end{equation*}
\begin{equation*}
\begin{aligned}
\Lambda_n^{(11)}=&\sum_{k=1}^{n-2}\sum_{p=k+1}^{n-1}\sum_{q=p+1}^n\left[\left({\beta_k}^2\beta_p\beta_q+{\beta_k}{\beta_p}^2\beta_q+{\beta_k}{\beta_p}{\beta_q}^2\right)C_kC_q\right],\\
\Lambda_n^{(12)}=&\sum_{k=1}^{n-3}\sum_{p=k+1}^{n-2}\sum_{q=p+1}^{n-1}\sum_{r=q+1}^n{\beta_k}\beta_p\beta_q\beta_rC_kC_r.
\end{aligned}
\end{equation*}
Based on the asymptotic expressions of $E$ and $L$ given in (\ref{eq:33}), the position vector $\mathbf{v}=\mathbf{x}_1^{(0)}$ can be calculated through a conventional approach, which represents the shape of the bottom surface $\mathcal{S}$ of the plate in the current configuration. By further using the iterative relations (\ref{eq:A1})-(\ref{eq:A3}) given in \ref{app:a}, the whole configuration of the plate sample can be recovered.

\begin{figure}[htp]
\begin{minipage}{0.49\textwidth}
\centering \includegraphics[width=0.95\textwidth]{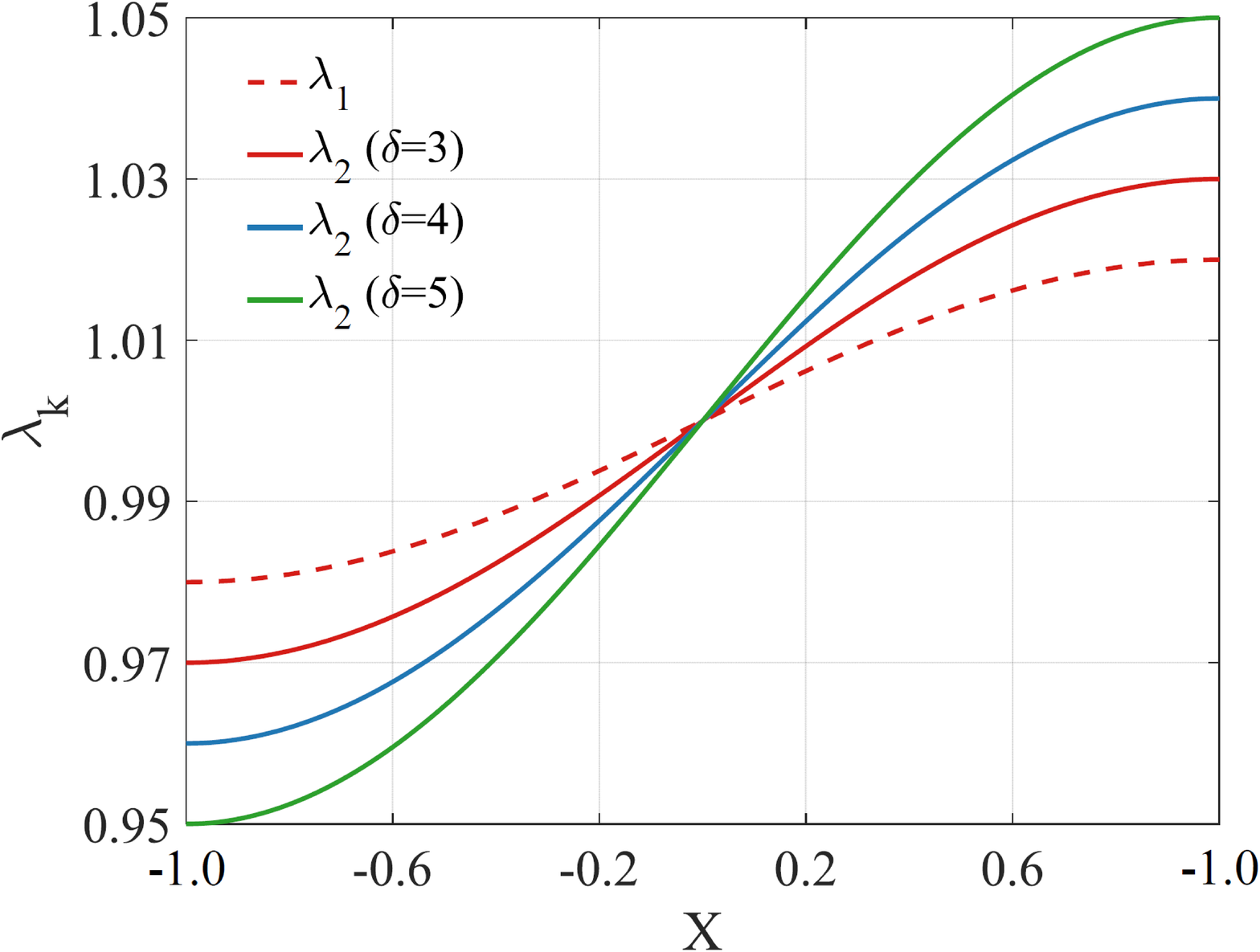}\\(a)\vspace{0.2cm}
\end{minipage}
\begin{minipage}{0.49\textwidth}
\centering \includegraphics[width=0.95\textwidth]{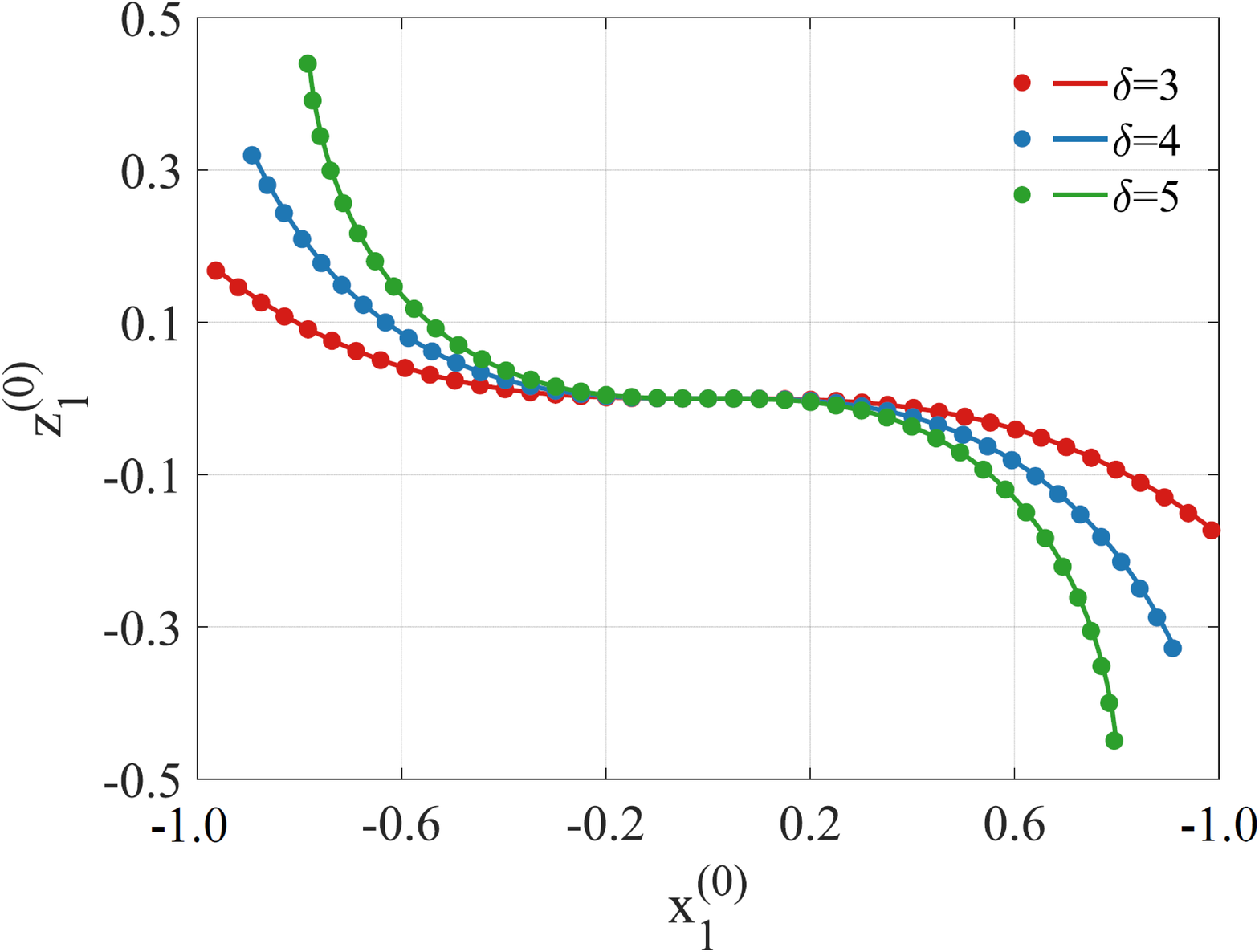}\\(b)\vspace{0.2cm}
\end{minipage}\\
\begin{minipage}{0.49\textwidth}
\centering \includegraphics[width=0.95\textwidth]{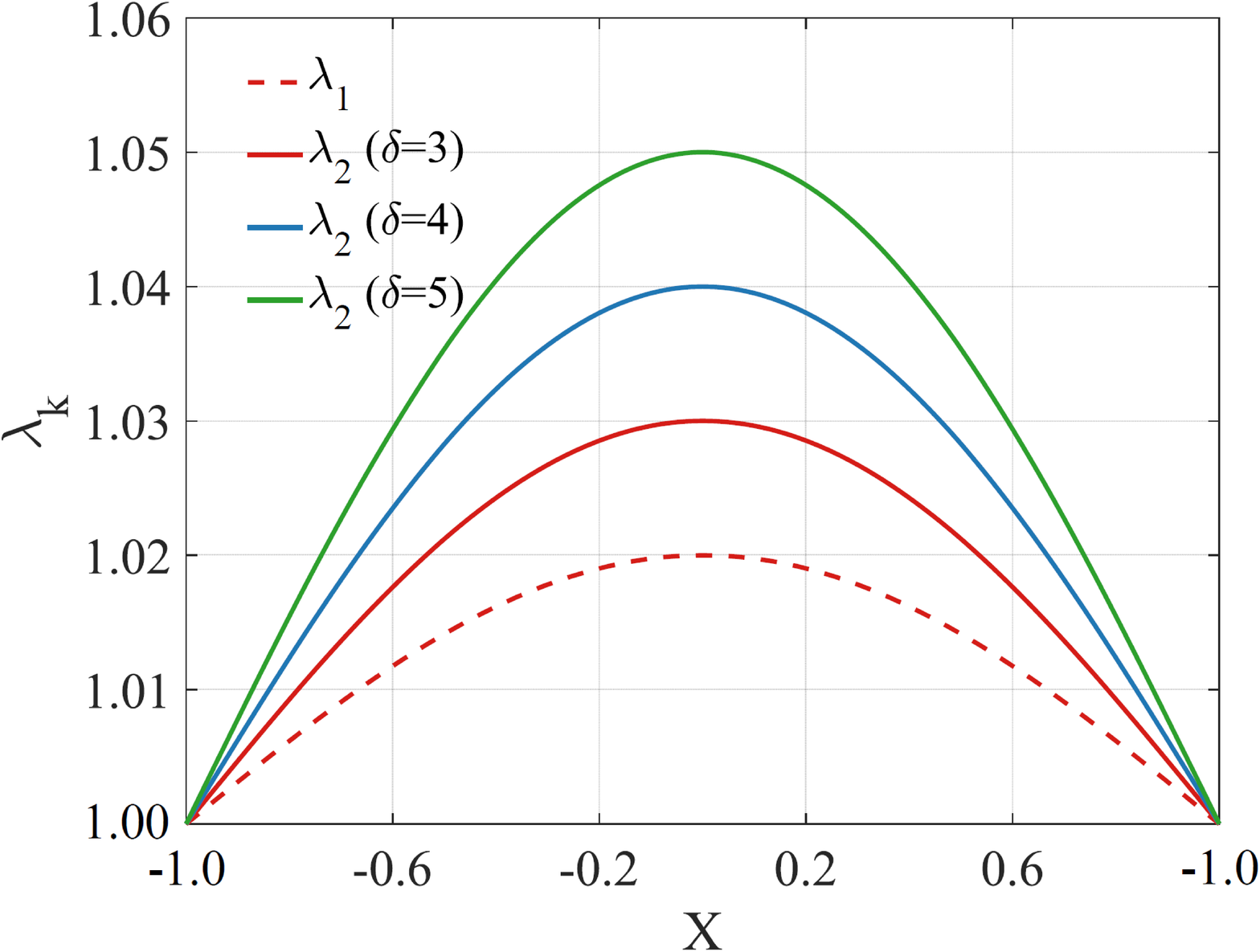}\\(c)\vspace{0.2cm}
\end{minipage}
\begin{minipage}{0.49\textwidth}
\centering \includegraphics[width=0.95\textwidth]{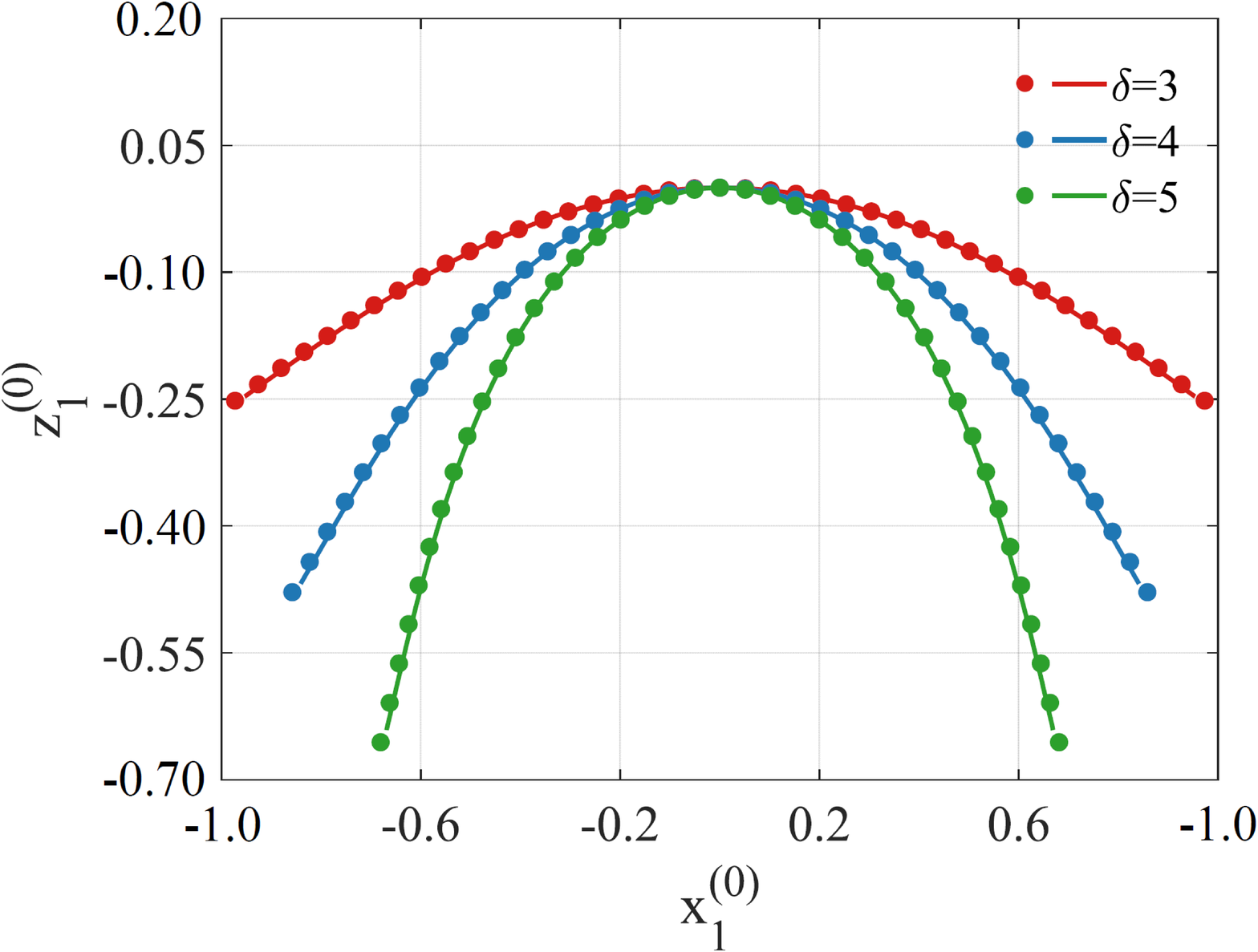}\\(d)\vspace{0.2cm}
\end{minipage}\\
\begin{minipage}{0.49\textwidth}
\centering \includegraphics[width=0.95\textwidth]{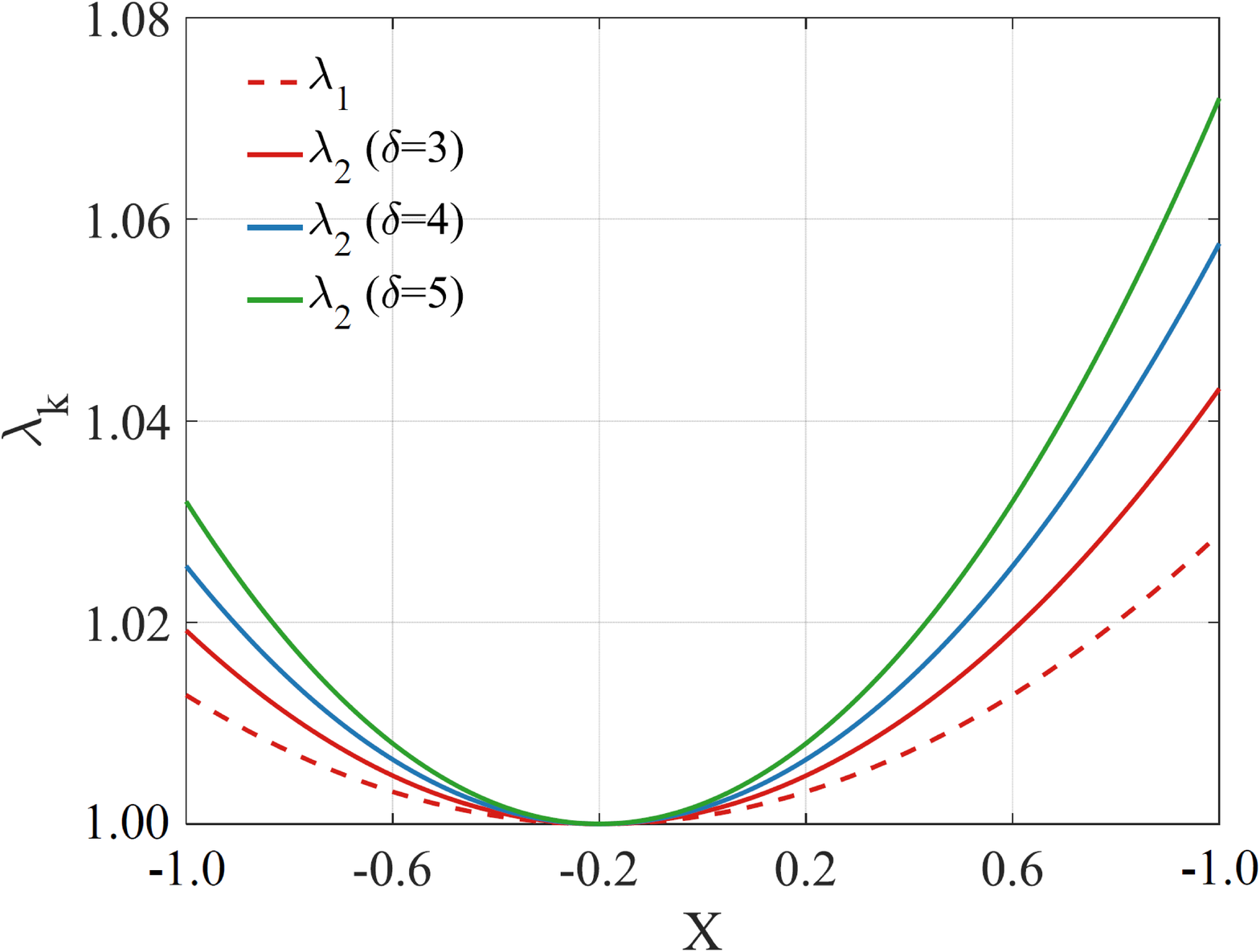}\\(e)\vspace{0.2cm}
\end{minipage}
\begin{minipage}{0.49\textwidth}
\centering \includegraphics[width=0.95\textwidth]{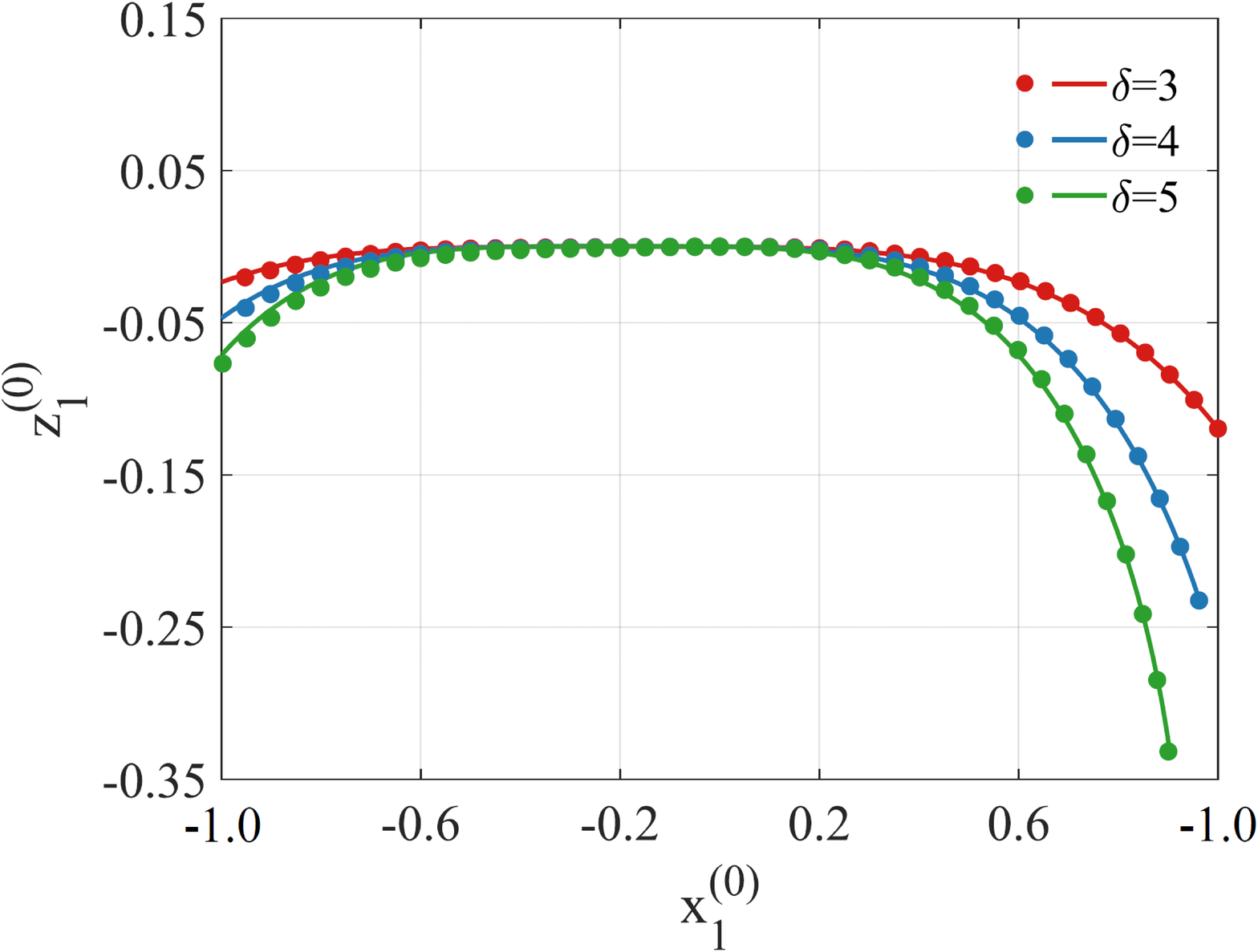}\\(f)\vspace{0.2cm}
\end{minipage}\\
\caption{The growth functions and deformed shapes of the bottom surface $\mathcal{S}$ of the bilayer hyperelastic plates: (a)-(b) case 1; (c)-(d) case 2; (e)-(f) case 3. (Solid curves: Analytical results; Dots: FE simulation results.)}
\label{fig:4}
\end{figure}

To demonstrate the efficiency of the analytical results (\ref{eq:33}), we study the growth-induced deformations of some bilayer hyperelastic plates, where the expressions of $E$ and $L$ are given by
\begin{equation}
\begin{aligned}
E=&\lambda^2+\Big[2h\big[\beta_1C_1\left[{\beta_1}^3C_1+\beta_2C_2\left(6{\beta_1}^2+9\beta_1\beta_2+4{\beta_2}^2\right)\right]\Delta\lambda_1\\
&+\beta_2C_2\left[{\beta_2}^3C_2-{\beta_1}^2C_1\left(2\beta_1+3\beta_2\right)\right]\Delta\lambda_2\big]\lambda\Big]/\Big[{\beta_1}^4{C_1}^2+ {\beta_2}^4{C_2}^2\\
&+2\beta_1\beta_2C_1C_2\big(2{\beta_1}^2+3\beta_1\beta_2+2{\beta_2}^2\big)\Big],\\
L=&\left[6\beta_1\beta_2C_1C_2(\beta_1+\beta_2)(\Delta\lambda_1-\Delta\lambda_2)\lambda\right]/\big[{\beta_1}^4{C_1}^2+{\beta_2}^4{C_2}^2\\
&+2\beta_1\beta_2C_1C_2\left(2{\beta_1}^2+3\beta_1\beta_2+2{\beta_2}^2\right)\big].
\end{aligned}
\label{eq:34}
\end{equation}
For the purpose of illustration, we consider three cases with the growth functions and the material and geometrical parameters given in Eqs. (\ref{eq:35})-(\ref{eq:37}). The curves of the growth functions are also shown in Figs. \ref{fig:4}a, \ref{fig:4}c and \ref{fig:4}e.
\begin{itemize}
\item \emph{Case 1}: $C_1/C_2=1$, $\beta_1/\beta_2=1$, $h=0.01$
\begin{equation}
\begin{aligned}
\left\{\begin{array}{l}\vspace{1.5ex}
\lambda_1 (X) = \displaystyle{1+\frac{1}{50}\mathrm{sin} \left(\frac{\pi X}{2} \right)},\\
\lambda_2 (X) = \displaystyle{1+\frac{\delta}{100} \mathrm{sin} \left(\frac{\pi X}{2} \right)},\ \ \delta=3,4,5.
\end{array}\right.
\end{aligned}
\label{eq:35}
\end{equation}

\item \emph{Case 2}: $C_1/C_2=1/5$, $\beta_1/\beta_2=1$, $h=0.01$
\begin{equation}
\begin{aligned}
\left\{\begin{array}{l}\vspace{1.5ex}
\lambda_1 (X) = \displaystyle{1+\frac{1}{50}\mathrm{cos} \left(\frac{\pi X}{2} \right)},\\
\lambda_2 (X) = \displaystyle{1+\frac{\delta}{100} \mathrm{cos} \left(\frac{\pi X}{2} \right)},\ \ \delta=3,4,5.
\end{array}\right.
\end{aligned}
\label{eq:36}
\end{equation}

\item \emph{Case 3}: $C_1/C_2=1$, $\beta_1/\beta_2=1/5$, $h_1=0.002$, $h_2=0.01$
\begin{equation}
\begin{aligned}
\left\{\begin{array}{l} \vspace{1.5ex}
\lambda_1 (X) = \displaystyle{1+\frac{1}{250} (5X+1)^2},\\
\lambda_2 (X) = \displaystyle{1+\frac{\delta}{500} (5X+1)^2},\ \ \delta=3,4,5.
\end{array}\right.
\end{aligned}
\label{eq:37}
\end{equation}
\end{itemize}

From the analytical results (\ref{eq:34}), we can determine the deformed shapes of the bottom surface $\mathcal{S}$ of the plates for the three cases, which are shown in Figs. \ref{fig:4}b, \ref{fig:4}d and \ref{fig:4}f. In order to verify the accuracy of analytical results, we also conduct finite element simulations on the growth-induced deformations of these bilayer hyperelastic plates. The user subroutine UMAT in ABAQUS is utilized for the numerical simulations, in which the compressible neo-Hookean constitutive relation $\tilde{\phi}_{k}(\mathbb{A}_k)=C_k\left[\mathrm{tr}(\mathbb{A}_k{\mathbb{A}_k}^T)-3\right]+\left[\mathrm{Det}(\mathbb{A}_k)-1\right]^2/D_k$ $(k=1,2)$ is adopted. The elastic deformation tensor is defined as $\mathbb{A}_k=\mathbb{F}_k{\mathbb{G}_k}^{-1}$, where the total deformation gradient tensor $\mathbb{F}_k$ is input from the finite element program and the growth functions in $\mathbb{G}_k$ are input as the state variables. The parameters $C_k$ and $D_k$ are related to the Young' modulus $E_k$ and the Poisson' ratio $\nu$ through $C_k=E_k/[4(1+\nu)]$ and $D_k=6(1-2\nu)/E_k$. The Poisson' ratio $\nu$ is chosen as $0.495$ to approximate the elastic incompressibility of soft materials. All the faces of the plate are set to be traction free and the rigid body motion of the plate is removed. For plane-strain problems, only one section of the plate sample perpendicular to the $Y_k$-axis needs to be taken into account. This section is meshed by using the CPE8H elements (8-node biquadratic plane strain quadrilateral hybrid elements). For all the three cases introduced in (\ref{eq:35})-(\ref{eq:37}), the element size is chosen to be $0.001\times0.001$. The obtained numerical simulation results are also shown in Figs. \ref{fig:4}b, \ref{fig:4}d and \ref{fig:4}f.

\begin{figure}[htp]
\begin{minipage}{0.49\textwidth}
\centering \includegraphics[width=0.93\textwidth]{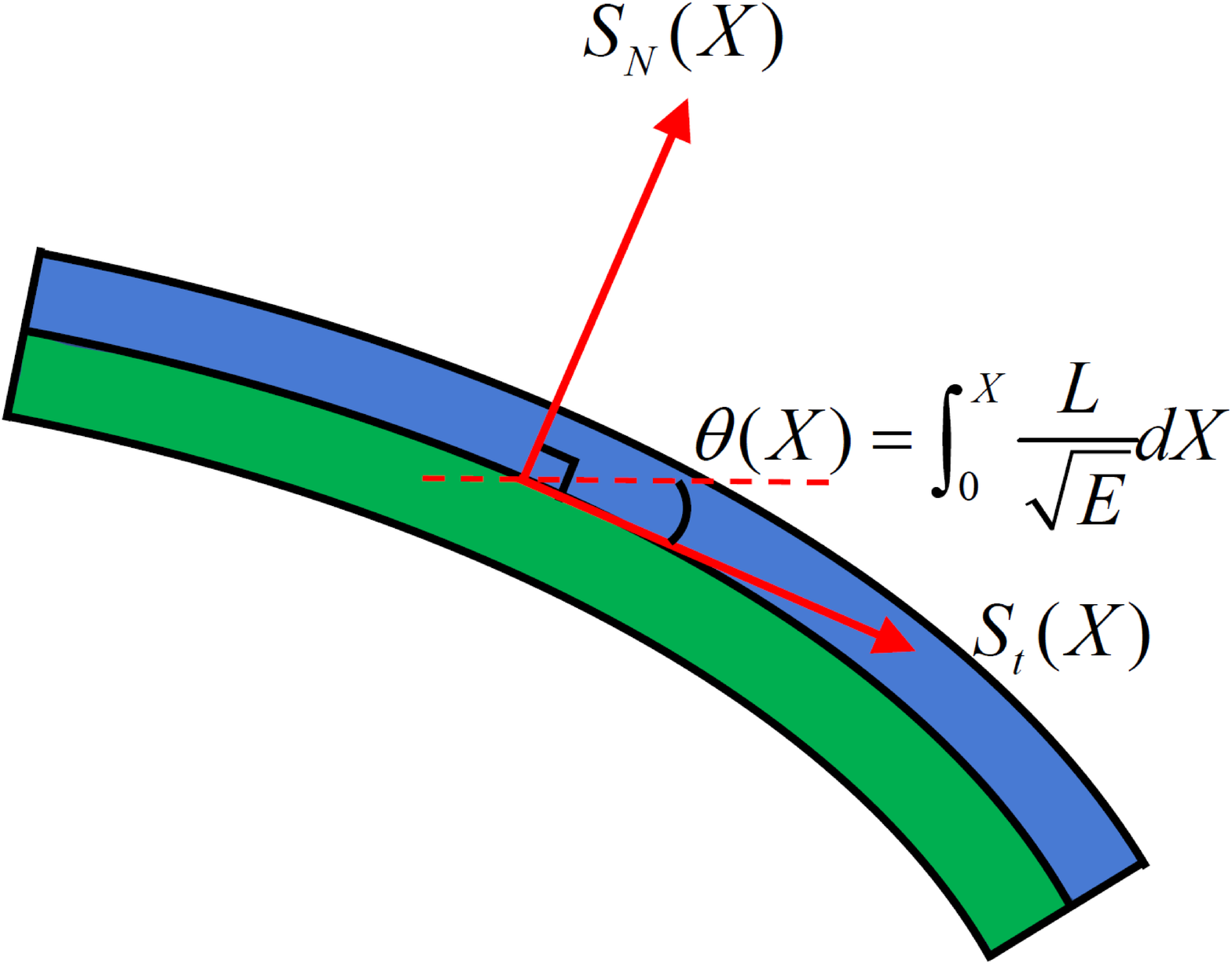}\\(a)\vspace{0.2cm}
\end{minipage}
\begin{minipage}{0.49\textwidth}
\centering \includegraphics[width=0.93\textwidth]{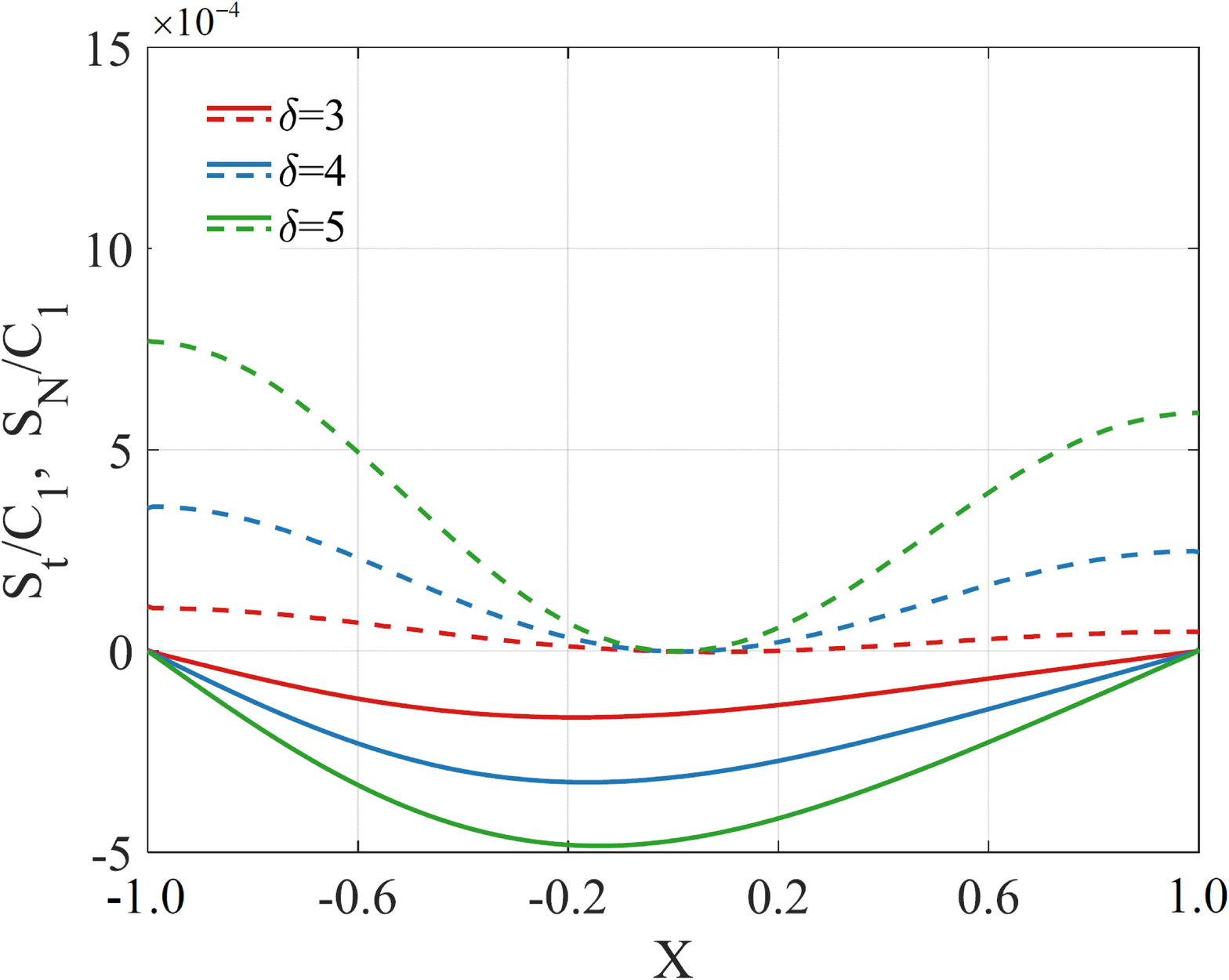}\\(b)\vspace{0.2cm}
\end{minipage}\\
\begin{minipage}{0.49\textwidth}
\centering \includegraphics[width=0.93\textwidth]{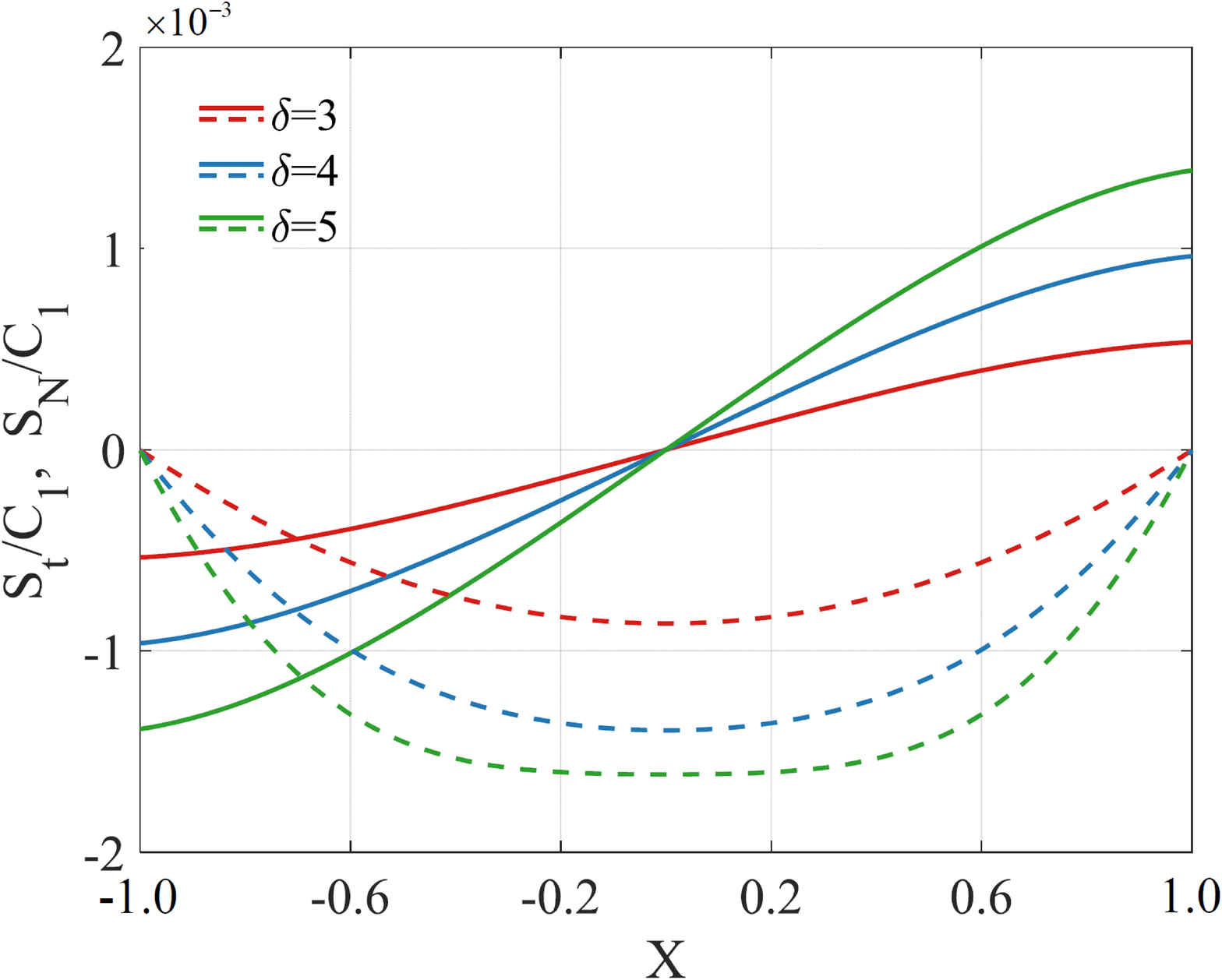}\\(c)\vspace{0.2cm}
\end{minipage}
\begin{minipage}{0.49\textwidth}
\centering \includegraphics[width=0.93\textwidth]{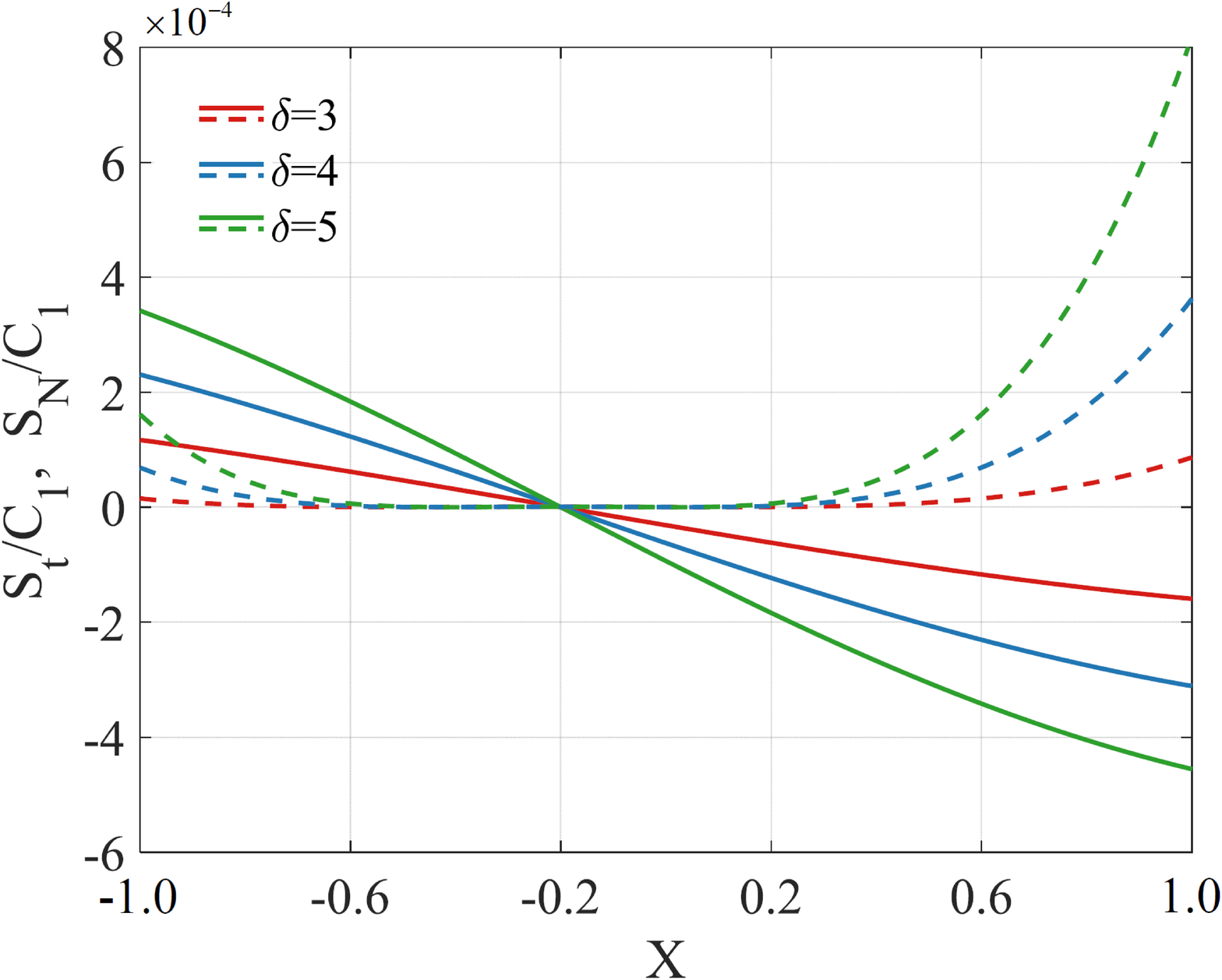}\\(d)\vspace{0.2cm}
\end{minipage}\\
\caption{Distributions of the normal stress $S_N$ and the shear stress $S_t$ on the interface between the two layers corresponding to the growth functions given in (\ref{eq:35})-(\ref{eq:37}): (a) illustration of $S_N$ and $S_t$; (b) case 1; (c) case 2; (d) case 3. (Solid curves: evolution curves of $S_t$; Dashed curves: evolution curves of $S_N$.)}
\label{fig:5}
\end{figure}

Through some comparisons, we found that the analytical results match the finite element simulation results quite well. Thus, the efficiency of the analytical results (\ref{eq:34}) can be verified. It is found that accompanying the increase of the growth function $\lambda_2(X)$ in the 2nd layer, the plate exhibits larger bending deformations, which should be attributed to the increases of residual stresses in the plate. Based on the analytical solutions $x_1^{(0)}$ and $z_1^{(0)}$ of the plate equation, the whole configuration of the plate can be recovered by virtue of the iterative relations given in \ref{app:a}. Furthermore, the residual stresses in the plate sample can be determined by using the expressions of the stress tensors given in (\ref{eq:17}). In Fig. \ref{fig:5}, the distributions of the normal stress $S_N$ and the shear stress $S_t$ on the interface between the two layers are shown. As the growth functions given in (\ref{eq:35})-(\ref{eq:37}) are not constants, both the normal and shear stress values change continuously along the interface, which are different from the cases studied in \citet{du2020,du2022}.

\subsection{Growth-induced instabilities of a bilayer hyperelastic plate}
\label{sec:3.2}

In Section \ref{sec:2}, the general multi-layered plate theory has been established without any assumptions on the magnitudes of displacement components. Under some specific loading conditions or boundary restrictions, the orders of displacement components may be identified in advance. In that case, the plate equation can be simplified through some further asymptotic analyses. In this subsection, under the assumption of plane-strain deformation, it will be shown that a FvK-type multi-layered plate model can be derived from the full-form plate equation (\ref{eq:26}). This FvK-type plate model will be used to study the growth-induced instabilities of a bilayer hyperelastic plate.

We still consider a multi-layered plate as shown in Fig. \ref{fig:3}, which has the half length $l=1$. Suppose the plate undergoes plane-strain deformations along the $Y_k$-axis, then the position vector of the bottom surface of the plate (i.e., the unknown in the plate equation (\ref{eq:26})) can be denoted as $\mathbf{x}_1^{(0)}=\{x_1^{(0)},Y_1,z_1^{(0)}\}$. The position components and the growth functions are rewritten into
\begin{equation}
\begin{aligned}
x_1^{(0)} = X_1 + U, \ \ z_1^{(0)}= Z_1 + W,\ \ \lambda_k = 1 + \Delta\lambda_k,
\end{aligned}
\label{eq:38}
\end{equation}
where $U$ and $W$ are the in-plane and transverse displacements, and $\Delta\lambda_k$ $(k=1,\cdots,n)$ is referred as the growth rate. To derive the FvK-type plate equations, we adopt the scaling relations $W \sim h$, $U \sim h^2$ and $\Delta\lambda_k \sim h^2$ \cite{derv2010,wang2018}. With these scalings, we substitute (\ref{eq:38}) into the plate equation (\ref{eq:26}) and dropping the terms of order higher than $O(h^3)$, the following two equations are obtained
\begin{equation}
\begin{aligned}
&8h\left[\Lambda_n^{(1)}(U_{,XX}+W_{,X}W_{,XX})-g_{\lambda,X}^{(1)}\right]-4h^2\Lambda_n^{(13)}W_{,3X}=-\tilde{q}_1,\\
&8h\left[W_{,X}\left(\Lambda_n^{(1)}U_{,XX}-g_{\lambda,X}^{(1)}\right) +\left(\Lambda_n^{(1)}U_{,X}-g_{\lambda}^{(1)}+\frac{3}{2}\Lambda_n^{(1)} {W_{,X}}^2\right)W_{,XX}\right]\\
&-\frac{4h^2}{\Lambda_n^{(1)}}\left[\Lambda_n^{(1)}\Lambda_n^{(13)} \left({W_{,XX}}^2+W_{,X}W_{,3X}\right)-g_{\lambda,XX}^{(4)}\right]\\
&-\frac{2h^3}{3\Lambda_n^{(1)}}\left(\Lambda_n^{(9)}+4\Lambda_n^{(10)}+ 12\Lambda_n^{(11)}+24\Lambda_n^{(12)}\right)W_{,4X}=-\tilde{q}_3,
\end{aligned}
\label{eq:39}
\end{equation}
where
\begin{equation*}
\begin{aligned}
\Lambda_n^{(13)}=\sum_{k=1}^n\beta_k\left(\beta_kC_k+2\sum_{p=k+1}^n\beta_pC_p\right),\\
\end{aligned}
\end{equation*}
and $\tilde{q}_1$ and $\tilde{q}_3$ are the in-plane and transverse components of the traction $\tilde{\mathbf{q}}$. From (\ref{eq:38}), it is known that the in-plane displacement is much smaller than the transverse displacement, thus the force component $\tilde{q}_1$ should be small and can be neglected. On the other hand, we consider the resultant of $\{11\}$-component of the second Piola-Kirchhoff stress tensor $\mathbb{T}_k=\mathbb{S}_k\mathbb{F}_k^{-T}$, which is denoted as $\bar{t}$. By using (\ref{eq:11})$_1$, (\ref{eq:14}) and (\ref{eq:A1})-(\ref{eq:A3}) in \ref{app:a}, it can be calculated that
\begin{equation}
\begin{aligned}
\bar{t}=&\sum_{k=1}^n\int_0^{\beta_k h}T_{k,11}+ZT_{k,11}^{(1)}dZ\\
=&4h\left[\Lambda_n^{(1)}(2U_{,X}+{W_{,X}}^2)-2g_{\lambda}^{(1)} \right]-4h^2\Lambda_n^{(13)}W_{,XX}.
\end{aligned}
\label{eq:40}
\end{equation}
By virtue of (\ref{eq:40}), the two equations given in (\ref{eq:39}) can be rewritten into
\begin{equation}
\begin{aligned}
&\frac{\partial\bar{t}}{\partial X}=0,\\
&\frac{2h^3\bar{\Lambda}_n}{3\Lambda_n^{(1)}}W_{,4X}-\frac{\partial\left(W_{,X}\bar{t}\right)}{\partial X}-\frac{4h^2 g_{\lambda,XX}^{(4)}}{\Lambda_n^{(1)}}=\tilde{q}_3.
\end{aligned}
\label{eq:41}
\end{equation}
where $\bar{\Lambda}_n=\Lambda_n^{(9)}+4\Lambda_n^{(10)} +12\Lambda_n^{(11)}+24\Lambda_n^{(12)}$. Eq. (\ref{eq:41}) is just the FvK-type plate equations for multi-layered hyperelastic plates. If we choose $n=1$ in (\ref{eq:41}), the FvK-type plate equations for single-layered hyperelastic plates can be recovered \cite{derv2010,wang2018}.

\begin{figure}[htp]
\centering \includegraphics[width=0.8\textwidth]{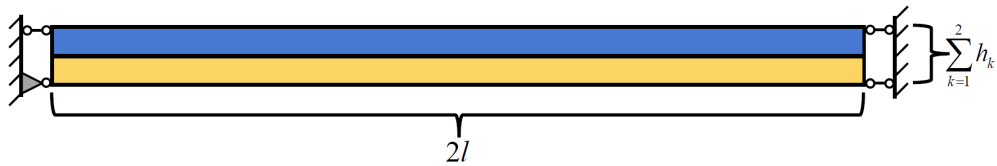}\\
\caption{The reference configuration and the boundary conditions of a bilayer hyperelastic plate.}
\label{fig:6}
\end{figure}

As an example of application of the FvK-type plate equations (\ref{eq:41}), we study the growth-induced instabilities of a bilayer hyperelastic plate. The reference configuration and the boundary conditions of the plate are shown in Fig. \ref{fig:6}. We suppose the top and bottom faces of the plate are traction free. At the left and right ends of the plate, the horizontal displacements are blocked. For simplicity, we suppose $\Delta\lambda_1(X)=0$ and $\Delta\lambda_2(X)$ is set to a constant. In fact, this problem has been studied in \citet{du2022}, where the instability analyses were conducted based on a full form plate equation system similar as that given in (\ref{eq:26}) (under the assumption of plane-strain deformation) and some numerical simulation results were also obtained. Here, we shall derive some analytical results based on the FvK-type plate equations (\ref{eq:41}) and compare them with the results obtained in \citet{du2022}.

For the bilayer plate sample considered in this example, the FvK-type plate equations (\ref{eq:41}) can be rewritten into
\begin{equation}
\begin{aligned}
&8h(C_1\beta_1+C_2\beta_2)(U_{,XX}+W_{,X}W_{,XX})\\
&-4h^2\big[C_1{\beta_1}^2+C_2\beta_2(2\beta_1+\beta_2)\big]W_{,3X}=0,
\end{aligned}
\label{eq:42}
\end{equation}
\begin{equation}
\begin{aligned}
&4h\big[(C_1\beta_1+C_2\beta_2)(2W_{,X}U_{,XX}+3{W_{,X}}^2W_{,XX}+2U_{,X}W_{,XX})\\
&-2C_2\beta_2\Delta\lambda_2W_{,XX}-4h^2\big[(C_1{\beta_1}^2+2C_1\beta_1\beta_2+C_2{\beta_2}^2)U_{,3X}\\
&+2(C_1\beta_1+C_2\beta_2)(\beta_1+\beta_2)({W_{,XX}}^2+W_{,X} W_{,3X})\big]\\
&+\frac{4h^3}{3}\big[C_1{\beta_1}^2(\beta_1+3\beta_2)+C_2{\beta_2}^2 (3\beta_1+\beta_2)\big]W_{,4X}=0,
\end{aligned}
\label{eq:43}
\end{equation}
In fact, (\ref{eq:42}) can be integrated once with respect to $X$ and yields (\ref{eq:40}). This plate equation system has the following homogeneous solution
\begin{equation}
\begin{aligned}
U_0=0,\ \ \ W_0=0,\ \ \bar{t}=-8\beta_2C_2h\Delta\lambda_2.
\end{aligned}
\label{eq:44}
\end{equation}
To study the growth-induced instabilities of this bilayer hyperelastic plate, we conduct some linear bifurcation analyses. For that purpose, we denote
\begin{equation}
\begin{aligned}
U=U_0+\varepsilon\Delta U,\ \ \ W=W_0+\varepsilon\Delta W.
\end{aligned}
\label{eq:45}
\end{equation}
By submitting (\ref{eq:44}) and (\ref{eq:45}) into (\ref{eq:42}) and (\ref{eq:43}) and dropping the nonlinear terms with respect to $\varepsilon$, the following equations are obtained
\begin{equation}
\begin{aligned}
8h(C_1\beta_1+C_2\beta_2)\Delta U_{,X}-4h^2\big[C_1{\beta_1}^2+C_2\beta_2(2\beta_1+\beta_2)\big]\Delta W_{,XX}=0,
\end{aligned}
\label{eq:46}
\end{equation}
\begin{equation}
\begin{aligned}
&-8\beta_2C_2h\Delta\lambda_2\Delta W_{,XX}-4h^2({\beta_1}^2C_1+2C_1\beta_1\beta_2+{\beta_2}^2C_2)\Delta U_{,3X}\\
&+\frac{4h^3}{3}\big[C_1{\beta_1}^2(\beta_1+3\beta_2)+C_2{\beta_2}^2 (3\beta_1+\beta_2)\big]\Delta W_{,4X}=0.
\end{aligned}
\label{eq:47}
\end{equation}
From (\ref{eq:46}), we have
\begin{equation}
\begin{aligned}
\Delta U_{,X}=\frac{h\big[C_1{\beta_1}^2+C_2\beta_2(2\beta_1+\beta_2)\big]\Delta W_{,XX}}{2(C_1\beta_1+C_2\beta_2)}.
\end{aligned}
\label{eq:48}
\end{equation}
By submitting (\ref{eq:48}) into (\ref{eq:47}), a single equation for $\Delta W$ can be obtained
\begin{equation}
\begin{aligned}
\phi_2\Delta W_{,XX}+\phi_4\Delta W_{,4X}=0,
\end{aligned}
\label{eq:49}
\end{equation}
where
\begin{equation*}
\begin{aligned}
\phi_2=&-4\beta_2C_2\Delta\lambda_2,\\
\phi_4=&-\frac{h^2[{\beta_1}^4{C_1}^2+{\beta_2}^4{C_2}^2+2\beta_1\beta_2C_1C_2 (2{\beta_1}^2+3\beta_1\beta_2+2{\beta_2}^2)]}{3(\beta_1C_1+\beta_2C_2)}.
\end{aligned}
\end{equation*}
Corresponding to the boundary restrictions of plate sample shown in Fig. \ref{fig:6}, we can also derive the following boundary conditions from (\ref{eq:27}) and (\ref{eq:28})
\begin{equation}
\begin{aligned}
\Delta W_{,X}(-1) = \Delta W_{,X}(1)=0, \ \ \ \Delta W_{,3X}(-1) = \Delta W_{,3X}(1)=0.
\end{aligned}
\label{eq:50}
\end{equation}
Next, we need to find the nontrivial solutions of (\ref{eq:49}) subject to the boundary conditions (\ref{eq:50}). The characteristic equation of (\ref{eq:49}) is given by
\begin{equation}
\begin{aligned}
\phi_2r^2+\phi_4r^4=0,
\end{aligned}
\label{eq:51}
\end{equation}
Through some conventional analyses, it is known that the nontrivial solutions of (\ref{eq:49}) only exist when characteristic equation (\ref{eq:51}) has purely imaginary roots $r=n\pi\mathbf{i}/2$ ($n=1,2,3,\cdots$), where $n$ is called the mode number and $\mathbf{i}$ is the imaginary unit. For $n=2k$ and $2k-1$, the nontrivial solutions are given by
\begin{equation}
\begin{aligned}
\begin{cases}
\Delta W=A\cos(k\pi X), \ \ \mathrm{if} \ n=2k,\\
\Delta W=A\sin[(k-1/2)\pi X], \ \ \mathrm{if} \ n=2k-1,\ \ k=1,2,\cdots
\end{cases}
\end{aligned}
\label{eq:52}
\end{equation}
On the other hand, by submitting $r=n\pi\mathbf{i}/2$ into (\ref{eq:51}), we can determine the bifurcation value of the growth parameter $\lambda_c$ $(\lambda_c=1+\Delta\lambda_2)$ corresponding to the mode number $n$, which is given by
\begin{equation}
\begin{aligned}
&\lambda_c = 1 + \frac{h^2n^2{\pi}^2[{\beta_1}^4{C_1}^2+{\beta_2}^4{C_2}^2+2\beta_1\beta_2C_1C_2(2{\beta_1}^2+3\beta_1\beta_2+2{\beta_2}^2)]}{48\beta_2C_2(\beta_1C_1+\beta_2C_2)},
\end{aligned}
\label{eq:53}
\end{equation}

\begin{figure}[htp]
\begin{minipage}{0.49\textwidth}
\centering \includegraphics[width=0.95\textwidth]{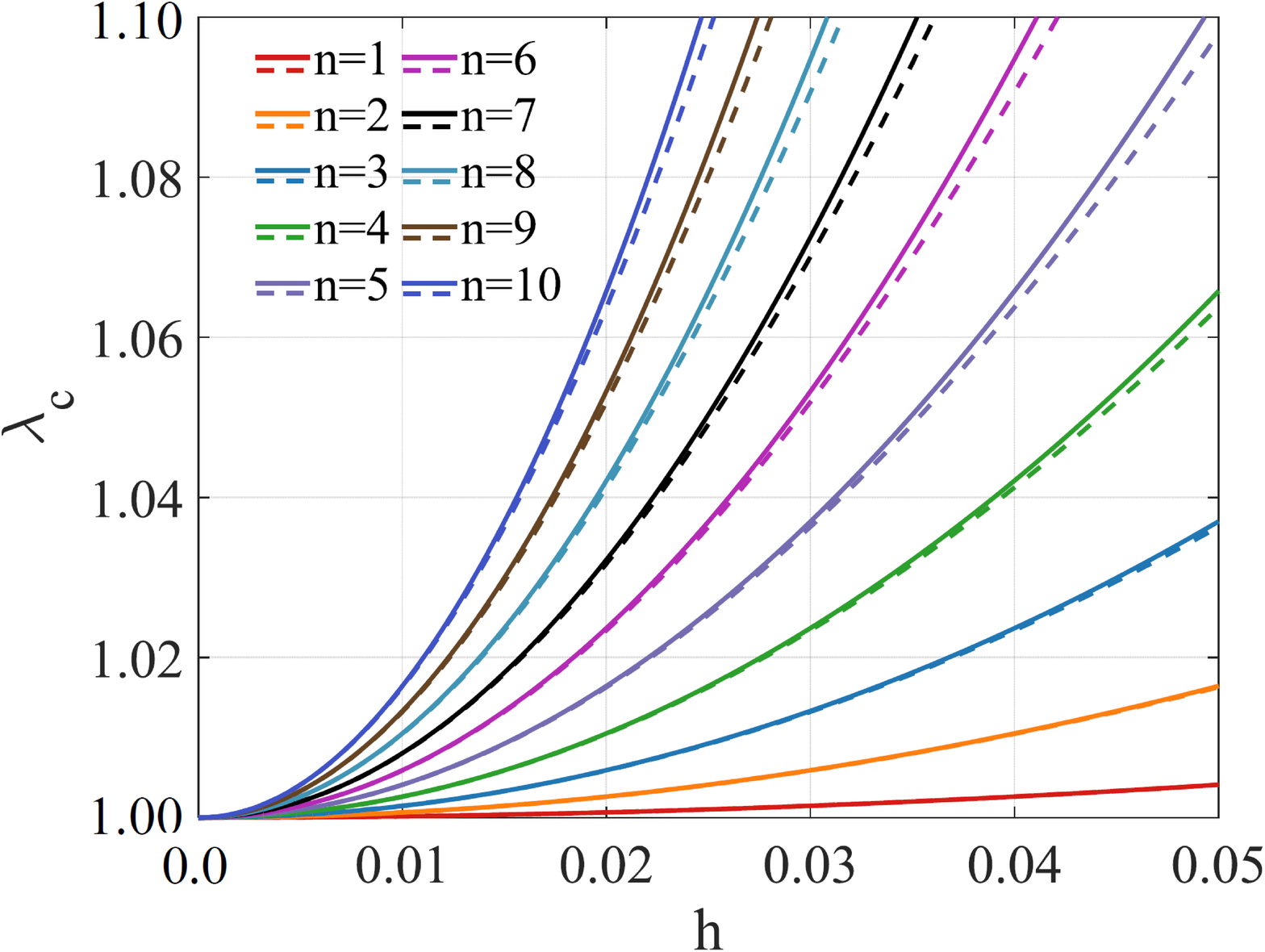}\\(a)\vspace{0.2cm}
\end{minipage}
\begin{minipage}{0.49\textwidth}
\centering \includegraphics[width=0.95\textwidth]{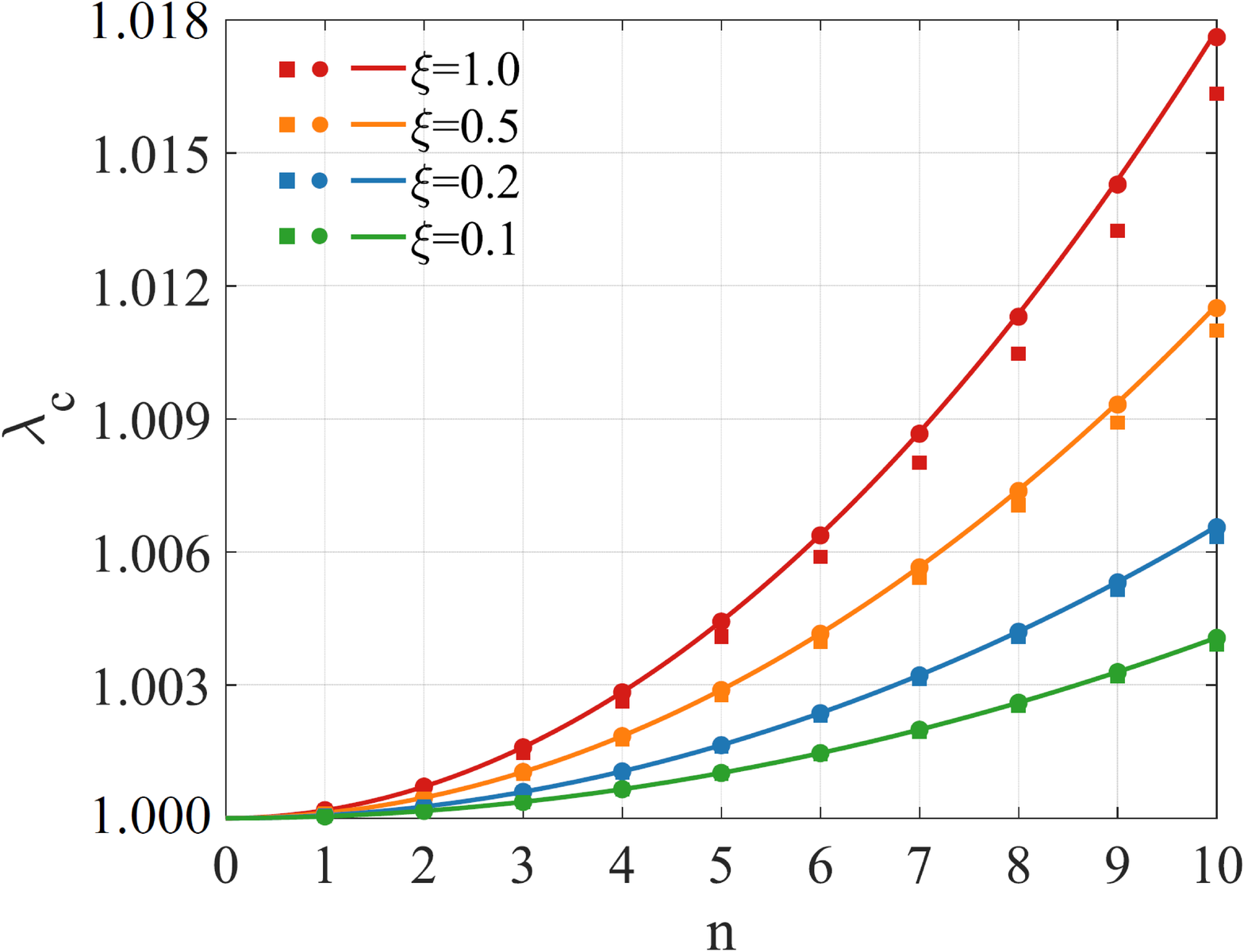}\\(b)\vspace{0.2cm}
\end{minipage}\\
\caption{(a) Comparisons of the bifurcation curves obtained from the FvK-type plate equations (solid curves) and those reported in \citet{du2022} (dash curves) ($C_1=C_2$, $\beta_1=\beta_2$); (b) Comparisons of the growth bifurcation parameter $\lambda_c$ obtained from the FvK-type plate equations (solid curves) and the analytical (circular dots) and numerical (square dots) results reported in \citet{du2022} ($h_1=0.01$, $h_2=0.002$, $\xi=C_1/C_2$).}
\label{fig:7}
\end{figure}

To show the accuracy of the analytical result (\ref{eq:53}), we plot some evolution curves of the growth bifurcation parameter $\lambda_c$ in Fig. \ref{fig:7}, which are compared with the analytical and numerical results reported in \citet{du2022}. It can be seen that the results obtained from the FvK-type plate equations are very closed to the results obtained from the full-form plate equation system and numerical simulations. Thus, the efficiency of the FvK-type plate model derived in this subsection is verified. The dependence of the growth bifurcation parameter $\lambda_c$ on the material and geometrical parameters can be seen clearly from (\ref{eq:53}). Some discussions on this issue can be found in \citet{du2022}.

\noindent\textbf{Remark:} The major aim of this subsection is to show that the general plate theory proposed in the current work can be reduced to a FvK-type plate theory under some assumptions of displacement components. As the FvK-type plate equations have simpler forms, they are easier to be solved. However, the application range of the FvK-type plate theory is limited. For example, it is not suitable for modeling the large bending deformations of plate samples as shown in Section \ref{sec:3.1} (where both the in-plane and transverse displacements have large magnitudes). On the other hand, by proposing some other assumptions on the displacements or applied tractions, the current general plate theory can be reduced to other classical plate or membrane theories (cf. \citet{wangff2019}).


\subsection{Growth-induced axisymmetric deformations of bilayer circular hyperelastic plates}
\label{sec:3.3}

\begin{figure}[htp]
\begin{minipage}{0.64\textwidth}
\centering \includegraphics[width=0.95\textwidth]{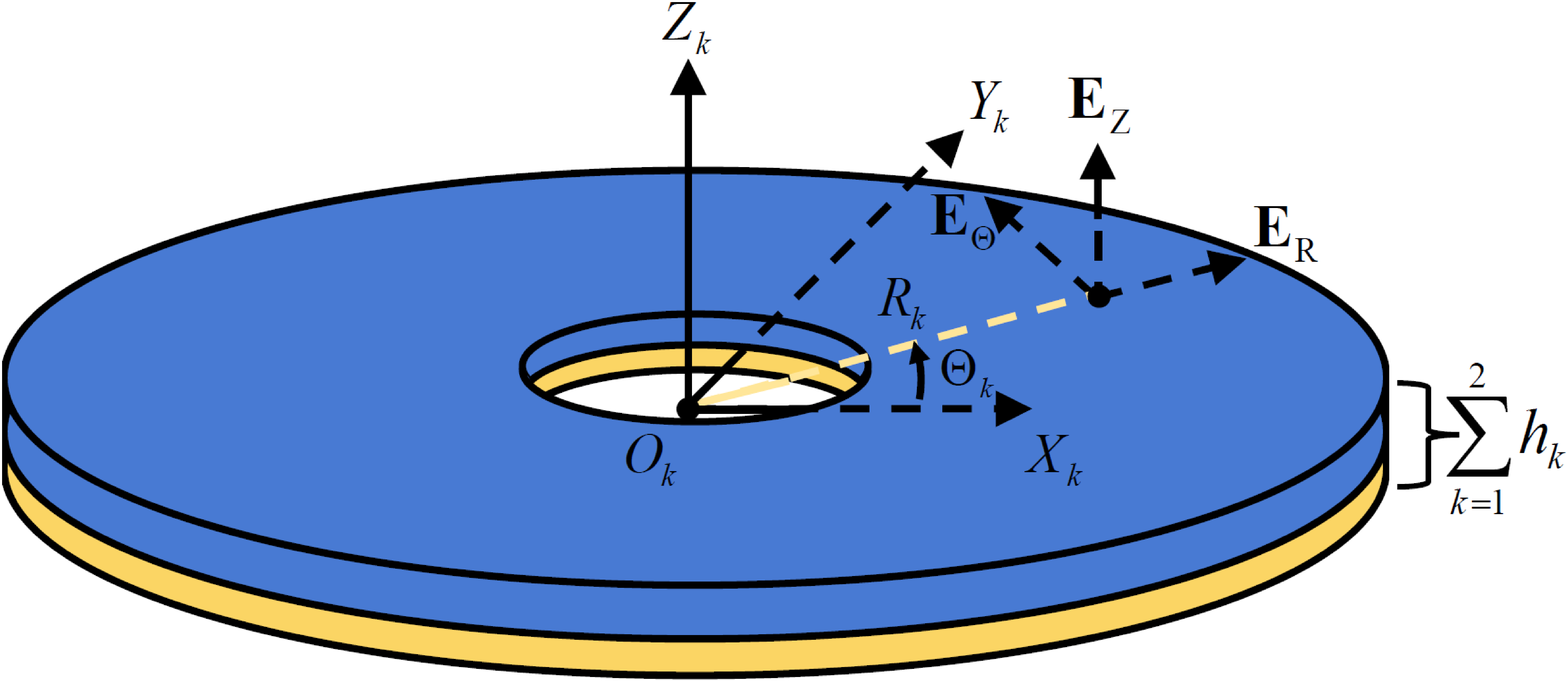}\\(a)\vspace{0.2cm}
\end{minipage}
\begin{minipage}{0.34\textwidth}
\centering \includegraphics[width=0.95\textwidth]{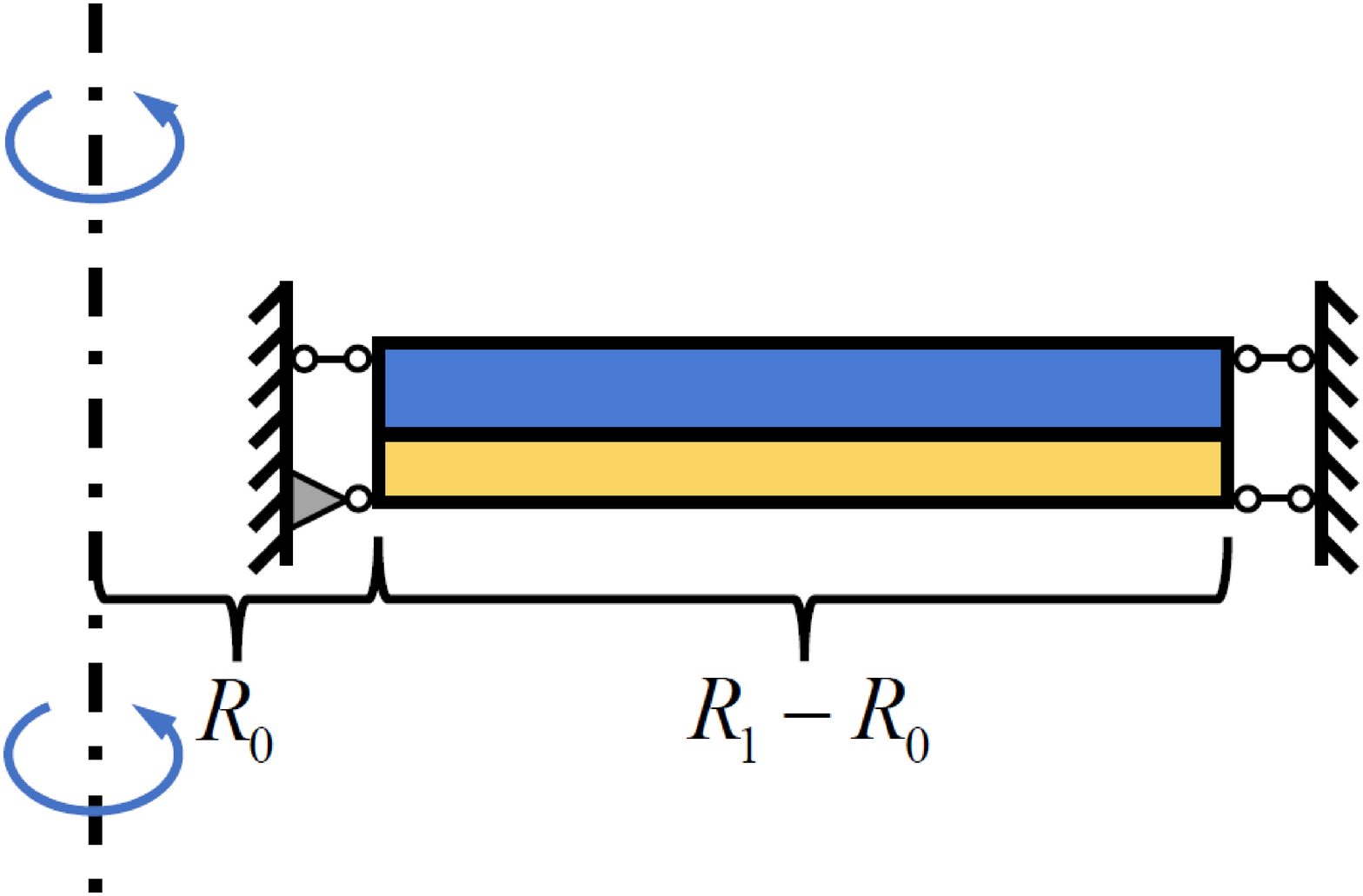}\\(b)\vspace{0.2cm}
\end{minipage}\\
\caption{(a) The reference configuration of a bilayer circular hyperelastic plate and the local cylindrical coordinate system; (b) illustration of the boundary restrictions proposed on the inner and outer lateral surfaces of the bilayer circular plate.}
\label{fig:8}
\end{figure}

In the third example, we study the growth-induced axisymmetric deformations of bilayer circular hyperelastic plates. The reference configuration of a bilayer circular plate sample is shown in Fig.\ref{fig:8}(a). For each layer, the local cylindrical coordinate system is established such that the reference configuration of the layer occupies the region $\kappa_k=\Omega\times[0,h_k]=[R_0,R_1]\times[0,2\pi]\times[0,h_k]$ $(k=1,2)$. The top and bottom faces of the plate are traction free. On the inner lateral surface $\partial\kappa_k^{(0)}={R_0}\times[0,2\pi]\times[0,h_k]$ and outer lateral surface $\partial\kappa_k^{(1)}={R_1}\times[0,2\pi]\times[0,h_k]$, the boundary restrictions are shown in Fig.\ref{fig:8}(b), where the horizontal displacements are blocked and the vertical shear stress components are supposed to be vanished. To remove the rigid body motion, the vertical displacement of the bottom curve of $\partial\kappa_1^{(0)}$ is set to be zero.

The coordinates of a material point in the reference and current configurations are denoted as $(R_k,\Theta_k,Z_k)$ and $(r_k,\theta_k,z_k)$, respectively. In each layer, the growth function along the radial direction ($\mathbf{E}_{R}$-axis) is denoted as $\lambda_{kR}(R_k)$ and that along the circumferential direction ($\mathbf{E}_{\Theta}$-axis) is denoted as $\lambda_{k\Theta}(R_k)$. Suppose these growth functions will induce axisymmetric deformation of the circular plate. Thus, we have $\theta_k=\Theta_k$ and the unknowns $r_k$, $z_k$ and $p_k$ only depend on the coordinates $(R_k,Z_k)$. The series expansions of $r_k$, $z_k$ and $p_k$ with respect to the thickness variable $Z$ are conducted as follow
\begin{equation}
\begin{aligned}
r_k(R_k,Z_k)=&\sum_{i=0}^3\frac{Z_k^i}{i!}r_k^{(i)}(R_k)+O(Z_k^4),\ \ z_k(R_k,Z_k)=\sum_{i=0}^3\frac{Z_k^i}{i!}z_k^{(i)}(R_k)+O(Z_k^4),\\
p_k(R_k,Z_k)=&\sum_{i=0}^3\frac{Z_k^i}{i!}p_k^{(i)}(R_k)+O(Z_k^3).
\end{aligned}
\label{eq:54}
\end{equation}
Through the similar derivation as that introduced in Section \ref{sec:2}, the iterative expressions of $r_k^{(1)}$-$r_k^{(3)}$, $z_k^{(1)}$-$z_k^{(3)}$, $p_k^{(0)}$-$p_k^{(2)}$ ($k=1,2$) and $\{r_2^{(0)},z_2^{(0)}\}$ can be obtained, which are listed in \ref{app:b}. By submitting these iterative relations into (\ref{eq:26}), the following plate equations can be derived
\begin{equation}
\begin{aligned}
&\frac{1}{R}(\bar{S}_{Rr}-\bar{S}_{\Theta\theta})+\bar{S}_{Rr,R}=0,\\
&\frac{1}{R}\bar{S}_{Rz}+\bar{S}_{Rz,R} = 0.
\end{aligned}
\label{eq:55}
\end{equation}
In Eq. (\ref{eq:55}), $\bar{S}_{Rr}$, $\bar{S}_{\Theta\theta}$ and $\bar{S}_{Rz}$ are the through-thickness resultants of the stress components, which are given by
\begin{equation*}
\begin{aligned}
\bar{S}_{Rr}=&2hr_{1,R}^{(0)}\Bigg[\frac{\beta_1C_1\lambda_{1\Theta}}{\lambda_{1R}}+\frac{\beta_2C_2\lambda_{2\Theta}}{\lambda_{2R}}-\frac{R^2\left(\beta_1C_1{\lambda_{1R}}^3{\lambda_{1\Theta}}^3+\beta_2C_2{\lambda_{2R}}^3{\lambda_{2\Theta}}^3\right)}{{r_1^{(0)}}^2\left({r_{1,R}^{(0)}}^2+{z_{1,R}^{(0)}}^2\right)}   \Bigg]\\
&+h^2\mathcal{W}_1(r_1^{(0)},r_{1,R}^{(0)},r_{1,RR}^{(0)},z_{1,R}^{(0)},z_{1,RR}^{(0)})\\ &+h^3\mathcal{W}_2(r_1^{(0)},r_{1,R}^{(0)},r_{1,RR}^{(0)},r_{1,3R}^{(0)},z_{1,R}^{(0)},z_{1,RR}^{(0)},z_{1,3R}^{(0)}),\\
\bar{S}_{Rz}=&2hz_{1,R}^{(0)}\Bigg[\frac{\beta_1C_1\lambda_{1\Theta}}{\lambda_{1R}}+\frac{\beta_2C_2\lambda_{2\Theta}}{\lambda_{2R}}-\frac{R^2\left(\beta_1C_1{\lambda_{1R}}^3{\lambda_{1\Theta}}^3+\beta_2C_2{\lambda_{2R}}^3{\lambda_{2\Theta}}^3\right)}{{r_1^{(0)}}^2\left({r_{1,R}^{(0)}}^2+{z_{1,R}^{(0)}}^2\right)}   \Bigg]\\
&+h^2\mathcal{W}_3(r_1^{(0)},r_{1,R}^{(0)},r_{1,RR}^{(0)},z_{1,R}^{(0)},z_{1,RR}^{(0)})\\ &+h^3\mathcal{W}_4(r_1^{(0)},r_{1,R}^{(0)},r_{1,RR}^{(0)},r_{1,3R}^{(0)},z_{1,R}^{(0)},z_{1,RR}^{(0)},z_{1,3R}^{(0)}),\\
\bar{S}_{\Theta\theta}=&2hR^3r_1^{(0)}\Bigg[\frac{\beta_1C_1\lambda_{1R}\lambda_{2\Theta}+\beta_2C_2\lambda_{2R}\lambda_{1\Theta}}{R^4\lambda_{1\Theta}\lambda_{2\Theta}}-\frac{\beta_1C_1{\lambda_{1R}}^3{\lambda_{1\Theta}}^3+\beta_2C_2{\lambda_{2R}}^3{\lambda_{2\Theta}}^3}{{r_1^{(0)}}^4\left({r_{1,R}^{(0)}}^2+{z_{1,R}^{(0)}}^2\right)} \Bigg]\\
&+h^2\mathcal{W}_5(r_1^{(0)},r_{1,R}^{(0)},r_{1,RR}^{(0)},z_{1,R}^{(0)},z_{1,RR}^{(0)})\\ &+h^3\mathcal{W}_6(r_1^{(0)},r_{1,R}^{(0)},r_{1,RR}^{(0)},r_{1,3R}^{(0)},z_{1,R}^{(0)},z_{1,RR}^{(0)},z_{1,3R}^{(0)}),
\end{aligned}
\end{equation*}
Here, the lengthy expressions of $\mathcal{W}_1-\mathcal{W}_6$ are omitted for brevity (the full expressions can be found in the supplementary document). By multiplying $R$ onto (\ref{eq:55})$_2$, this equation can be integrated once with respect to $R$. Further from the condition that the vertical shear stresses on the lateral surfaces $\partial\kappa_k^{(0)}$ and $\partial\kappa_k^{(1)}$ are vanished, it is known that (\ref{eq:55})$_2$ is equivalent to the equation
\begin{equation}
\begin{aligned}
\bar{S}_{Rz} = 0.
\end{aligned}
\label{eq:56}
\end{equation}

To simplify the plate equations (\ref{eq:55})$_1$ and (\ref{eq:56}), we rewrite the spatial position components and the growth functions as follow
\begin{equation}
\begin{aligned}
r_1^{(0)}=&R_1 + U, \ \ z_1^{(0)}= W,\\
\lambda_{kR}=&1+\Delta\lambda_{kR},\ \ \lambda_{k\Theta}=1+\Delta\lambda_{k\Theta},\ \ \ k=1,2.
\end{aligned}
\label{eq:57}
\end{equation}
Due to the boundary restrictions shown in Fig. \ref{fig:8}(b), the in-plane displacement should be much smaller than the transverse displacement in this circular plate. Thus, we adopt the scaling relations $W \sim h$, $U \sim h^2$, $\Delta\lambda_{kR} \sim h^2$ and $\Delta\lambda_{k\Theta} \sim h^2$. By submitting (\ref{eq:57}) into (\ref{eq:55})$_1$ and (\ref{eq:56}) and dropping the high-order terms, the following two equations can be obtained
\begin{equation}
\begin{aligned}
&-\frac{2h}{R^2}\bigg\{4(C_1\beta_1+C_2\beta_2)U+R\Big[2C_1\beta_1\big[\Delta\lambda_{1R}-\Delta\lambda_{1\Theta}+R(2\Delta\lambda_{1R,R}\\
&+\Delta\lambda_{1\Theta,R})\big]+2C_2\beta_2[\Delta\lambda_{2R}-\Delta\lambda_{2\Theta}+R(2\Delta\lambda_{2R,R}+\Delta\lambda_{2\Theta,R})]\\
&-(C_1\beta_1+C_2\beta_2)\big[(4U_{,R}+{W_{,R}}^2)+4R(U_{,RR}+W_{,R} W_{,RR})\big]\Big]\bigg\}\\
&+\frac{4h^2}{R^2}\big[C_1{\beta_1}^2+C_2\beta_2(2\beta_1+\beta_2 )\big]\big[{W_{,R}}-R({W_{,RR}}+R{W_{,3R}})\big]=0,
\end{aligned}
\label{eq:58}
\end{equation}
\begin{equation}
\begin{aligned}
&4h \Big\{(C_1\beta_1+C_2\beta_2) U-R\big[C_1\beta_1(2\Delta\lambda_{1R}+\Delta\lambda_{1\Theta})\\
&+C_2\beta_2(2\Delta\lambda_{2R}+\Delta\lambda_{2\Theta})-(C_1\beta_1+C_2\beta_2)(2U_{,R}+{W_{,R}}^2)\big]\Big\}W_{,R}\\
&+\frac{h^2}{R}\bigg\{4\big[C_2{\beta_2}^2+C_1\beta_1(\beta_1+ 2\beta_2)\big]U+R\Big[2C_1\beta_1(\beta_1+2\beta_2)\big[\Delta\lambda_{1R}\\
&-\Delta\lambda_{1\Theta}+R(2\Delta\lambda_{1R,R}+\Delta\lambda_{1\Theta,R})\big]+2C_2{\beta_2}^2\big[(\Delta\lambda_{2R}-\Delta\lambda_{2\Theta})\\
&+R(2\Delta\lambda_{2R,R}+\Delta\lambda_{2\Theta,R})\big]-4(C_1{\beta_1}^2+C_1\beta_1\beta_2+C_2{\beta_2}^2)\\
&\times(U_{,R}+RU_{,RR})-\big[C_1\beta_1(3\beta_1+2\beta_2)+C_2\beta_2(4\beta_1+3\beta_2)\big]{W_{,R}}^2\\
&-8R(\beta_1+\beta_2)(C_1\beta_1+C_2\beta_2)W_{,R}W_{,RR}\Big]\bigg\}-\frac{{4{h^3}}}{{3R}}\big[C_2\beta{2^2}(3\beta_1+\beta_2)\\
&+C_1{\beta_1}^2(\beta_1+3\beta_2)\big]\big[W_{,R}-R(W_{,RR}+R W_{,3R})\big] = 0.
\end{aligned}
\label{eq:59}
\end{equation}
Corresponding to the boundary restrictions on the inner and outer lateral surfaces of the plate (cf. Fig. \ref{fig:8}), we can also derive the following boundary conditions
\begin{equation}
\begin{aligned}
U(R_0)=U(R_1)=0,\ \ W(R_0)=0,\ \ W_{,R}(R_0)=W_{,R}(R_1)=0.
\end{aligned}
\label{eq:60}
\end{equation}
To solve the plate equation system (\ref{eq:58})-(\ref{eq:60}), one needs to use the singular perturbation method. This task seems to be difficult and will be further investigated in our future work. However, this ODE system is easy to be solved numerically. In the current work, the ODE package 'bvp4c' in Matlab is used to solve this ODE system.

\begin{figure}[htp]
\begin{minipage}{0.20\textwidth}
\centering \includegraphics[width=1\textwidth]{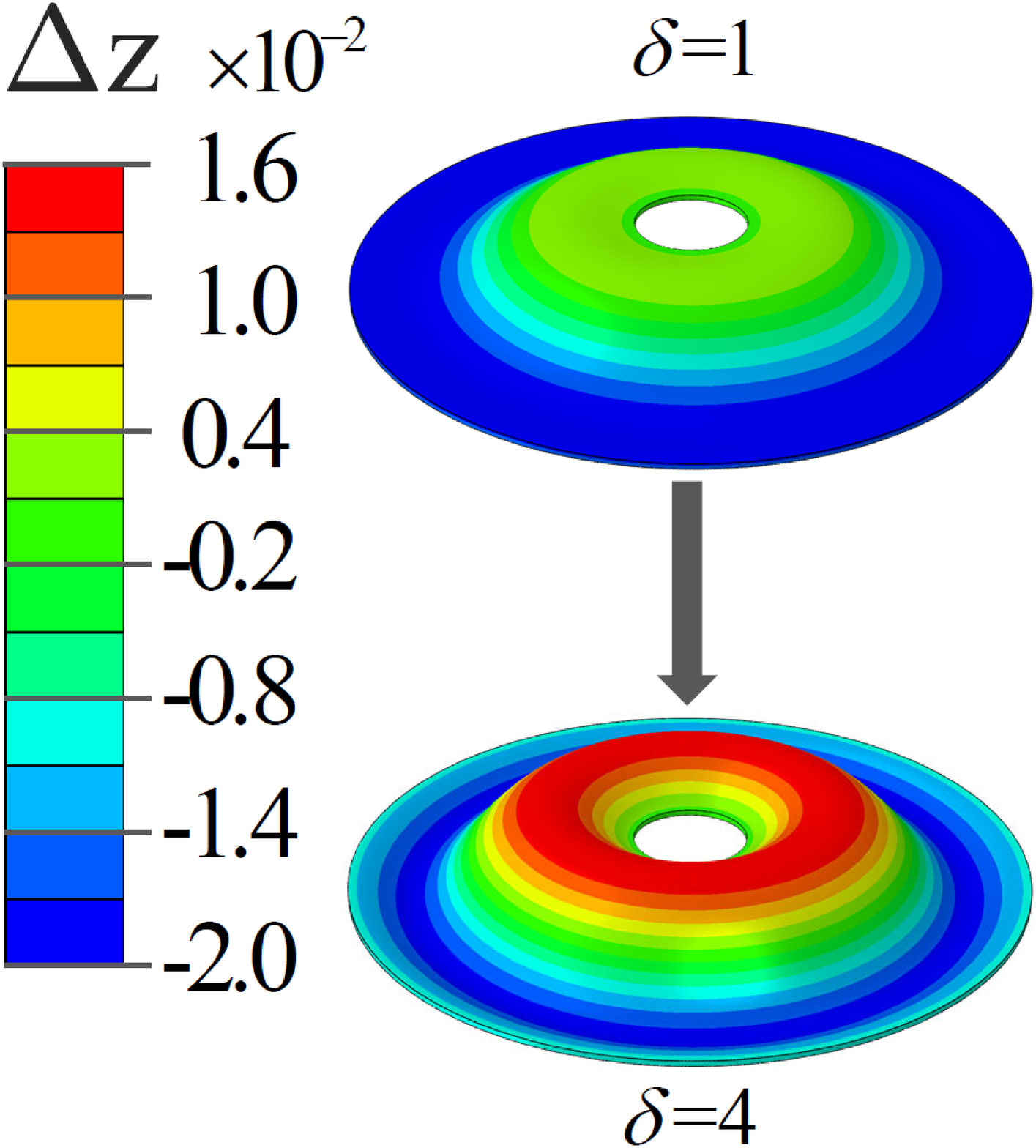}\\(a)\vspace{0.2cm}
\end{minipage}
\begin{minipage}{0.39\textwidth}
\centering \includegraphics[width=1\textwidth]{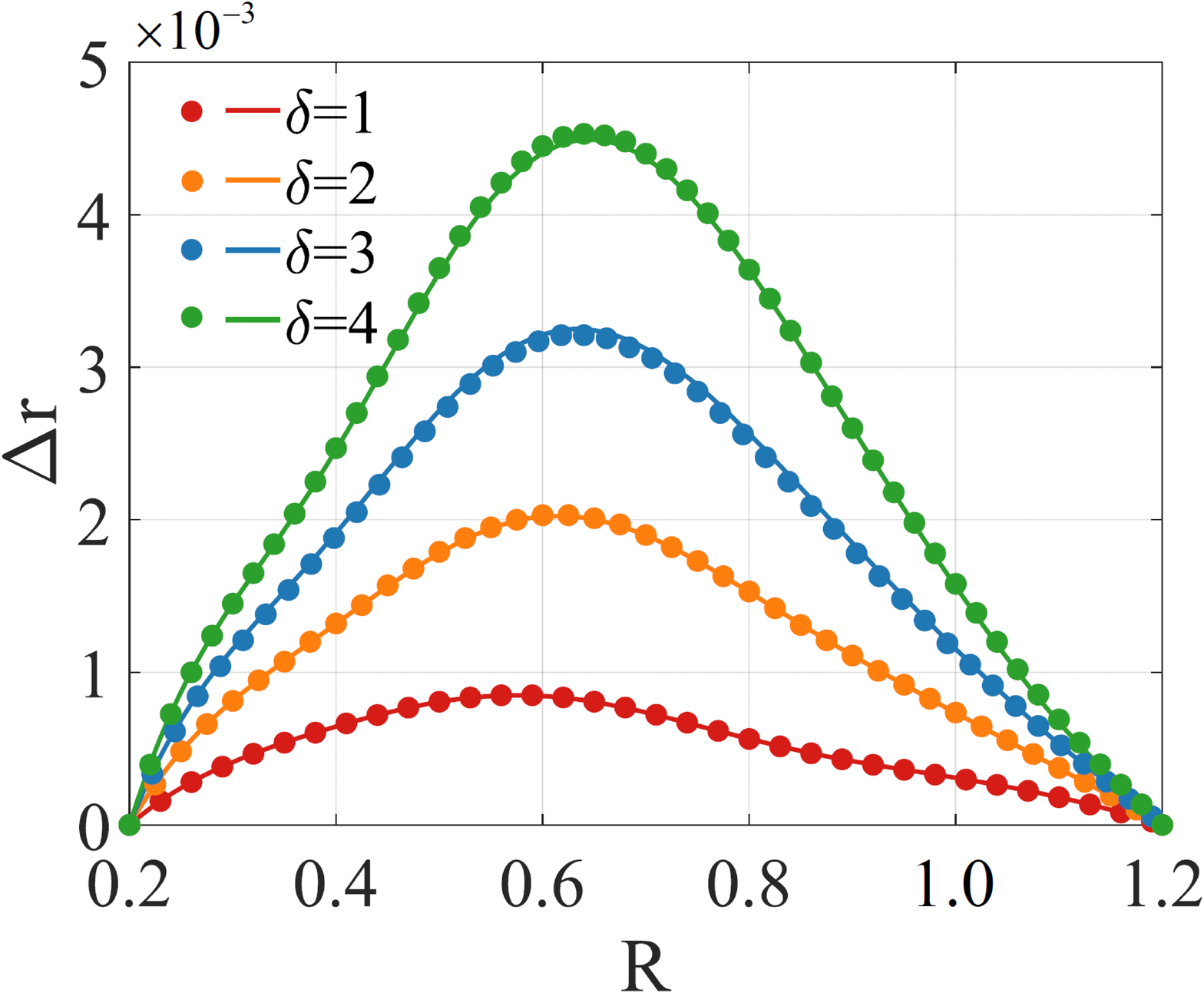}\\(b)\vspace{0.2cm}
\end{minipage}
\begin{minipage}{0.39\textwidth}
\centering \includegraphics[width=1\textwidth]{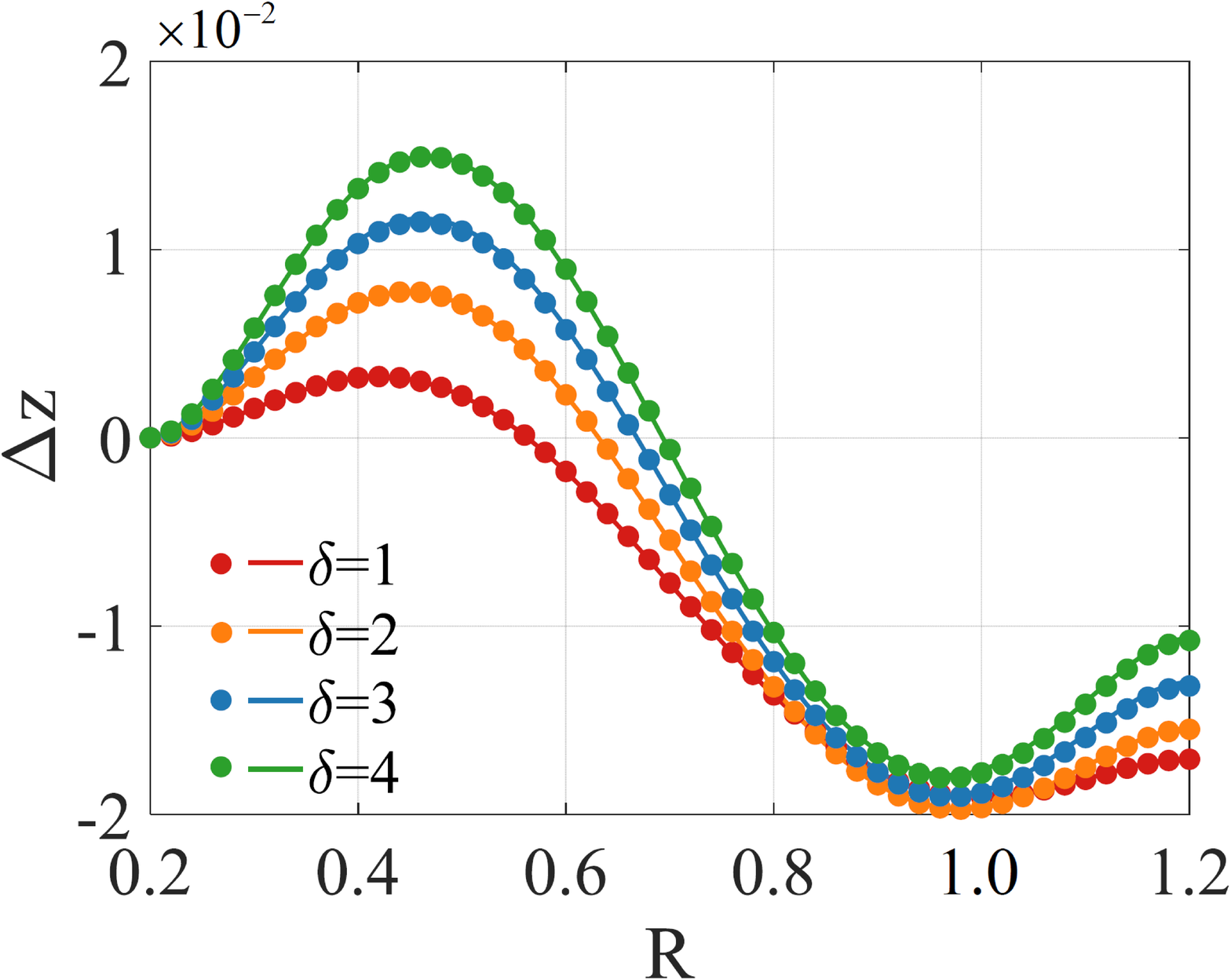}\\(c)\vspace{0.2cm}
\end{minipage}\\
\begin{minipage}{0.20\textwidth}
\centering \includegraphics[width=1\textwidth]{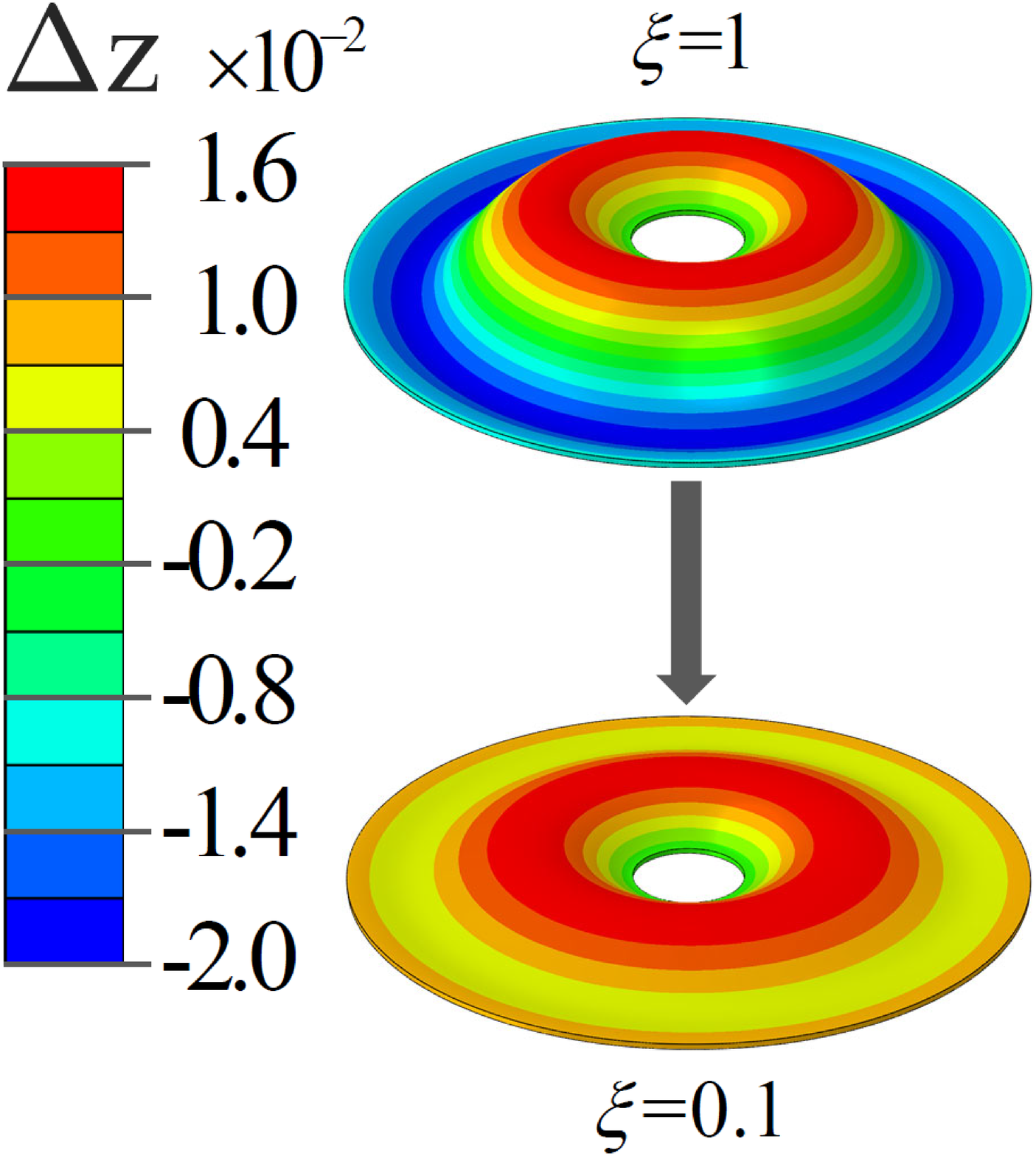}\\(d)\vspace{0.2cm}
\end{minipage}
\begin{minipage}{0.39\textwidth}
\centering \includegraphics[width=1\textwidth]{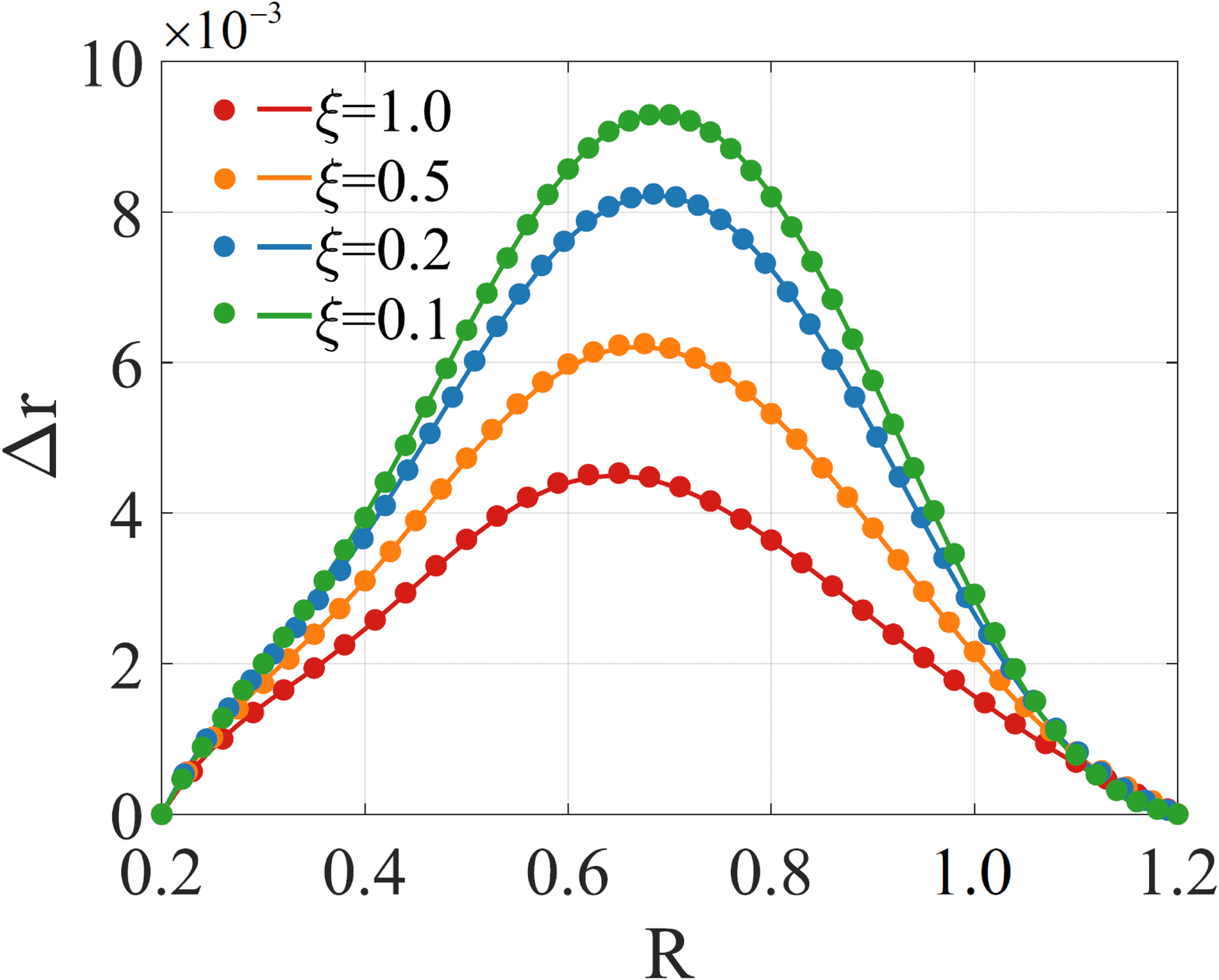}\\(e)\vspace{0.2cm}
\end{minipage}
\begin{minipage}{0.39\textwidth}
\centering \includegraphics[width=1\textwidth]{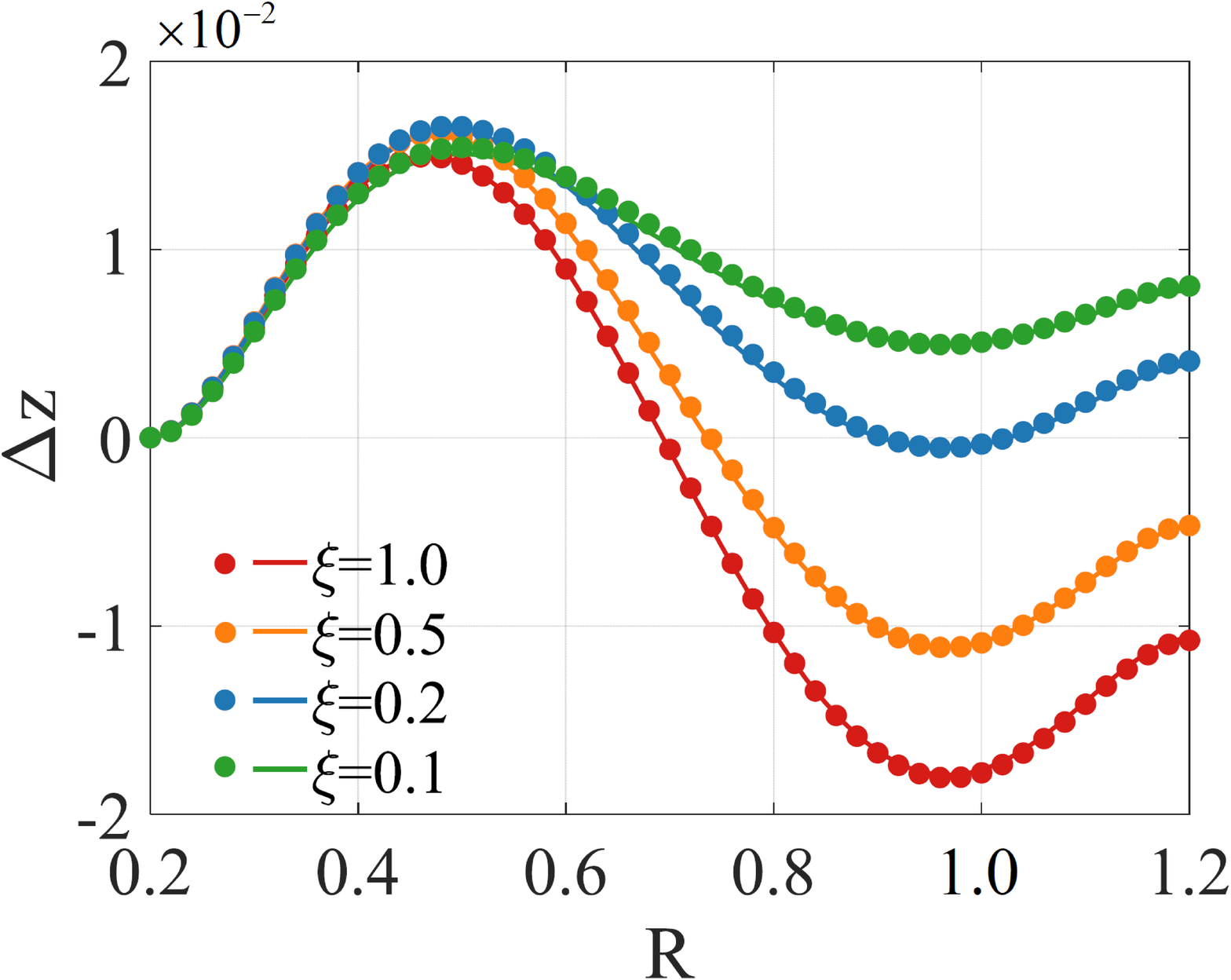}\\(f)\vspace{0.2cm}
\end{minipage}\\
\begin{minipage}{0.20\textwidth}
\centering \includegraphics[width=1\textwidth]{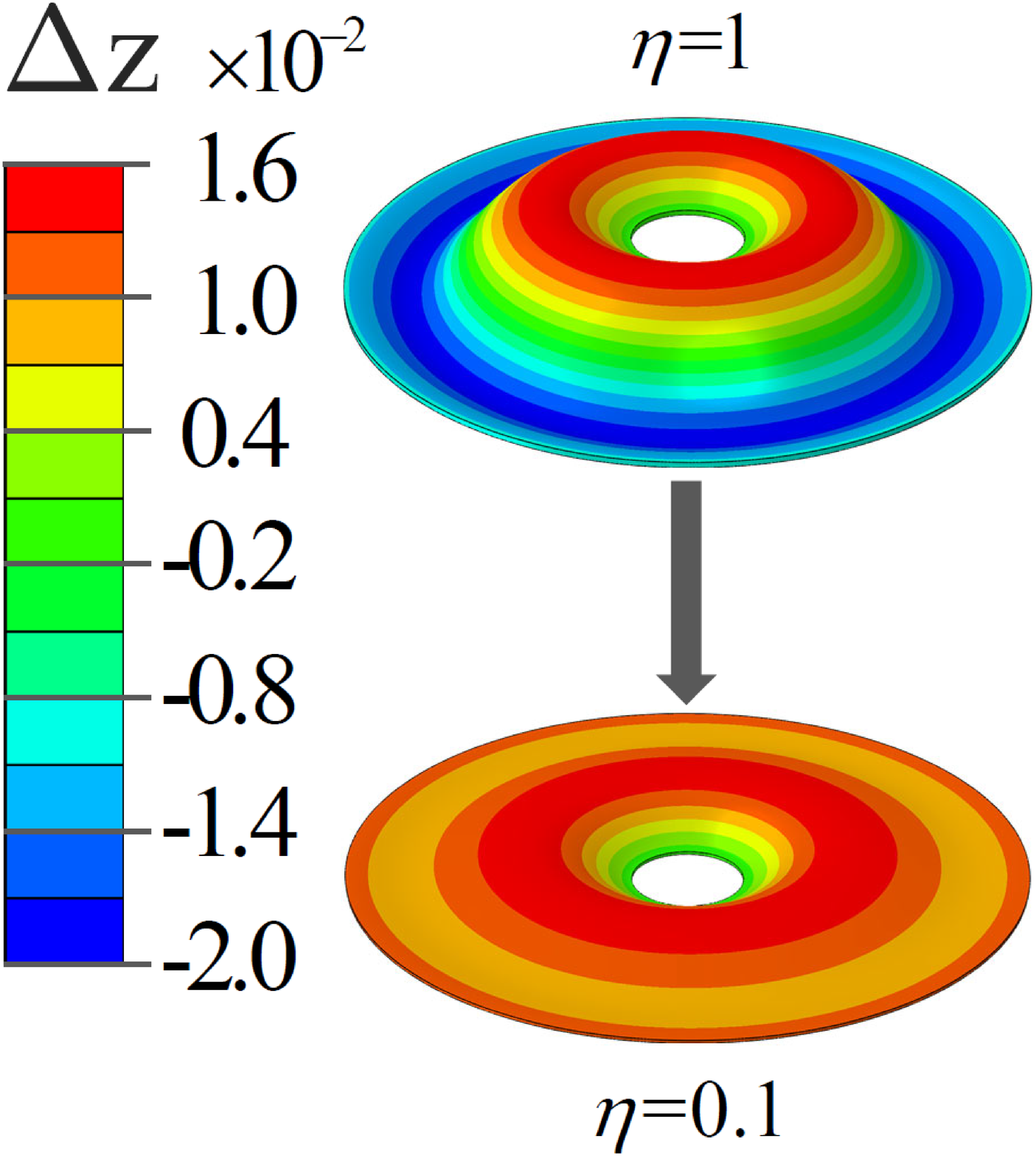}\\(g)\vspace{0.2cm}
\end{minipage}
\begin{minipage}{0.39\textwidth}
\centering \includegraphics[width=1\textwidth]{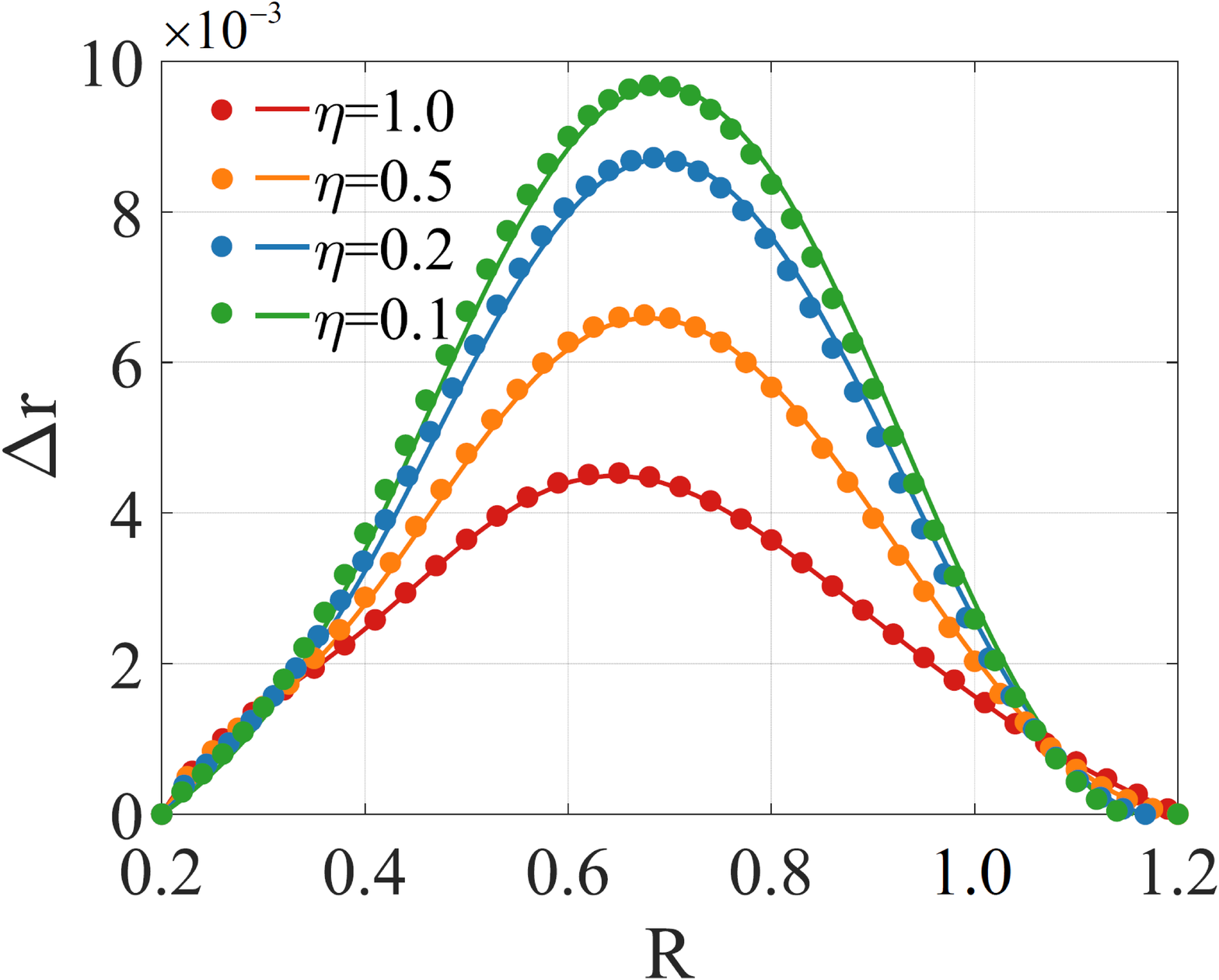}\\(h)\vspace{0.2cm}
\end{minipage}
\begin{minipage}{0.39\textwidth}
\centering \includegraphics[width=1\textwidth]{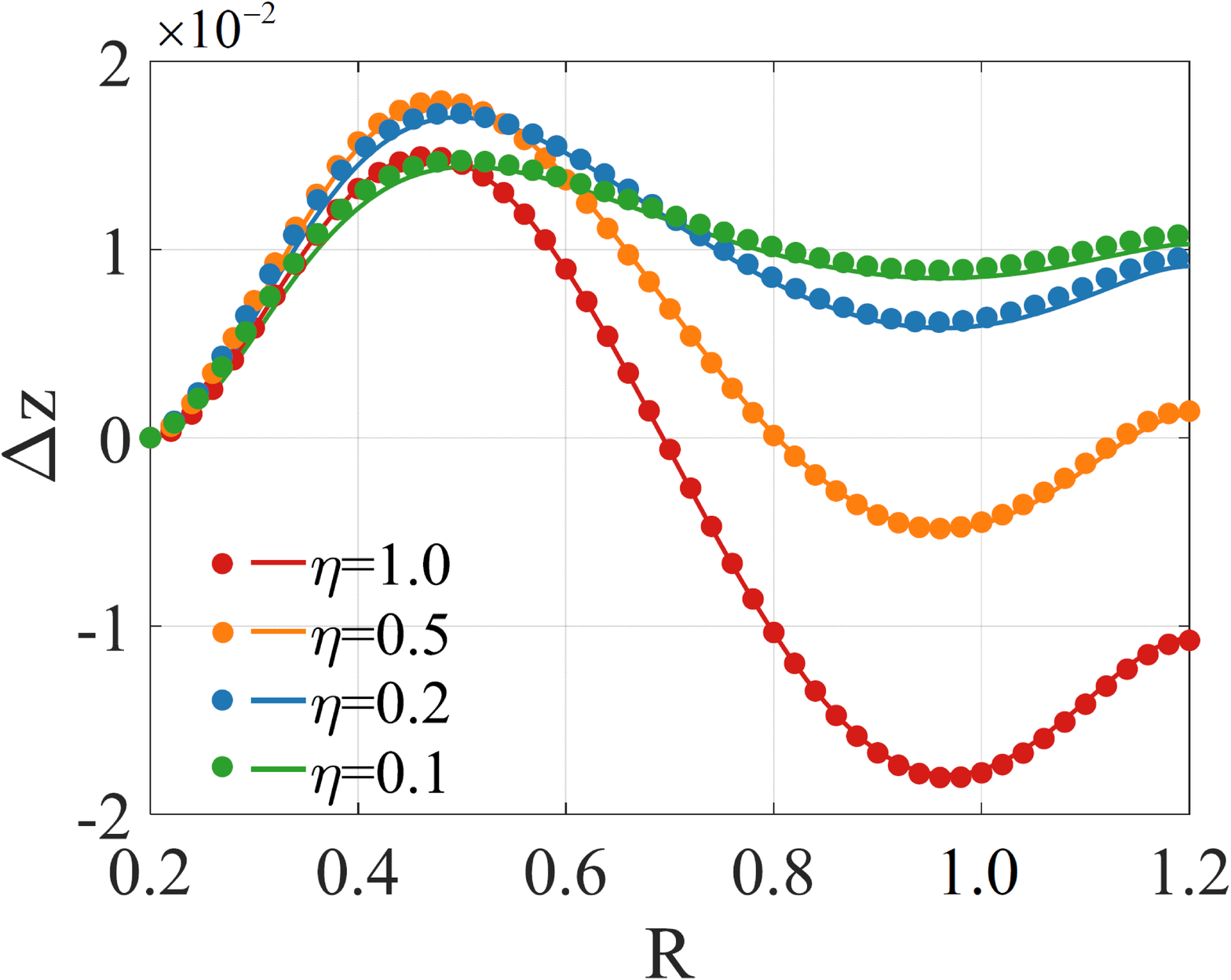}\\(i)\vspace{0.2cm}
\end{minipage}\\
\caption{Numerical results for growth-induced axisymmetric deformations of bilayer circular hyperelastic plates, which are obtained from the plate equations (solid curves) and finite element simulations (circular dots): (a)-(c) $C_1=C_2$, $\beta_1=\beta_2$ and $\delta=1,2,3,4$; (d)-(f) $\beta_1=\beta_2$, $\delta=4$ and $C_1/C_2=1,0.5,0.2,0.1$ ($\xi=C_1/C_2$); (g)-(i) $C_1=C_2$, $\delta=4$ and $\beta_1/\beta_2=1,0.5,0.2,0.1$ ($\eta=\beta_1/\beta_2$). }
\label{fig:9}
\end{figure}

For the purpose of illustration, we choose the following growth functions
\begin{equation}
\begin{aligned}
\left\{\begin{array}{l}\vspace{1.5ex}
\Delta\lambda_{1R}=\Delta\lambda_{1\Theta}=\Delta\lambda_{2\Theta}=0 \\
\Delta\lambda_{2R}=\displaystyle{1+\frac{\delta}{100}\mathrm{sin}\left[\frac{2\pi(5R-1)}{5}\right]},\ \ \delta=1,2,3,4.
\end{array}\right.
\end{aligned}
\label{eq:61}
\end{equation}
The values of inner and outer radii of the plate are set to be $R_0=0.2$ and $R_1=1.2$. Corresponding to some different material and geometrical parameters, the plate equation system (\ref{eq:58})-(\ref{eq:60}) is solved numerically, from which the in-plane and transverse displacements of the plate can be determined, which are shown in Fig. \ref{fig:9}. On the other hand, the growth-induced deformations of the 3D plate sample are simulated by using the finite element method, where the UMAT subroutine in ABAQUS with the compressible neo-Hookean constitutive relations is adopted. The settings of finite element simulations are similar as those introduced in section \ref{sec:3.1}. For axisymmetric deformations of the circular plate, only one section along the radial direction needs to be taken into account. This section is meshed by using the CAX8H elements (8-node biquadratic axisymmetric quadrilateral hybrid elements) with the size $0.001\times0.001$. For the purpose of comparison, the finite element simulation results are also shown in Fig. \ref{fig:9}. It can be seen that the numerical results obtained from the  different approaches shown very good consistencies, which further verify the efficiency of the multi-layered plate theory.

\section{Shape-programming of multi-layered hyperelastic plates}
\label{sec:4}

In section \ref{sec:3}, some direct problems of growth-induced deformations of multi-layered hyperelastic plates have been studied, where the growth functions in the plate samples are specified in advance. In this section, we begin to study the inverse problem. That is, to generate certain 3D target shapes, how to arrange the growth functions in the multi-layered hyperelastic plates? This problem is known as `shape-programming' \cite{liu2016} and its solutions have wide potential applications in the engineering fields (cf. \citet{glad2016,yuk2017,sief2019}). During the past years, the problem of shape-programming of soft material samples has attracted extensive research interests (e.g. \citet{dias2011,jone2015,acha2019,nojo2021}). In our previous works \cite{wang2019,li2022,wang2022}, some explicit formulas for shape-programming of single-layered hyperelastic plates through differential growth have been derived. Here, based on the multi-layered plate theory proposed in section \ref{sec:3}, we aim to derive some explicit formulas for shape-programming of multi-layered hyperelastic plates.

To obtained some concrete results, we only consider the bilayer hyperelastic plates in this section and suppose that the plates are made of incompressible neo-Hookean materials. It should be pointed out that we do not aim to control the whole configuration of the plate samples. As the plate equation system (\ref{eq:26})-(\ref{eq:29}) is established based on the bottom face $\Omega_1^-$ of the plate, shape programming will also be conducted by only considering the 3D target shape $\mathcal{S}$ of the bottom face in the current configuration.


\subsection{Shape-programming of bilayer hyperelastic plates under plane-strain deformations}
\label{sec:4.1}

In this subsection, we study the shape-programming of a bilayer hyperelastic plate under the assumption of plane-strain deformations. The plate sample still has the reference configuration shown in Fig. \ref{fig:3}, where the number of layers is $n=2$ and the half-length is $l=1$. All the faces of the sample are traction free, only except that some restrictions are applied to remove the rigid body motion. Growth in the plate only occurs along the $X_k$-axis and the growth functions are given by $\lambda_k(X) = \lambda(X) + h \Delta\lambda_k(X)$ $(k=1,2)$. By setting $n=2$, Eqs. (\ref{eq:31}) and (\ref{eq:32}) can be rewritten into
\begin{equation}
\begin{aligned}
&\frac{2(\beta_1C_1+\beta_2C_2)(E^2-{\lambda}^4)}{E^2\lambda} \mathbf{v}_{,X}\\
&+h\bigg\{-\frac{2}{E^{\frac{7}{2}}{\lambda}^2}\Big[(\beta_1C_1\Delta\lambda_1+\beta_2C_2\Delta\lambda_2)E^{\frac{7}{2}}+(\beta_1+\beta_2)(\beta_1C_1+\beta_2C_2)\\
&\times E^2L{\lambda}^2+3(\beta_1C_1\Delta\lambda_1+\beta_2C_2\Delta\lambda_2)E^{\frac{3}{2}}{\lambda}^4+\big[\beta_1C_1(\beta_1-\beta_2)\\
&+\beta_2C_2(3\beta_1+\beta_2)\big]L{\lambda}^6\Big]\mathbf{v}_{,X}+\frac{1}{E^4 \lambda}\Big[(\beta_1+\beta_2)(\beta_1C_1+\beta_2C_2)E^2\\
&+\big[\beta_1C_1(\beta_1+3\beta_2)-\beta_2C_2(\beta_1-\beta_2)\big]{\lambda}^4(2E\lambda_{,X}-\lambda E_{,X})\Big]\mathbf{v}_N\bigg\}=\mathbf{0}.
\end{aligned}
\label{eq:62}
\end{equation}
\begin{equation}
\begin{aligned}
&2h^2\beta_1\beta_2(C_1-C_2)\left(1-\frac{{\lambda}^4}{E^2}\right)\\
&-h^3\Big[\frac{2\beta_1\beta_2(\beta_1C_2+\beta_2C_1)(\Delta\lambda_1-\Delta\lambda_2)}{(\beta_1+\beta_2)\lambda}-\frac{L\lambda}{3E^{\frac{3}{2}}}\big[2{\beta_1}^3C_1\\
&+2{\beta_1}^2\beta_2(C_1+2C_2)+\beta_1{\beta_2}^2(C_1+5C_2)+2{\beta_2}^3C_2\big]\\
&+\frac{2\beta_1\beta_2{\lambda}^3}{(\beta_1+\beta_2)E^2}\big[(4\beta_1C_1-\beta_1C_2+3\beta_2C_1)\Delta\lambda_1-(3\beta_1C_2-\beta_2C_1\\
&+4\beta_2C_2)\Delta\lambda_2\big]-\frac{L{\lambda}^5}{3E^{\frac{7}{2}}}\big[\beta_1C_1(2{\beta_1}^2-14\beta_1\beta_2-{\beta_2}^2)+\beta_2C_2(20{\beta_1}^2\\
&+7\beta_1\beta_2+2{\beta_2}^2)\big]\Big]=0.
\end{aligned}
\label{eq:63}
\end{equation}

For any given target surface $\mathcal{S}$, the fundamental quantities $E$ and $L$ can be calculated from its parametric equation. Then, by solving Eqs. (\ref{eq:62}) and (\ref{eq:63}), the following explicit expressions of $\lambda_1$ and $\lambda_2$ are obtained
\begin{equation}
\begin{aligned}
&\lambda_1=\sqrt{E}+\frac{h\big[{\beta_2}^3C_2-{\beta_1}^2C_1 (2\beta_1+3\beta_2)\big]L}{6\beta_1C_1(\beta_1+\beta_2)\sqrt{E}},\\
&\lambda_2=\sqrt{E}+\frac{h\big[{\beta_1}^3C_1+\beta_2C_2(6{\beta_1}^2+9\beta_1\beta_2+4{\beta_2}^2)\big]L}{6\beta_2 C_2(\beta_1+\beta_2)\sqrt{E}}.
\end{aligned}
\label{eq:64}
\end{equation}
Eq. (\ref{eq:64}) provides the analytical formulas for shape-programming of bilayer hyperelastic plates under plane-strain deformations. From these formulas, we found that the growth functions $\lambda(X)$ and $\Delta\lambda_k(X)$ are closely related to the in-plane stretching and out-of-plane bending of the plate, respectively.

\begin{figure}[htp]
\begin{minipage}{0.49\textwidth}
\centering \includegraphics[width=0.95\textwidth]{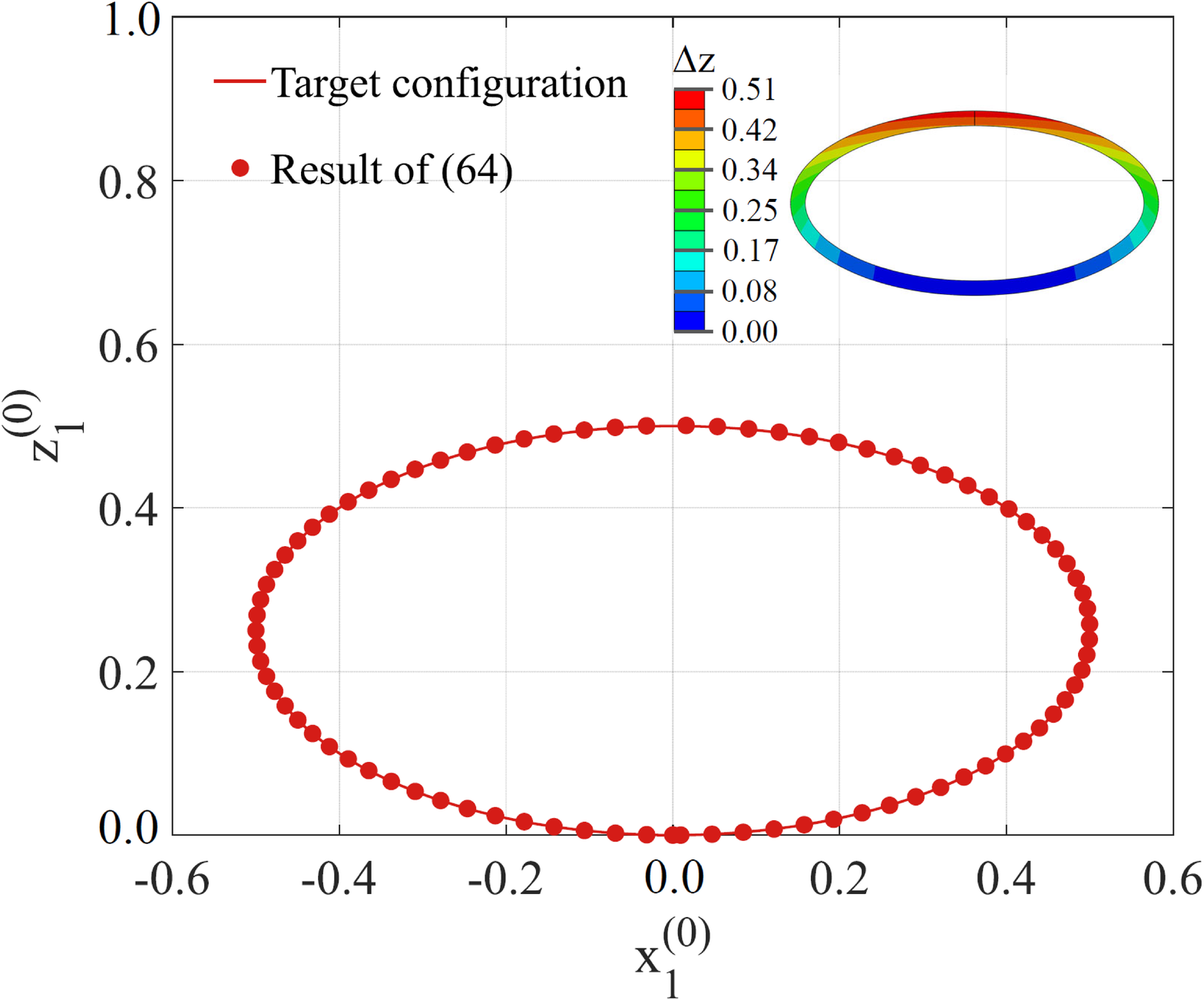}\\(a)\vspace{0.2cm}
\end{minipage}
\begin{minipage}{0.49\textwidth}
\centering \includegraphics[width=0.95\textwidth]{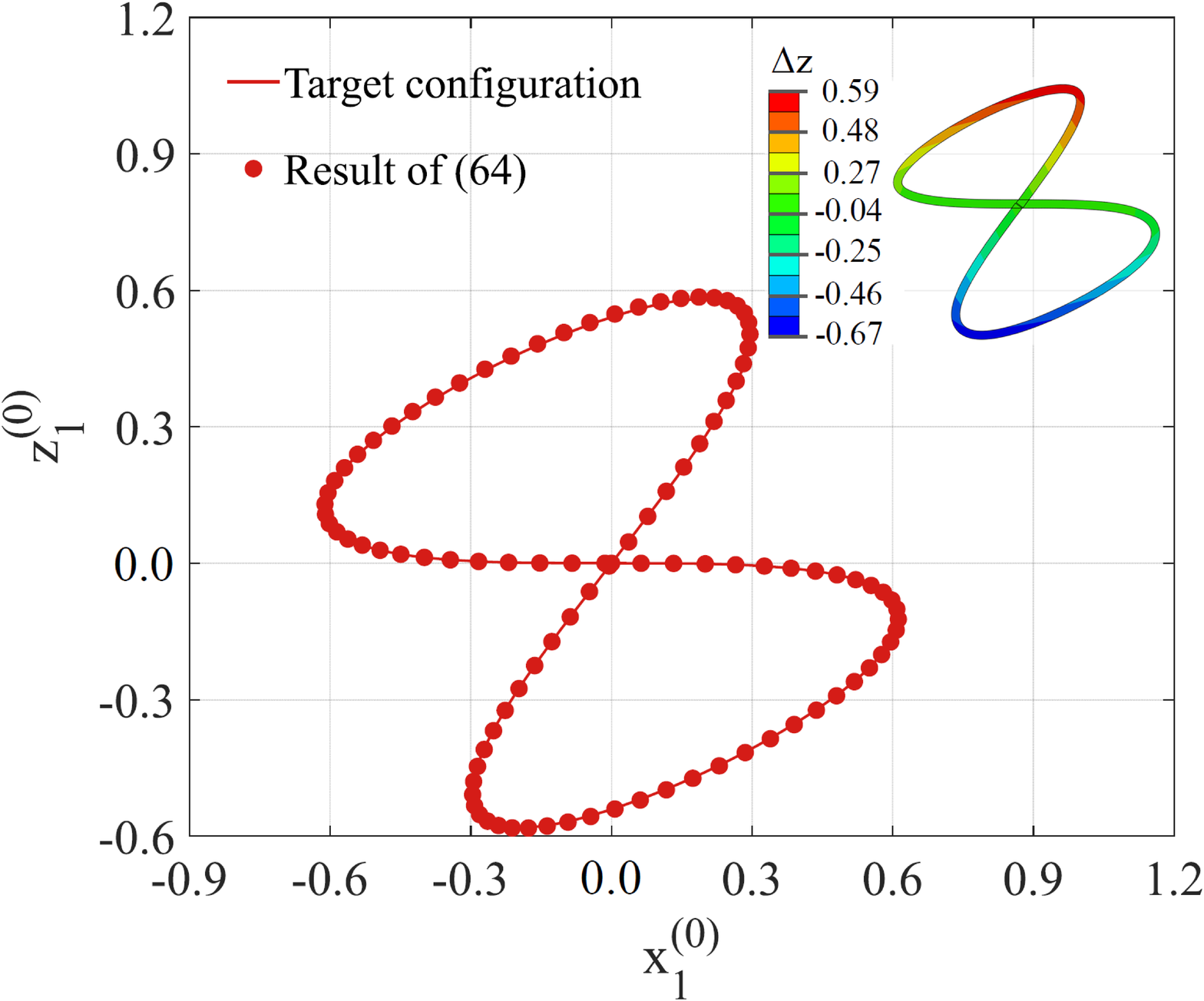}\\(b)\vspace{0.2cm}
\end{minipage}\\
\begin{minipage}{0.49\textwidth}
\centering \includegraphics[width=0.95\textwidth]{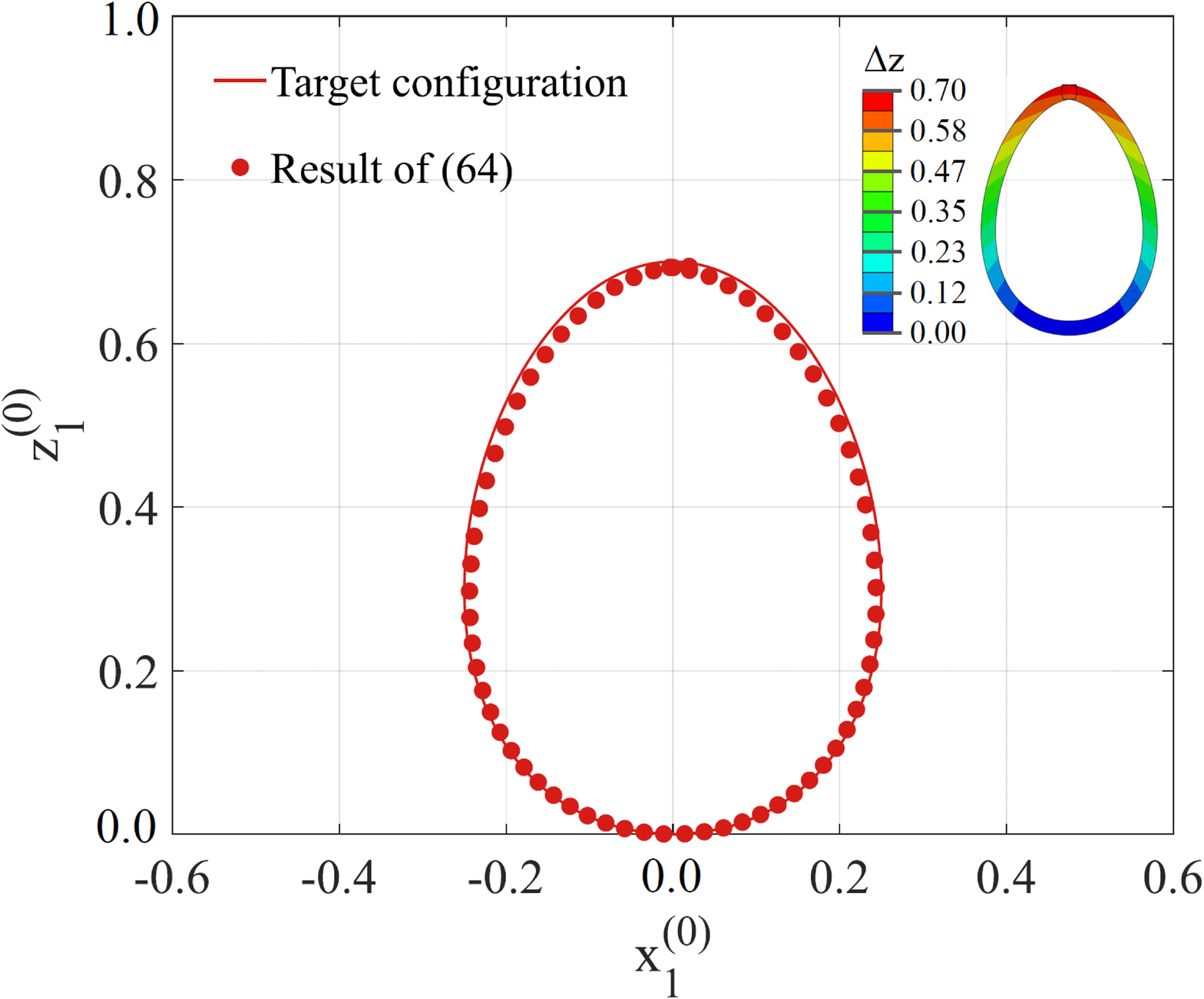}\\(c)\vspace{0.2cm}
\end{minipage}
\begin{minipage}{0.49\textwidth}
\centering \includegraphics[width=0.95\textwidth]{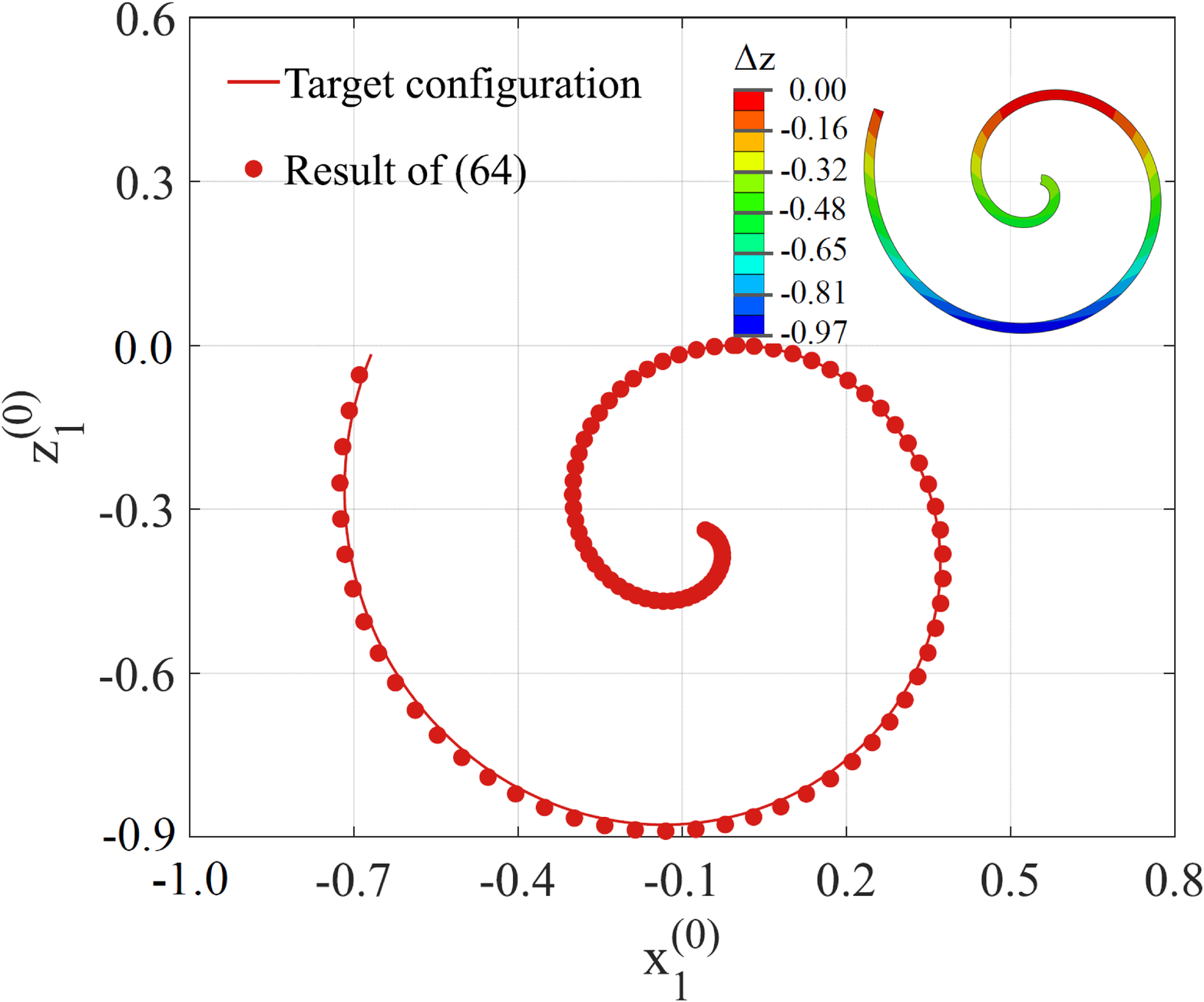}\\(d)\vspace{0.2cm}
\end{minipage}\\
\caption{Comparisons of the target surfaces $\mathcal{S}$ (solid curves) and the finite element simulation results with growth functions obtained from the formulas (\ref{eq:64}) (circular dots): (a) an elliptic curve; (b) a butterfly curve; (c) an oval curve; (d) a spiral curve. (The corresponding growth functions and the material and geometrical parameters are given in \ref{app:c}.)}
\label{fig:10}
\end{figure}

To demonstrate the efficiency of formulas (\ref{eq:64}), we introduce some illustrative examples. In these examples, the target shapes of $\mathcal{S}$ are selected to be an elliptic curve, a butterfly curve, an oval curve and a spiral curve. The parametric equations of these target shapes and the corresponding growth functions are listed in Eqs. (\ref{eq:C1})-(\ref{eq:C4}) of \ref{app:c}, where the total thickness of the plate sample is set to be $0.04$. To verify the accuracy of the analytical formulas (\ref{eq:64}), we also conduct finite element simulations on the growth behaviors of the bilayer plate, where the growth functions given in \ref{app:c} are adopted. For plane-strain deformations, only one section of the plate sample perpendicular to the $Y_k$-axis is considered in the numerical simulations, which is meshed by using the CPE8H elements with the size $0.002\times0.002$. The other settings of numerical calculations are same as that introduced in Section \ref{sec:3.1}. In Fig. \ref{fig:10}, we show the comparisons of the target shapes of $\mathcal{S}$ and the results obtained from finite element simulations (only the bottom lines of the section are plotted). It can be seen that the growth functions obtained from formula (\ref{eq:64}) can generate the target shapes of $\mathcal{S}$ accurately.

\subsection{Shape-programming of bilayer circular hyperelastic plates under axisymmetric deformations}
\label{sec:4.2}

In this subsection, we study the shape-programming of a bilayer circular hyperelastic plate under the assumption of axisymmetric deformations. The plate sample still has the reference configuration shown in Fig. \ref{fig:8}a. The two layers in the plate have inner radius $R_0$, outer radius $R_1$ and thickness $h_k$ $(k=1,2)$. To achieve the goal of arbitrary shape changes, the surface of the sample is supposed to be traction free, only except that some restrictions are applied to remove the rigid body motion. Growth in the plate occurs along the radial and the circumferential directions, which have the growth functions $\lambda_{kR}=\lambda_R+h\Delta\lambda_{kR}$ and $\lambda_{k\Theta}=\lambda_{\Theta}+h\Delta\lambda_{k\Theta}$ $(k=1,2)$.

The plate equations for axisymmetric deformations of bilayer circular hyperelastic plates have been derived in (\ref{eq:55})$_1$ and (\ref{eq:56}). To facilitate the following derivations, the components of the position vector $\mathbf{v}=\mathbf{x}_1^{(0)}$ in the plate equations will be replaced by the fundamental quantities of the deformed surface $\mathcal{S}$. Under the cylindrical coordinate system, these fundamental quantities are given by
\begin{equation}
\begin{aligned}
E=&{r_{1,R}^{(0)}}^2+{z_{1,R}^{(0)}}^2,\ \ \ \ F=0,\ \ \ \ G={r_1^{(0)}}^2,\\
L=&\frac{-z_{1,R}^{(0)}r_{1,RR}^{(0)}+r_{1,R}^{(0)}z_{1,RR}^{(0)}}{\sqrt{{r_{1,R}^{(0)}}^2+{z_{1,R}^{(0)}}^2}},\ \ \ M=0,\ \ \  N=\frac{r_1^{(0)}z_{1,R}^{(0)}}{\sqrt{{r_{1,R}^{(0)}}^2+{z_{1,R}^{(0)}}^2}}.
\end{aligned}
\label{eq:65}
\end{equation}
After the replacements, the plate equations (\ref{eq:55})$_1$ and (\ref{eq:56}) can be rewritten into
\begin{equation}
\begin{aligned}
&\frac{(\beta_1C_1+\beta_2C_2)h}{2RE^3G^{\frac{5}{2}}{\lambda_R}^2\lambda_{\Theta}} \Bigg\{3R^4G{\lambda_R}^5{\lambda_{\Theta}}^4E_{,R}G_{,R}+E^2\bigg[R^4{N}^2{\lambda_R}^5{\lambda_{\Theta}}^4\\
&+R^2G^2\lambda_R{\lambda_{\Theta}}^2E_{,R}G_{,R}\bigg]-2E^3G^2\bigg[2G{\lambda_R}^3+R\lambda_{\Theta}\Big[RLN\lambda_R\lambda_{\Theta}\\
&-G_{,R}\big[\lambda_{\Theta}(\lambda_R-R\lambda_{R,R})+R\lambda_R\lambda_{\Theta,R}\big]\Big]\bigg]+R^3E{\lambda_R}^4{\lambda_{\Theta}}^3\\
&\times\bigg[3R\lambda_R\lambda_{\Theta}{G_{,R}}^2+2G\Big[RLN\lambda_R\lambda_{\Theta}-3G_{,R}\big[\lambda_{\Theta}(\lambda_R+R\lambda_{R,R})\\
&+R\lambda_R\lambda_{\Theta,R}\big]\Big]\bigg]\Bigg\}+h^2\mathcal{W}_7(\lambda_R,\lambda_{\Theta},\Delta\lambda_{kR},\Delta\lambda_{k\Theta}) = 0,
\end{aligned}
\label{eq:66}
\end{equation}
\begin{equation}
\begin{aligned}
\frac{(\beta_1C_1+\beta_2C_2)hR\lambda_{\Theta}}{E^{\frac{3}{2}}G^{\frac{3}{2}}\lambda_R}N\big(E^2G-R^2{\lambda_R}^4{\lambda_{\Theta}}^2\big)+h^2\mathcal{W}_8(\lambda_R,\lambda_{\Theta},\Delta\lambda_{kR},\Delta\lambda_{k\Theta})= 0,
\end{aligned}
\label{eq:67}
\end{equation}
Besides the plate equations, from the moment boundary condition (\ref{eq:29}), we have another two equations
\begin{equation}
\begin{aligned}
M_{\Theta}=&\frac{2h^2R\beta_1\beta_2(C_1-C_2){\lambda_{\Theta}}^2(E^2G-R^2{\lambda_R}^4{\lambda_{\Theta}}^2)}{E^2G^{\frac{3}{2}}}\\
&+h^3\mathcal{W}_9(\lambda_R,\lambda_{\Theta},\Delta\lambda_{kR},\Delta\lambda_{k\Theta})=0,\\
M_R=&\frac{h^2\beta_1\beta_2(C_1-C_2){\lambda_R}^2G_{,R}[R^4{\lambda_R}^2{\lambda_{\Theta}}^4-EG^2]}{E^2G^{\frac{5}{2}}}\\
&+h^3\mathcal{W}_{10}(\lambda_R,\lambda_{\Theta},\Delta\lambda_{kR},\Delta\lambda_{k\Theta})=0.
\end{aligned}
\label{eq:68}
\end{equation}
Eq. (\ref{eq:68}) should be satisfied not only at the inner and outer lateral surfaces, but also in the whole region $R_0\leq R\leq R_1$. The lengthy expressions of the terms $\mathcal{W}_7$-$\mathcal{W}_{10}$ are omitted here for brevity, which can be found in the supplementary document.

For any given target surface $\mathcal{S}$, the fundamental quantities $E$, $G$, $L$ and $N$ can be calculated from the its parametric equation. Then, by setting the coefficients of $h^i$ in Eqs. (\ref{eq:66})-(\ref{eq:68}) to be zero, the growth functions $\lambda_R$, $\lambda_{\Theta}$, $\Delta\lambda_{kR}$ and $\Delta\lambda_{k\Theta}$ $(k=1,2)$ can be determined. Through this approach, we have
\begin{equation}
\begin{aligned}
&\lambda_{1R}=\sqrt{E}+\frac{h\big[{\beta_2}^3C_2-{\beta_1}^2C_1(2\beta_1+3\beta_2)\big]L}{6\beta_1C_1(\beta_1+\beta_2)\sqrt{E}},\\
&\lambda_{2R}=\sqrt{E}-\frac{h\big[{\beta_1}^3C_1+\beta_2C_2(6{\beta_1}^2+9\beta_1\beta_2+4{\beta_2}^2)\big]L}{6\beta_2C_2(\beta_1+\beta_2)\sqrt{E}},\\
&\lambda_{1\Theta}=\frac{\sqrt{G}}{R}+\frac{h\big[{\beta_2}^3C_2-{\beta_1}^2C_1(2\beta_1+3\beta_2)\big]N}{12\beta_1C_1R(\beta_1+\beta_2)\sqrt{G}},\\
&\lambda_{2\Theta}=\frac{\sqrt{G}}{R}-\frac{h\big[{\beta_1}^3C_1+\beta_2C_2(6{\beta_1}^2+9\beta_1\beta_2+4{\beta_2}^2)\big]N}{12\beta_2C_2R(\beta_1+\beta_2)\sqrt{G}}.
\end{aligned}
\label{eq:69}
\end{equation}
Eq. (\ref{eq:69}) provides the analytical formulas of shape-programming for the axisymmetric deformations of bilayer circular hyperelastic plates. From (\ref{eq:69}), it can be found that $\lambda_R$ and $\lambda_{\Theta}$ only depend on the first fundamental quantities $E$ and $G$, while $\Delta\lambda_{kR}$ and $\Delta\lambda_{k\Theta}$ also depend on the second fundamental quantities $L$ and $N$, as well as the material and geometrical parameters.

To demonstrate the efficiency of the formulas (\ref{eq:69}), we introduce some illustrative examples. In these examples, the dimensions of the circular plate are set to be $R_0=0.2$, $R_1=1.2$ and $h_1+h_2=0.06$. The target shapes of $\mathcal{S}$ are selected to be an ellipsoid, an oval surface, an Ipomoea cairica and a pot surface. The parametric equations for these target surfaces are listed in Eqs. (\ref{eq:C5})-(\ref{eq:C8}) of \ref{app:c}. By using the formulas given in (\ref{eq:69}), the growth functions corresponding to these target surfaces can be calculated, which are also listed in \ref{app:c}. To verify the accuracies of these growth functions, we conduct finite element simulations on the growth behaviors of the bilayer circular plate, where the growth functions given in \ref{app:c} are adopted. For axisymmetric deformations, only one section cutting along the radius direction of the circular plate is considered in the numerical simulations, which is meshed by using the CAX8H elements with the size $0.001\times0.001$. The other settings of numerical calculations are same as that introduced in Section 3.3.

\begin{figure}[htp]
\begin{minipage}{0.49\textwidth}
\centering \includegraphics[width=0.95\textwidth]{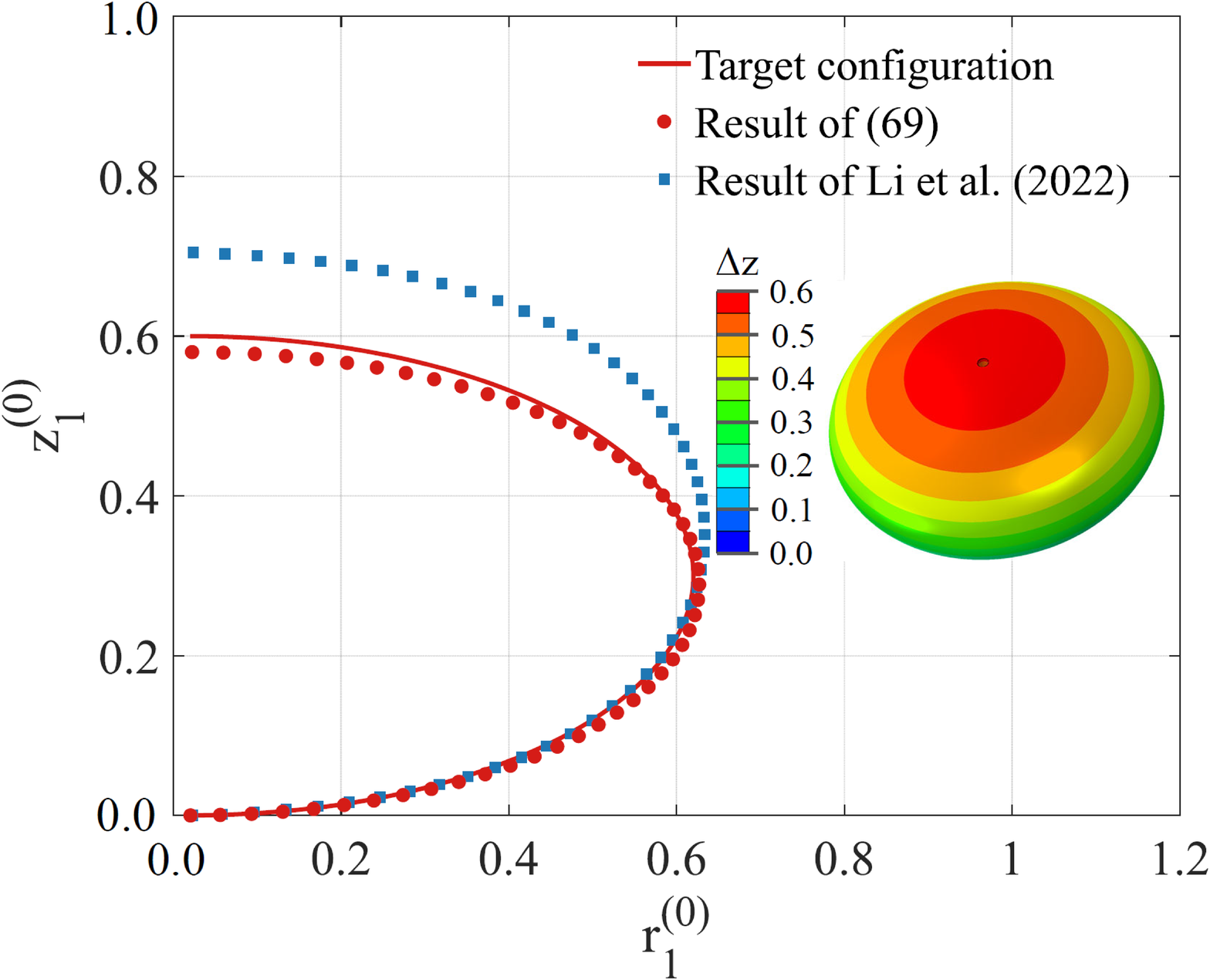}\\(a)\vspace{0.2cm}
\end{minipage}
\begin{minipage}{0.49\textwidth}
\centering \includegraphics[width=0.95\textwidth]{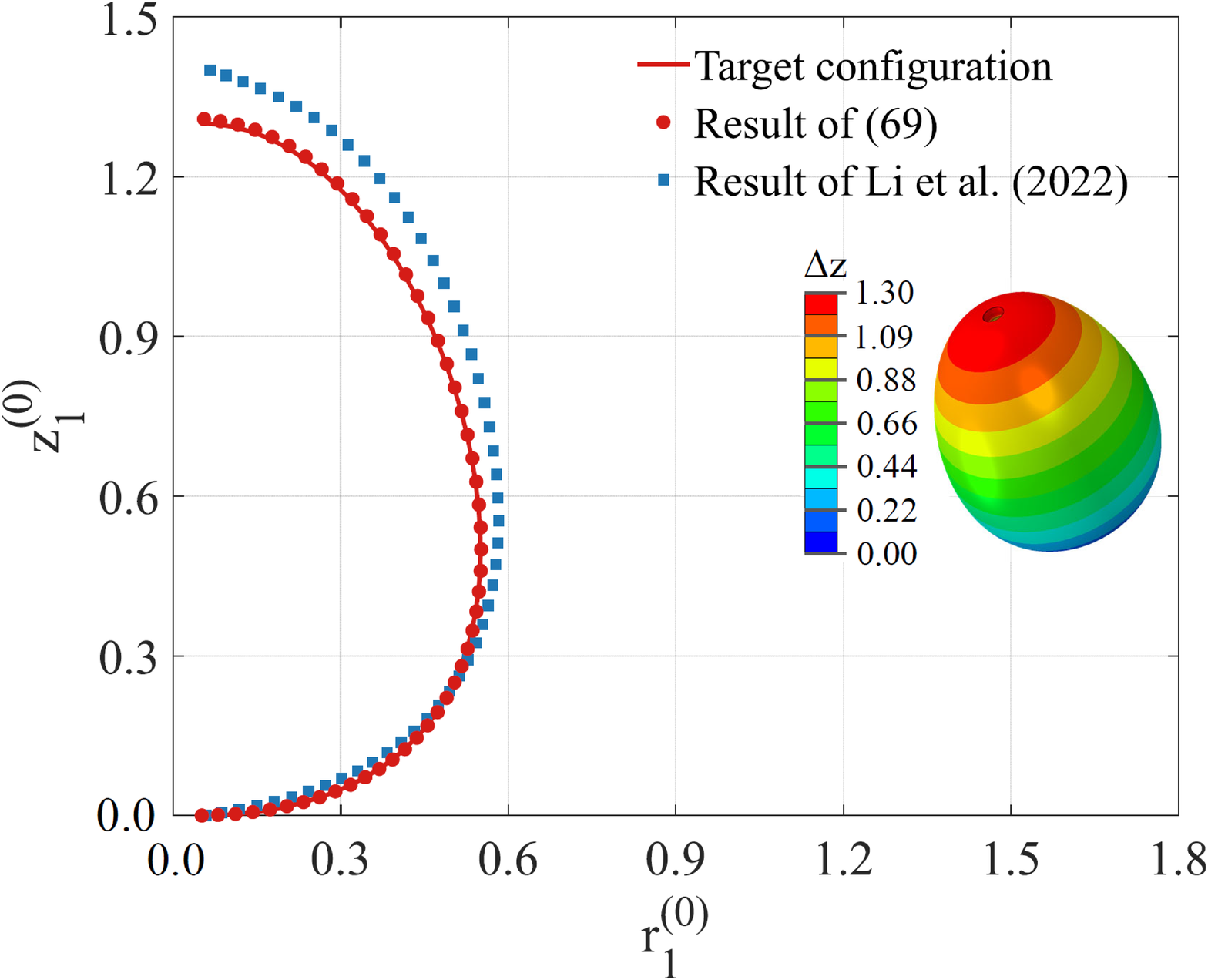}\\(b)\vspace{0.2cm}
\end{minipage}\\
\begin{minipage}{0.49\textwidth}
\centering \includegraphics[width=0.95\textwidth]{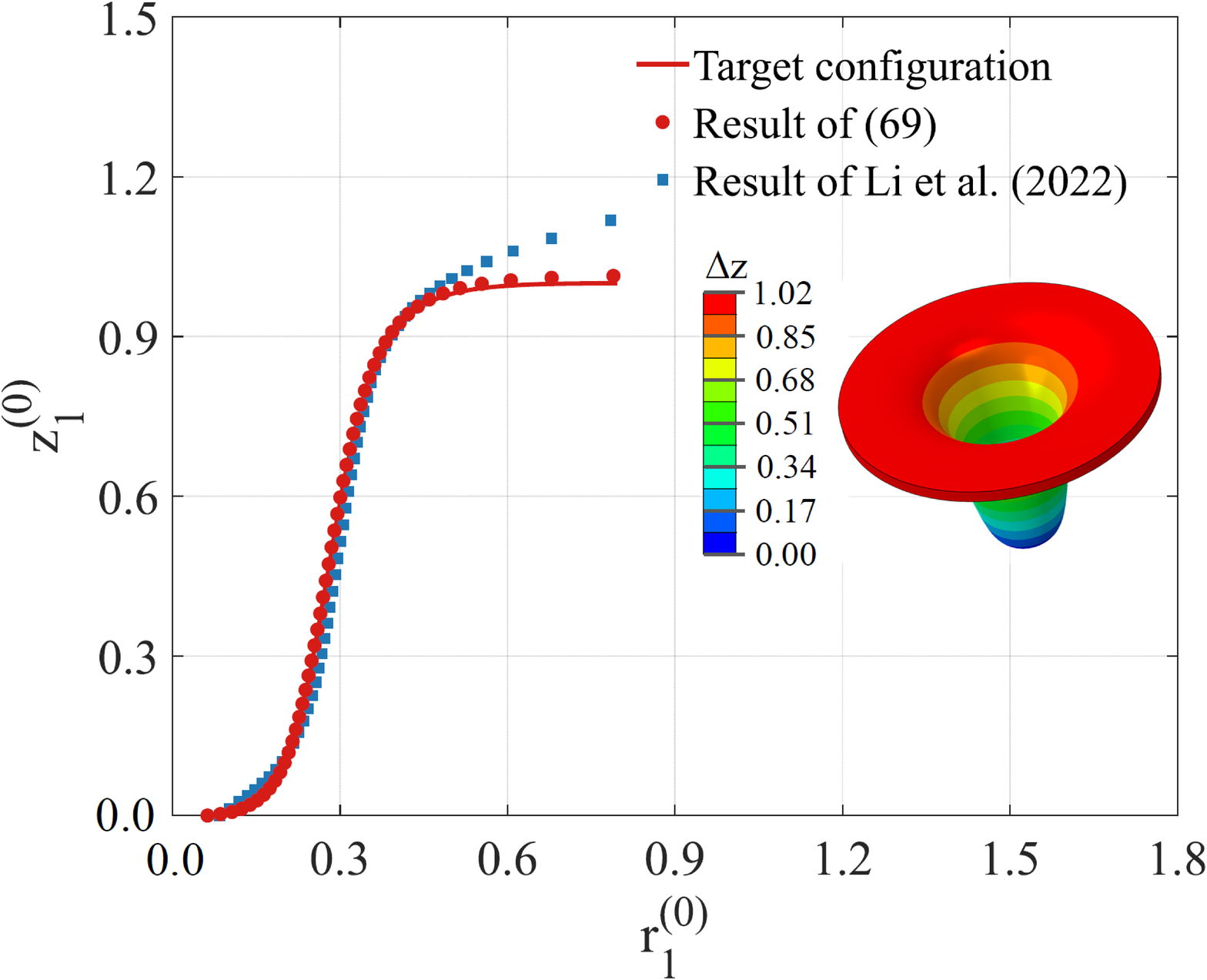}\\(c)\vspace{0.2cm}
\end{minipage}
\begin{minipage}{0.49\textwidth}
\centering \includegraphics[width=0.95\textwidth]{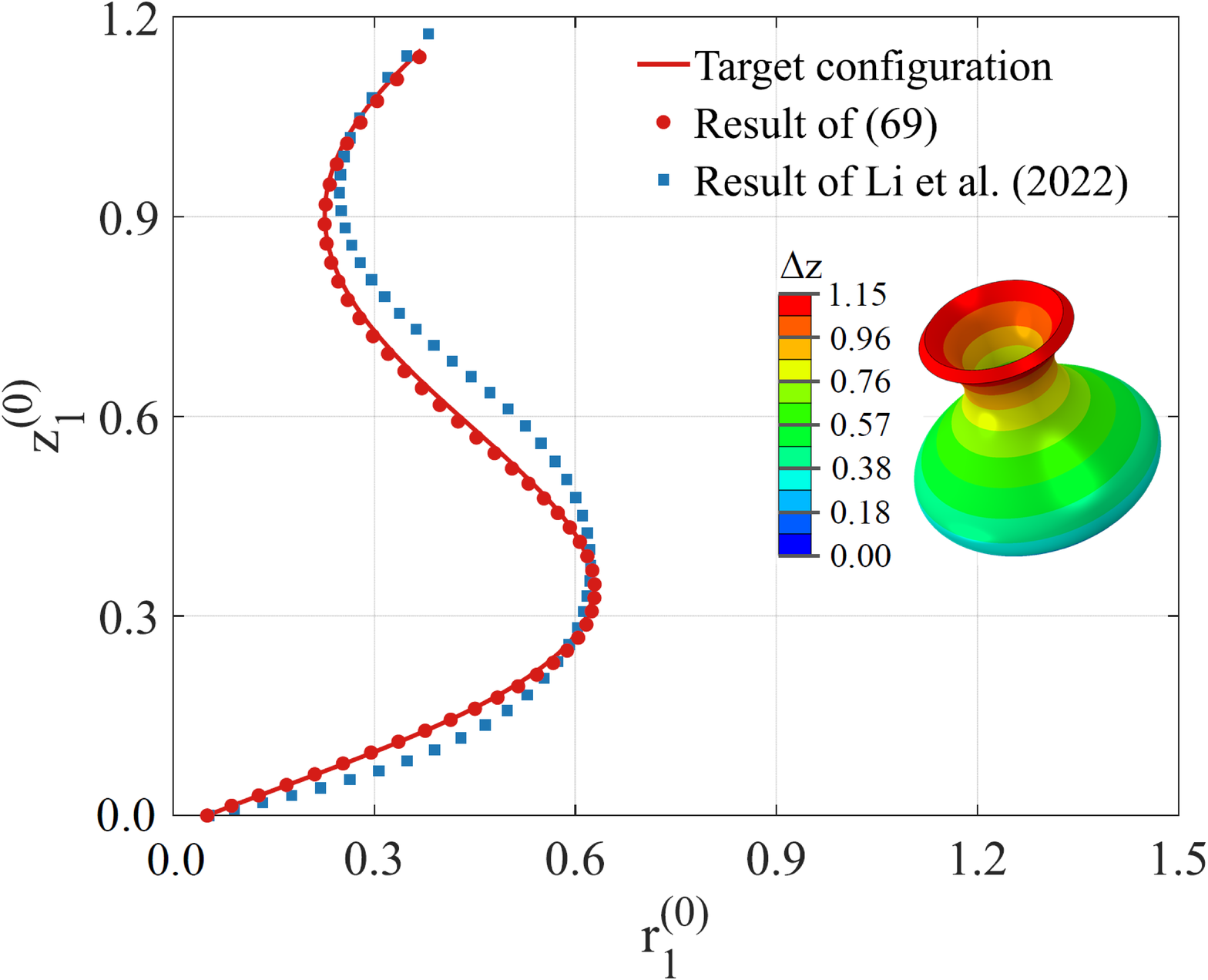}\\(d)\vspace{0.2cm}
\end{minipage}\\
\caption{Comparisons of the target surfaces of $\mathcal{S}$ (solid curves) and finite element simulation results obtained with growth functions derived from (\ref{eq:69}) (circular dots) and that given in \citet{li2022} (square dots): (a) an ellipsoid; (b) an oval surface; (c) an Ipomoea cairica; (d) a pot surface. (The corresponding growth functions and the material and geometrical parameters are given in \ref{app:c}.)}
\label{fig:11}
\end{figure}

In Fig. \ref{fig:11}, we show the comparisons of the target shapes of $\mathcal{S}$ and the results obtained from finite element simulations (only the bottom lines of the section are plotted). It can be seen that in these examples, the growth functions obtained from (\ref{eq:69}) can generate the target shapes $\mathcal{S}$ accurately. So, the efficiency of the formulas (\ref{eq:69}) is verified. In \citet{li2022}, the shape-programming of a single-layered circular plate was studied, where only the leading-order terms of the plate equations were taken into account. The formulas of shape-programming obtained in \citet{li2022} contain the leading-order terms of (\ref{eq:69}), which can already yield good results for the single-layered circular plate. However, these formulas are not sufficient for multi-layered plate samples. In fact, it can be seen that leading-order terms of (\ref{eq:69}) do not contain the material and geometrical parameters of the plate. Thus, the influences of these parameters cannot be reflected. In Fig. \ref{fig:11}, we also show the numerical results obtained with the growth functions derived from the formulas given in \citet{li2022}. It is found that the target shapes $\mathcal{S}$ and the numerical results show obvious differences.

\section{Conclusion}
\label{sec:6}

In this paper, we proposed a general multi-layered hyperelastic plate theory within the framework of nonlinear elasticity, which can be applied to study the growth-induced deformations of soft material plates with multi-layered structures. The following tasks have been accomplished in the current work:
\begin{itemize}

\item Starting from the 3D governing system and through a series expansion-truncation approach, the plate equation system for modeling the growth-induced deformations of multi-layered hyperelastic plates was established, which is applicable for the plate samples with general material properties, geometrical shapes and number of layers. There is also no restrictions on the growth functions and the deformation styles of plate samples.

\item The plate theory was applied to study the growth-induced deformations and instabilities of several typical multi-layered plate samples. Some analytical and numerical solutions of the plate equation system were derived, which show good consistencies with the finite element simulation results. It was also found that the current plate theory can be reduced to the multi-layered FvK-type plate theory under some assumptions of the displacement components.

\item Based on the current plate theory, the problems of shape-programming of some typical multi-layered hyperelastic plate samples were solved. The explicit analytical formulas of shape-programming were derived, which reveal the dependance of the growth functions on the geometrical quantities of 3D target surfaces and the properties of the plate samples. The accuracy of these formulas has been verified through finite element simulations.

\end{itemize}

In our opinion, the plate theory established in the current work are helpful for the design of intelligent soft devices with multi-layered plate structures, which would have wide potential applications in the engineering fields. In the future works, the current plate theory will be applied to develop new types of plate elements, then some much more complicated problems can be solved through the numerical approach.

\section*{Supplementary materials}

Supplementary materials associated with this paper include: (1) the full-form expressions of some quantities in the paper; (2) a movie to illustrate the growth process of the examples in Section \ref{sec:4}.


\appendix

\section{The iterative relations of some unknowns in deriving the 2D vector plate equation}
\label{app:a}

During the derivation procedure of the 2D vector plate equation in Section \ref{sec:2.2}, the iterative relations of some unknowns are derived, which are listed in this appendix.

In step one, the $8n$ unknowns $\mathbf{x}_k^{(2)}-\mathbf{x}_k^{(3)}$ and $p_k^{(1)}-p_k^{(2)}$ ($k=1,2,\cdots,n$) are given by
\begin{equation}
\begin{aligned}
\mathbf{x}_k^{(2)}=&-{\mathbb{B}_k}^{-1} \mathbf{f}_k^{(2)}+p_k^{(1)}{\mathbb{B}_k}^{-1}(\widehat{\mathbb{G}}_k\mathcal{R}_k^{(0)})^T\mathbf{k},\\
p_k^{(1)}=&-\frac{1}{D_k}\mathcal{R}_k^{(0)}:\left[\nabla_r \mathbf{x}_k^{(1)} \bar{\mathbb{G}}_k - {\mathbb{B}_k}^{-1} \mathbf{f}^{(2)} \otimes (\bar{\mathbb{G}}_k)^T\mathbf{k}\right],\\
\mathbf{x}_k^{(3)}=&-{\mathbb{B}_k}^{-1}\mathbf{f}_k^{(3)}+2p_k^{(1)}{\mathbb{B}_k}^{-1}(\mathcal{R}_k^{(1)}:\mathbb{A}_k^{(1)})^T(\widehat{\mathbb{G}}_k)^T\mathbf{k}\\
&+p_k^{(2)}{\mathbb{B}_k}^{-1}(\mathcal{R}_k^{(0)})^T(\widehat{\mathbb{G}}_k)^T\mathbf{k},\\
p_k^{(2)}=&-\frac{1}{D_k}\bigg\{\mathbb{R}_k^{(1)}[\mathbb{A}_k^{(1)},\mathbb{A}_k^{(1)}]+\mathcal{R}_k^{(0)}:\nabla_r\mathbf{x}_k^{(2)}\bar{\mathbb{G}}_k\\
&-\mathcal{R}_k^{(0)}:{\mathbb{B}_k}^{-1}\mathbf{f}_k^{(3)}\otimes(\bar{\mathbb{G}}_k)^T\mathbf{k}\\
&+2p_k^{(1)}\mathcal{R}_k^{(0)}:\left[{\mathbb{B}_k}^{-1}\left(\mathcal{R}_k^{(1)}[\mathbb{A}_k^{(1)}]\right)^T(\widehat{\mathbb{G}}_k)^T\mathbf{k} \right]\otimes(\bar{\mathbb{G}}_k)^T\mathbf{k}\bigg\},
\end{aligned}
\label{eq:A1}
\end{equation}
where $k=1,2,\cdots,n$ and
\begin{equation*}
\begin{aligned}
\mathbf{f}_k^{(2)}=&\nabla_r\cdot\mathbb{S}_k^{(0)}+\left[\left(\mathcal{A}_k^{(1)}-p_k^{(0)}\mathcal{R}_k^{(1)}\right):\left(\nabla_r\mathbf{x}_k^{(1)}\bar{\mathbb{G}}_k\right)\right]^T(\widehat{\mathbb{G}}_k)^T\mathbf{k},\\
\mathbf{f}_k^{(3)}=&\nabla_r\cdot\mathbb{S}_k^{(1)}+\left[\left(\mathcal{A}_k^{(1)}-p_k^{(0)}\mathcal{R}_k^{(1)}\right):\left(\nabla_r\mathbf{x}_k^{(2)}\bar{\mathbb{G}}_k\right)\right]^T(\widehat{\mathbb{G}}_k)^T\mathbf{k}\\
&+\left[\left(\mathcal{A}_k^{(2)}-p_k^{(0)}\mathcal{R}_k^{(2)}\right)[\mathbb{A}_k^{(1)},\mathbb{A}_k^{(1)}]\right]^T(\widehat{\mathbb{G}}_k)^T \mathbf{k},\\
D_k=&\mathcal{R}_k^{(0)}:\left[{\mathbb{B}_k}^{-1}(\widehat{\mathbb{G}}_k\mathcal{R}_k^{(0)})^T\mathbf{k}\right]\otimes(\bar{\mathbb{G}}_k)^T\mathbf{k},\\
B_{kij}=&\left(\mathcal{A}_k^{(1)}-p_k^{(0)}\mathcal{R}_k^{(1)}\right)_{minj}\left[(\widehat{\mathbb{G}}_k)^T\mathbf{k} \right]_m\left[(\bar{\mathbb{G}}_k)^T\mathbf{k}\right]_n.
\end{aligned}
\end{equation*}

In step two, the explicit expressions of $\mathbf{x}_1^{(1)}$ and $p_1^{(0)}$ in terms of $\mathbf{x}_1^{(0)}$ for incompressible neo-Hookean materials are given by
\begin{equation}
\begin{aligned}
\mathbf{x}_1^{(1)}=&\frac{-\mathbf{q}^--2C_k\nabla_r\mathbf{x}_t^{(0)}\bar{\mathbb{G}}_k(\widehat{\mathbb{G}}_k)^T\mathbf{k}+p_k^{(0)}(\mathbf{x}_{t,X}^{(0)}\wedge\mathbf{x}_{t,Y}^{(0)})}{2C_k\left[(\bar{\mathbb{G}}_k)^T\mathbf{k}(\widehat{\mathbb{G}}_k)^T\mathbf{k}\right]},\\
p_1^{(0)}=&\frac{2C_k\left[(\bar{\mathbb{G}}_k)^T\mathbf{k}(\widehat{\mathbb{G}}_k)^T\mathbf{k}\right]}{\mathrm{Det}\left(\bar{\mathbb{G}}_k\right) {\left|\mathbf{x}_{t,X}^{(0)}\wedge\mathbf{x}_{t,Y}^{(0)}\right|}^2}\\
&+\frac{\left[\mathbf{q}^-+2C_k\nabla_r\mathbf{x}_t^{(0)}\bar{\mathbb{G}}_k(\widehat{\mathbb{G}}_k)^T\mathbf{k}\right](\mathbf{x}_{t,X}^{(0)}\wedge\mathbf{x}_{t,Y}^{(0)})}{{\left|\mathbf{x}_{t,X}^{(0)}\wedge\mathbf{x}_{t,Y}^{(0)}\right|}^2}.
\end{aligned}
\label{eq:A2}
\end{equation}

In step three, the asymptotic expressions of $\mathbf{x}_{k+1}^{(0)}$, $\mathbf{x}_{k+1}^{(1)}$ and $p_{k+1}^{(0)}$ ($k=1,2,\cdots,n-1$) for incompressible neo-Hookean materials are given by
\begin{equation}
\begin{aligned}
\mathbf{x}_{k+1}^{(0)}=&\sum_{i=0}^{4}\tilde{\mathbf{x}}_k,\\
\mathbf{x}_{k+1}^{(1)}=&\frac{\tilde{\mathbf{s}}_k-2C_{k+1}\nabla_r\mathbf{x}_{k+1}^{(0)}\bar{\mathbb{G}}_{k+1}(\widehat{\mathbb{G}}_{k+1})^T\mathbf{k}+p_{k+1}^{(0)}(\mathbf{x}_{k+1,X}^{(0)}\wedge\mathbf{x}_{k+1,Y}^{(0)})}{2C_{k+1}\left[(\bar{\mathbb{G}}_{k+1})^T\mathbf{k}(\widehat{\mathbb{G}}_{k+1})^T\mathbf{k}\right]},\\
p_{k+1}^{(0)}=&\frac{2C_{k+1}\left[(\bar{\mathbb{G}}_{k+1})^T\mathbf{k}(\widehat{\mathbb{G}}_{k+1})^T\mathbf{k}\right]}{\mathrm{Det}\left(\bar{\mathbb{G}}_{k+1}\right){\left|\mathbf{x}_{k+1,X}^{(0)}\wedge\mathbf{x}_{k+1,Y}^{(0)}\right|}^2}\\
&-\frac{\left[\tilde{\mathbf{s}}_k-2 C_{k+1}\nabla_r\mathbf{x}_{k+1}^{(0)} \bar{\mathbb{G}}_{k+1} (\widehat{\mathbb{G}}_{k+1})^T\mathbf{k}\right] (\mathbf{x}_{k+1,X}^{(0)}\wedge\mathbf{x}_{k+1,Y}^{(0)})}{{\left|\mathbf{x}_{k+1,X}^{(0)}\wedge\mathbf{x}_{k+1,Y}^{(0)}\right|}^2},
\end{aligned}
\label{eq:A3}
\end{equation}
where $k=1,2,\cdots,n-1$ and
\begin{equation*}
\begin{aligned}
&\tilde{\mathbf{x}}_k=\displaystyle{\frac{{h_k}^i}{i!}}\mathbf{x}_k^{(i)},\ \ k=1,2,\cdots,n-1,\\
&\tilde{\mathbf{s}}_k=\displaystyle{\sum_{i=0}^{3} \left[\frac{{h_k}^i}{i!} \mathbb{S}_k^{(i)}\right]^T \mathbf{k}},\ \ k=1,2,\cdots,n-1.
\end{aligned}
\end{equation*}


\section{Some iterative relations for deriving the plate equations of bilayer circular plates}
\label{app:b}

In Section \ref{sec:3.3}, some iterative relations are obtained for deriving the plate equations of bilayer circular hyperelastic plates, which are listed in this appendix.

The expressions of $r_1^{(1)}$, $z_1^{(1)}$, $p_1^{(0)}$, $r_2^{(0)}$, $z_2^{(0)}$, $r_2^{(1)}$, $z_1^{(1)}$ and $p_2^{(0)}$ in terms of $r_1^{(0)}$ and $z_1^{(0)}$ are given by
\begin{small}
\begin{equation}
\begin{aligned}
r_1^{(1)}=&-\frac{R\lambda_{1R}\lambda_{1\Theta}z_{1,R}^{(0)}}{r_1^{(0)}\Big({r_{1,R}^{(0)}}^2+{z_{1,R}^{(0)}}^2\Big)},\ \ z_1^{(1)}=&\frac{R\lambda_{1R}\lambda_{1\Theta}r_{1,R}^{(0)}}{r_1^{(0)}\Big({r_{1,R}^{(0)}}^2+{z_{1,R}^{(0)}}^2\Big)},\\
p_1^{(0)}=&\frac{2C_1R^2{\lambda_{1R}}^2{\lambda_{1\Theta}}^2}{{r_1^{(0)}}^2\Big({r_{1,R}^{(0)}}^2+{z_{1,R}^{(0)}}^2\Big)},
\end{aligned}
\label{eq:B1}
\end{equation}

\begin{equation}
\begin{aligned}
r_2^{(0)}=&r_1^{(0)}-\frac{hR\beta_1\lambda_{1R}\lambda_{1\Theta}z_{1,R}^{(0)}}{r_1^{(0)}\Big({r_{1,R}^{(0)}}^2+{z_{1,R}^{(0)}}^2\Big)}+O(h^2),\\
z_2^{(0)}=&z_1^{(0)}+\frac{hR\beta_1\lambda_{1R}\lambda_{1\Theta}r_{1,R}^{(0)}}{r_1^{(0)}\Big({r_{1,R}^{(0)}}^2+{z_{1,R}^{(0)}}^2\Big)}+O(h^2),
\end{aligned}
\label{eq:B2}
\end{equation}

\begin{equation}
\begin{aligned}
r_2^{(1)}=&-\frac{R\lambda_{2R}\lambda_{2\Theta}z_{1,R}^{(0)}}{r_1^{(0)}\Big({r_{1,R}^{(0)}}^2+{z_{1,R}^{(0)}}^2\Big)}
-\frac{h\beta_1}{C_2R^2{\lambda_{1R}}^2\lambda_{1\Theta}\lambda_{2R}\lambda_{2\Theta}{r_1^{(0)}}^3\Big({r_{1,R}^{(0)}}^2+{z_{1,R}^{(0)}}^2\Big)^3}\\
&\times\bigg[C_1R^2\lambda_{1R} {\lambda_{1\Theta}}^2\Big[\Big({r_{1,R}^{(0)}}^2+{z_{1,R}^{(0)}}^2\Big)^2{r_1^{(0)}}^2+3R^2{\lambda_{1R}}^4{\lambda_{1\Theta}}^2\Big] r_1^{(0)}r_{1,R}^{(0)}\\
&\times\big(r_{1,R}^{(0)}r_{1,RR}^{(0)}+z_{1,R}^{(0)}z_{1,RR}^{(0)}\big)+R^3{\lambda_{1R}}^2\lambda_{1\Theta}\Big[3C_1R{\lambda_{1R}}^3{\lambda_{1\Theta}}^3 {r_{1,R}^{(0)}}^2+C_2R\\
&\times\lambda_{1R}\lambda_{1\Theta}{\lambda_{2R}}^2{\lambda_{2\Theta}}^2\Big({z_{1,R}^{(0)}}^2-{r_{1,R}^{(0)}}^2\Big)-\big(3C_1{\lambda_{1R}}^2{\lambda_{1\Theta}}^2 -C_2{\lambda_{2R}}^2{\lambda_{2\Theta}}^2\big)\\
&\times\big[R\lambda_{1\Theta}\lambda_{1R,R}+\lambda_{1R}(\lambda_{1\Theta}+R\lambda_{1\Theta,R})\big]r_1^{(0)}r_{1,R}^{(0)} -C_2R\lambda_{1R}\lambda_{1\Theta}{\lambda_{2R}}^2\\
&\times{\lambda_{2\Theta}}^2r_1^{(0)}r_{1,RR}^{(0)}\Big]\Big({r_{1,R}^{(0)}}^2+{z_{1,R}^{(0)}}^2\Big)-C_1\big[R\lambda_{1\Theta} [R\lambda_{1\Theta}\lambda_{1R,R}-\lambda_{1R}(\lambda_{1\Theta}\\
&+R\lambda_{1\Theta,R})]+{\lambda_{1R}}^3r_1^{(0)}r_{1,R}^{(0)}\big]{r_1^{(0)}}^3r_{1,R}^{(0)}\Big({r_{1,R}^{(0)}}^2+{z_{1,R}^{(0)}}^2\Big)^2\bigg]+O(h^2),
\end{aligned}
\label{eq:B3}
\end{equation}

\begin{equation}
\begin{aligned}
z_2^{(1)}=&\frac{R\lambda_{2R}\lambda_{2\Theta}r_{1,R}^{(0)}}{r_1^{(0)}\Big({r_{1,R}^{(0)}}^2+{z_{1,R}^{(0)}}^2\Big)}-\frac{h\beta_1}{C_2R^2{\lambda_{1R}}^2 \lambda_{1\Theta}\lambda_{2R}\lambda_{2\Theta}{r_1^{(0)}}^3\Big({r_{1,R}^{(0)}}^2+{z_{1,R}^{(0)}}^2\Big)^3}\\
&\times\bigg[C_1R^2\lambda_{1R}{\lambda_{1\Theta}}^2\Big[\Big({r_{1,R}^{(0)}}^2+{z_{1,R}^{(0)}}^2\Big)^2{r_1^{(0)}}^2+3R^2{\lambda_{1R}}^4{\lambda_{1\Theta}}^2\Big] r_1^{(0)}z_{1,R}^{(0)}\\
&\times\Big(r_{1,R}^{(0)}r_{1,RR}^{(0)}+z_{1,R}^{(0)}z_{1,RR}^{(0)}\Big)+R^3{\lambda_{1R}}^2\lambda_{1\Theta}\Big[R\lambda_{1R}\lambda_{1\Theta}\big(3C_1R {\lambda_{1R}}^2{\lambda_{1\Theta}}^2\\
&-2C_2{\lambda_{2R}}^2{\lambda_{2\Theta}}^2 \big)r_{1,R}^{(0)}z_{1,R}^{(0)} - \Big(3C_1{\lambda_{1R}}^2{\lambda_{1\Theta}}^2 -C_2{\lambda_{2R}}^2{\lambda_{2\Theta}}^2\Big) \big[R\lambda_{1\Theta}\\
&\times\lambda_{1R,R}+\lambda_{1R}(\lambda_{1\Theta}+R\lambda_{1\Theta,R})\big]r_1^{(0)}z_{1,R}^{(0)}-C_2R\lambda_{1R}\lambda_{1\Theta} {\lambda_{2R}}^2{\lambda_{2\Theta}}^2r_1^{(0)}\\
&\times z_{1,RR}^{(0)}\Big]-C_1\Big[R\lambda_{1\Theta}[R\lambda_{1\Theta}\lambda_{1R,R}-\lambda_{1R}(\lambda_{1\Theta}+R\lambda_{1\Theta,R})]+{\lambda_{1R}}^3r_1^{(0)}\\
&\times r_{1,R}^{(0)}\Big] {r_1^{(0)}}^3z_{1,R}^{(0)}\Big({r_{1,R}^{(0)}}^2+{z_{1,R}^{(0)}}^2\Big)^2\bigg]+O(h^2),
\end{aligned}
\label{eq:B4}
\end{equation}

\begin{equation}
\begin{aligned}
p_2^{(0)}=&\frac{2C_2R^2{\lambda_{2R}}^2{\lambda_{2\Theta}}^2}{{r_1^{(0)}}^2\Big({r_{1,R}^{(0)}}^2+{z_{1,R}^{(0)}}^2\Big)}-\frac{2h\beta_1}{R\lambda_{1R}\lambda_{1\Theta} {r_1^{(0)}}^4\Big({r_{1,R}^{(0)}}^2+{z_{1,R}^{(0)}}^2\Big)^3}\bigg[R^4{\lambda_{1R}}^2\\
&\times{\lambda_{1\Theta}}^2\big(C_1{\lambda_{1R}}^2{\lambda_{1\Theta}}^2-2C_2{\lambda_{2R}}^2{\lambda_{2\Theta}}^2\big)\Big[ \Big({r_{1,R}^{(0)}}^2+{z_{1,R}^{(0)}}^2\Big)z_{1,R}^{(0)}+r_1^{(0)}\\
&\times\Big(r_{1,R}^{(0)}z_{1,RR}^{(0)}-z_{1,R}^{(0)}r_{1,RR}^{(0)}\Big)\Big]-C_1\Big[R^2{\lambda_{1\Theta}}^2 \Big(r_{1,R}^{(0)}z_{1,RR}^{(0)}-z_{1,R}^{(0)}r_{1,RR}^{(0)}\Big)\\
&+{\lambda_{1R}}^2r_1^{(0)}z_{1,R}^{(0)}\Big]{r_1^{(0)}}^3\Big({r_{1,R}^{(0)}}^2+{z_{1,R}^{(0)}}^2\Big)^2\bigg]+O(h^2).
\end{aligned}
\label{eq:B5}
\end{equation}
\end{small}

The expressions of $r_k^{(2)}-r_k^{(3)}$, $z_k^{(2)}-z_k^{(3)}$, $p_k^{(1)}-p_k^{(2)}$ ($k=1,2$) are listed in Eqs. (\ref{eq:B6})-(\ref{eq:B11}). By using (\ref{eq:B1})-(\ref{eq:B5}), all the iterative relations can be expressed in terms of $r_1^{(0)}$ and $z_1^{(0)}$.
\begin{small}
\begin{equation}
\begin{aligned}
r_k^{(2)}=&\frac{1}{2C_kR^2{\lambda_{kR}}^3{\lambda_{k\Theta}}^2 r_k^{(0)}\Big({r_{k,R}^{(0)}}^2+{z_{k,R}^{(0)}}^2\Big)}\bigg[R{\lambda_{kR}}^2\lambda_{k\Theta}\Big({r_k^{(0)}}^2p_{k,R}^{(0)}r_{k,R}^{(0)}+2C_k\\
&\times R\lambda_{kR}\lambda_{k\Theta}r_k^{(1)}z_{k,R}^{(0)}\Big)\Big(z_k^{(1)}r_{k,R}^{(0)}-r_k^{(1)}z_{k,R}^{(0)}\Big)+2C_kr_k^{(0)}\Big[{\lambda_{kR}}^3 r_k^{(0)}{r_{k,R}^{(0)}}^2\\
&+R\lambda_{k\Theta}\Big[R{\lambda_{kR}}^3\lambda_{k\Theta}z_{k,R}^{(0)} \Big(z_k^{(1)}r_{k,R}^{(1)}-r_k^{(1)}z_{k,R}^{(1)}\Big)+R\lambda_{k\Theta}\lambda_{kR,R}r_{k,R}^{(0)}\\
&\times \Big({r_{k,R}^{(0)}}^2+{z_{k,R}^{(0)}}^2\Big)-\lambda_{kR}r_{k,R}^{(0)} \Big[(\lambda_{k\Theta} +R\lambda_{k\Theta,R})\Big({r_{k,R}^{(0)}}^2+{z_{k,R}^{(0)}}^2\Big)\\
&+R\lambda_{k\Theta}\Big(r_{k,R}^{(0)}r_{k,RR}^{(0)}+z_{k,R}^{(0)}z_{k,RR}^{(0)}\Big)\Big]\Big]\Big]\bigg]
\end{aligned}
\label{eq:B6}
\end{equation}

\begin{equation}
\begin{aligned}
z_k^{(2)}=&\frac{1}{2C_kR^2{\lambda_{kR}}^3{\lambda_{k\Theta}}^2 r_k^{(0)}\Big({r_{k,R}^{(0)}}^2+{z_{k,R}^{(0)}}^2\Big)}\bigg[2C_kR^2{\lambda_{kR}}^3{\lambda_{k\Theta}}^2r_k^{(1)}r_{k,R}^{(0)}z_{k,R}^{(0)}\\
&-R{\lambda_{kR}}^2\lambda_{k\Theta}r_k^{(1)}\Big[{r_k^{(0)}}^2p_{k,R}^{(0)}{z_{k,R}^{(0)}}^2+2C_kR\lambda_{kR}\lambda_{k\Theta}r_{k,R}^{(0)}\Big(z_k^{(1)}r_{k,R}^{(0)}-r_k^{(0)}\\
&\times z_{k,R}^{(1)}\Big)\Big]+r_k^{(0)}\Big[R{\lambda_{kR}}^2\lambda_{k\Theta}z_k^{(1)} r_{k,R}^{(0)}\Big(r_k^{(0)}p_{k,R}^{(0)}z_{k,R}^{(0)}-2C_kR\lambda_{kR}\lambda_{k\Theta}\\
&\times r_{k,R}^{(1)}\Big)+2C_kz_{k,R}^{(0)}\Big[{\lambda_{kR}}^3r_k^{(0)}r_{k,R}^{(0)}-R \lambda_{k\Theta} \Big[(\lambda_{kR}\lambda_{k\Theta}+R\lambda_{kR}\lambda_{k\Theta,R}\\
&-R\lambda_{k\Theta}\lambda_{kR,R})\Big({r_{k,R}^{(0)}}^2+{z_{k,R}^{(0)}}^2\Big)+R\lambda_{kR}\lambda_{k\Theta}\Big(r_{k,R}^{(0)}r_{k,RR}^{(0)}+z_{k,R}^{(0)}z_{k,RR}^{(0)}\Big)\Big]\Big]\Big]\bigg]
\end{aligned}
\label{eq:B7}
\end{equation}

\begin{equation}
\begin{aligned}
p_k^{(1)}=&\frac{1}{R\lambda_{kR}\lambda_{k\Theta}{r_k^{(0)}}^2\Big({r_{k,R}^{(0)}}^2+{z_{k,R}^{(0)}}^2\Big)}\bigg[2C_kR^2{\lambda_{kR}}^2{\lambda_{k\Theta}}^2 {r_k^{(1)}}^2z_{k,R}^{(0)}+R\lambda_{kR}\\
&\times\lambda_{k\Theta} r_k^{(1)}\Big[{r_k^{(0)}}^2p_{k,R}^{(0)}r_{k,R}^{(0)}+2C_kR\lambda_{kR}\lambda_{k\Theta}\Big(r_k^{(0)}z_{k,R}^{(1)}-z_k^{(1)}r_{k,R}^{(0)}\Big)\Big]\\
&+r_k^{(0)}\Big[R\lambda_{kR}\lambda_{k\Theta}z_k^{(1)}(r_k^{(0)} p_{k,R}^{(0)}z_{k,R}^{(0)}-2C_kR\lambda_{kR}\lambda_{k\Theta}r_{k,R}^{(1)})+2C_k\\
&\times\Big[{\lambda_{kR}}^2 r_k^{(0)}z_{k,R}^{(0)}+R^2{\lambda_{k\Theta}}^2\Big(r_{k,R}^{(0)}z_{k,RR}^{(0)}-z_{k,R}^{(0)}r_{k,RR}^{(0)}\Big)\Big]\Big]\bigg]
\end{aligned}
\label{eq:B8}
\end{equation}

\begin{equation}
\begin{aligned}
r_k^{(3)}=&\frac{1}{2C_kR^2{\lambda_{kR}}^3{\lambda_{k\Theta}}^2r_k^{(0)}\Big({r_{k,R}^{(0)}}^2+{z_{k,R}^{(0)}}^2\Big)}\Bigg[R{\lambda_{kR}}^2 \lambda_{k\Theta}{r_k^{(0)}}^2r_{k,R}^{(0)}\Big(z_k^{(2)}p_{k,R}^{(0)}\\
&\times r_{k,R}^{(0)}+z_k^{(1)}p_{k,R}^{(1)}r_{k,R}^{(0)}-r_k^{(2)}p_{k,R}^{(0)}z_{k,R}^{(0)}-r_k^{(1)} p_{k,R}^{(1)}z_{k,R}^{(0)}+p_k^{(1)} r_{k,R}^{(1)}z_{k,R}^{(0)}\\
&-p_k^{(1)}r_{k,R}^{(0)}z_{k,R}^{(1)}\Big)+2C_kR^2{\lambda_{kR}}^3{\lambda_{k\Theta}}^2 z_{k,R}^{(0)}\Big[\Big(z_k^{(1)}r_{k,R}^{(0)}-3r_k^{(1)}z_{k,R}^{(0)}\Big)r_k^{(2)}\\
&+2\Big(z_k^{(2)}r_{k,R}^{(0)}+z_k^{(1)}r_{k,R}^{(1)}-r_k^{(1)}z_{k,R}^{(1)}\Big)r_k^{(1)}\Big]-r_k^{(0)}\Big[R{\lambda_{kR}}^2\lambda_{k\Theta}{r_k^{(1)}}^2p_{k,R}^{(0)}\\
&\times r_{k,R}^{(0)}z_{k,R}^{(0)}-{\lambda_{kR}}^2r_k^{(1)}\Big[R\lambda_{k\Theta }z_k^{(1)} p_{k,R}^{(0)}{r_{k,R}^{(0)}}^2+2C_k\lambda_{kR}\Big({r_{k,R}^{(0)}}^2-R^2{\lambda_{k\Theta}}^2\\
&\times z_{k,R}^{(0)}z_{k,R}^{(2)}\Big)\Big]-2C_kR\lambda_{k\Theta}\Big[R{\lambda_{kR}}^3 \lambda_{k\Theta}z_{k,R}^{(0)}\Big(2z_k^{(2)}r_{k,R}^{(1)}+z_k^{(1)}r_{k,R}^{(2)}\\
&-2r_k^{(2)}z_{k,R}^{(1)}\Big) + R\lambda_{k\Theta}\lambda_{kR,R}r_{k,R}^{(0)}\Big(r_{k,R}^{(0)}r_{k,R}^{(1)}+z_{k,R}^{(0)}z_{k,R}^{(1)}\Big)-\lambda_{kR}r_{k,R}^{(0)}(R\\
&\times\lambda_{k\Theta,R}\Big(r_{k,R}^{(0)}r_{k,R}^{(1)}+z_{k,R}^{(0)} z_{k,R}^{(1)}\Big)+\lambda_{k\Theta}\Big[r_{k,R}^{(0)}\Big(r_{k,R}^{(1)}+Rr_{k,RR}^{(1)}\Big)+z_{k,R}^{(0)}\\
&\times\Big(z_{k,R}^{(1)}+Rz_{k,RR}^{(1)}\Big)\Big]\Big]\Big]\Bigg]
\end{aligned}
\label{eq:B9}
\end{equation}

\begin{equation}
\begin{aligned}
z_k^{(3)}=&\frac{1}{2C_kR^2{\lambda_{kR}}^3{\lambda_{k\Theta}}^2r_k^{(0)}\Big({r_{k,R}^{(0)}}^2+{z_{k,R}^{(0)}}^2\Big)}\bigg[R{\lambda_{kR}}^2\lambda_{k\Theta}{r_k^{(1)}}^2\Big(4C_kR\lambda_{kR}\lambda_{k\Theta}\\
&\times r_{k,R}^{(0)}z_{k,R}^{(1)}-r_k^{(0)}p_{k,R}^{(0)}{z_{k,R}^{(0)}}^2\Big)+R{\lambda_{kR}}^2\lambda_{k\Theta} r_k^{(2)}\Big[2C_kR\lambda_{kR}\lambda_{k\Theta}\Big(3r_k^{(1)}\\
&\times r_{k,R}^{(0)}z_{k,R}^{(0)}+2r_{k,R}^{(0)}z_{k,R}^{(1)}-z_k^{(1)}{r_{k,R}^{(0)}}^2\Big)-{r_k^{(0)}}^2p_{k,R}^{(0)}{z_{k,R}^{(0)}}^2\Big]+{\lambda_{kR}}^2r_k^{(1)}\Big[R\\
&\times\lambda_{k\Theta} z_k^{(1)}r_{k,R}^{(0)}\Big(r_k^{(0)}p_{k,R}^{(0)}z_{k,R}^{(0)}-4C_kR\lambda_{kR}\lambda_{k\Theta}r_{k,R}^{(1)}\Big)-4C_kR^2\lambda_{kR}{\lambda_{k\Theta}}^2\\
&\times z_k^{(2)}{r_{k,R}^{(0)}}^2+r_k^{(0)}\Big[2C_k\lambda_{kR} r_{k,R}^{(0)}\Big(z_{k,R}^{(0)}+R^2{\lambda_{k\Theta}}^2z_{k,R}^{(2)}\Big)-R\lambda_{k\Theta} r_k^{(0)}\\
&\times p_{k,R}^{(1)}{z_{k,R}^{(0)}}^2\Big]\Big]-R\lambda_{k\Theta}r_k^{(0)}\Big[{\lambda_{kR}}^2z_k^{(2)}r_{k,R}^{(0)}\Big(4C_kR\lambda_{kR}\lambda_{k\Theta}r_{k,R}^{(1)}-r_k^{(0)}\\
&\times p_{k,R}^{(0)}z_{k,R}^{(0)}\Big)+{\lambda_{kR}}^2z_k^{(1)} r_{k,R}^{(0)}\Big(2C_kR\lambda_{kR}\lambda_{k\Theta}r_{k,R}^{(2)}-r_k^{(0)}p_{k,R}^{(1)}z_{k,R}^{(0)}\Big)\\
&+z_{k,R}^{(0)}\Big[{\lambda_{kR}}^2p_k^{(1)}r_k^{(0)}\Big(r_{k,R}^{(0)}z_{k,R}^{(1)}-r_{k,R}^{(1)}z_{k,R}^{(0)}\Big)-2C_kR\lambda_{k\Theta} \lambda_{1R,R}\Big(r_{k,R}^{(0)}\\
&\times r_{k,R}^{(1)}+z_{k,R}^{(0)}z_{k,R}^{(1)}\Big)+2C_k\lambda_{kR}\Big[R\lambda_{k\Theta,R}(r_{k,R}^{(0)}r_{k,R}^{(1)}+z_{k,R}^{(0)} z_{k,R}^{(1)})+\lambda_{k\Theta}\\
&\times\Big[r_{k,R}^{(0)}(r_{k,R}^{(1)}+Rr_{k,RR}^{(1)})+z_{k,R}^{(0)}\Big(z_{k,R}^{(1)}+Rz_{k,RR}^{(1)}\Big)\Big]\Big]\Big]\Big]\bigg]
\end{aligned}
\label{eq:B10}
\end{equation}

\begin{equation}
\begin{aligned}
p_k^{(2)}=&\frac{1}{R{\lambda_{kR}}^2\lambda_{k\Theta}{r_k^{(0)}}^2\Big({r_{k,R}^{(0)}}^2+{z_{k,R}^{(0)}}^2\Big)}\bigg[R{\lambda_{kR}}^2\lambda_{k\Theta} {r_k^{(1)}}^2(r_k^{(0)}p_{k,R}^{(0)}r_{k,R}^{(0)}+4C_k\\
&R\lambda_{kR}\lambda_{k\Theta}z_{k,R}^{(1)})+R{\lambda_{kR}}^2\lambda_{k\Theta}r_k^{(2)}\Big[{r_k^{(0)}}^2 p_{k,R}^{(0)}r_{k,R}^{(0)}+2C_kR\lambda_{kR}\lambda_{k\Theta}\Big(2r_k^{(1)}\\
&\times z_{k,R}^{(0)}+2r_k^{(0)}z_{k,R}^{(1)}-z_k^{(1)}r_{k,R}^{(0)}\Big)\Big]+{\lambda_{kR}}^2r_k^{(1)}\Big[R\lambda_{k\Theta}{r_k^{(0)}}^2p_{k,R}^{(1)}r_{k,R}^{(0)}\\
&-4C_kR^2\lambda_{kR} {\lambda_{k\Theta}}^2(z_k^{(1)}r_{k,R}^{(1)}+z_k^{(2)}r_{k,R}^{(0)})+r_k^{(0)}\Big[R\lambda_{k\Theta} \Big[z_k^{(1)}p_{k,R}^{(0)}z_{k,R}^{(0)}\\
&-p_k^{(1)}\Big({r_{k,R}^{(0)}}^2+{z_{k,R}^{(0)}}^2\Big)\Big]+2C_k\lambda_{kR}\Big(z_{k,R}^{(0)}+ R^2{\lambda_{k\Theta}}^2z_{k,R}^{(2)}\Big)\Big]\Big]-R\lambda_{k\Theta}\\
&\times r_k^{(0)}\Big[2C_kR{\lambda_{kR}}^3\lambda_{k\Theta}z_k^{(1)}r_{k,R}^{(2)}-{\lambda_{kR}}^2 r_k^{(0)}z_k^{(1)}p_{k,R}^{(1)}z_{k,R}^{(0)}+2C_k\lambda_{kR}\lambda_{k\Theta}\\
&\times r_{k,R}^{(1)}z_{k,R}^{(0)}+{\lambda_{kR}}^2z_k^{(2)}\Big(4C_kR\lambda_{kR} \lambda_{k\Theta}r_{k,R}^{(1)}-r_k^{(0)}p_{k,R}^{(0)}z_{k,R}^{(0)}\Big)-2C_k\lambda_{kR}\\
&\times\lambda_{k\Theta}r_{k,R}^{(0)}z_{k,R}^{(1)}+{\lambda_{kR}}^2p_k^{(1)} r_k^{(0)}\Big(r_{k,R}^{(0)}r_{k,R}^{(1)}+z_{k,R}^{(0)}z_{k,R}^{(1)}\Big)-2C_kR(\lambda_{k\Theta}\\
&\times\lambda_{kR,R}-\lambda_{kR}\lambda_{k\Theta,R})\left(r_{k,R}^{(1)}z_{k,R}^{(0)}-r_{k,R}^{(0)}z_{k,R}^{(1)}\right)+2C_kR\lambda_{kR}\lambda_{k\Theta}\Big(z_{k,R}^{(0)}\\
&\times r_{k,RR}^{(1)}-r_{k,R}^{(0)}z_{k,RR}^{(1)}\Big)\Big]\bigg]
\end{aligned}
\label{eq:B11}
\end{equation}
\end{small}


\section{The position fields and growth functions for the examples of shape-programming}
\label{app:c}

The position fields and growth fields for the examples of shape-programming in Section \ref{sec:4.1} are listed below.
\begin{itemize}
\item \emph{Case\ 1}: an elliptic curve ($C_1/C_2=1$, $\beta_1/\beta_2=1$, $h_1=0.02$, $h_2=0.02$)
\begin{small}
\begin{equation}
\begin{aligned}
\left\{\begin{array}{l}\vspace{1.5ex}
x_1^{(0)}(X)=\displaystyle{\frac{1}{2}\mathrm{sin}(\pi X)},\\\vspace{1.5ex}
z_1^{(0)}(X)=\displaystyle{-\frac{1}{4}\left[\mathrm{cos} (\pi X )-1\right]},\\\vspace{1.5ex}
\lambda_1(X)=\displaystyle{\frac{\pi\sqrt{2C_{\lambda}^{(1)}(X)}}{8}-\frac{2\pi}{75C_{\lambda}^{(1)}(X)}},\\\vspace{1.5ex}
\lambda_2(X)=\displaystyle{\frac{\pi\sqrt{2C_{\lambda}^{(1)}(X)}}{8}-\frac{2\pi}{15C_{\lambda}^{(1)}(X)}},\\
C_{\lambda}^{(1)} (X) =3\mathrm{cos} (2\pi X) + 5.
\end{array}\right.
\end{aligned}
\label{eq:C1}
\end{equation}
\end{small}

\item \emph{Case\ 2}: a butterfly curve ($C_1/C_2=1$, $\beta_1/\beta_2=1$, $h_1=0.02$, $h_2=0.02$)
\begin{small}
\begin{equation}
\begin{aligned}
\left\{\begin{array}{l}\vspace{1.5ex}
x_1^{(0)}(X)=\displaystyle{\frac{\sqrt{5}}{10}\mathrm{sin}(\pi X)\left[4\mathrm{cos}(\pi X)+1 \right]},\\\vspace{1.5ex}
z_1^{(0)}(X)=\displaystyle{\frac{\sqrt{5}}{5}\mathrm{sin}(\pi X)\left[\mathrm{cos}(\pi X)-1\right]},\\\vspace{1.5ex}
\lambda_1(X)=\displaystyle{\frac{\pi\sqrt{2C_{\lambda}^{(2)}(X)}}{4}+\frac{\pi C_{\lambda}^{(3)}(X)}{75C_{\lambda}^{(2)}(X)}},\\\vspace{1.5ex}
\lambda_2(X)=\displaystyle{\frac{\pi\sqrt{2C_{\lambda}^{(2)}(X)}}{4}+\frac{\pi C_{\lambda}^{(3)}(X)}{15C_{\lambda}^{(2)}(X)}},\\\vspace{1.5ex}
C_{\lambda}^{(2)}(X)=\mathrm{cos}(2\pi X)+4\mathrm{cos}(4\pi X)+ 5,\\
C_{\lambda}^{(3)}(X)=3\mathrm{sin}(\pi X)+\mathrm{sin}(3\pi X).
\end{array}\right.
\end{aligned}
\label{eq:C2}
\end{equation}
\end{small}

\item \emph{Case\ 3}: an oval curve ($C_1/C_2=1$, $\beta_1/\beta_2=3$, $h_1=0.03$, $h_2=0.01$)
\begin{small}
\begin{equation}
\begin{aligned}
\left\{\begin{array}{l}\vspace{1.5ex}
x_1^{(0)}(X)=\displaystyle{\frac{1}{4}\mathrm{sin}(\pi X)},\\\vspace{1.5ex}
z_1^{(0)}(X)=\displaystyle{\frac{1}{20}\left[\mathrm{cos}(\pi X)-6\right]\left[\mathrm{cos}(\pi X)-1\right]},\\\vspace{1.5ex}
\lambda_1(X)=\displaystyle{\frac{\pi\sqrt{2C_{\lambda}^{(4)}(X)}}{40}+\frac{\pi C_{\lambda}^{(5)}(X)}{18C_{\lambda}^{(4)}(X)}},\\\vspace{1.5ex}
\lambda_2(X)=\displaystyle{\frac{\pi\sqrt{2C_{\lambda}^{(4)}(X)}}{40}+\frac{7\pi C_{\lambda}^{(5)}(X)}{30C_{\lambda}^{(4)}(X)}},\\\vspace{1ex}
C_{\lambda}^{(4)}(X)=75-14\mathrm{cos}(\pi X)-24\mathrm{cos}(2\pi X)+14\mathrm{cos}(3\pi X)\\\vspace{1.5ex}
\ \ \ \ \ \ \ \ \ \ \ \ \ \ \ -\mathrm{cos}(4\pi X),\\
C_{\lambda}^{(5)}(X)=3\mathrm{cos}(\pi X) + \mathrm{cos}(3\pi X) - 14.
\end{array}\right.
\end{aligned}
\label{eq:C3}
\end{equation}
\end{small}

\item \emph{Case\ 4}: a spiral curve ($C_1/C_2=3$, $\beta_1/\beta_2=1$, $h_1=0.02$, $h_2=0.02$)
\begin{small}
\begin{equation}
\begin{aligned}
\left\{\begin{array}{l}\vspace{1.5ex}
x_1^{(0)}(X)=\displaystyle{\frac{1}{15}\left[(X+1)\left[\mathrm{cos}(5\pi X)+5\mathrm{sin}(5X)\right]-1\right]},\\\vspace{1.5ex}
z_1^{(0)}(X)=\displaystyle{\frac{1}{15}\left[(X+1)\left[5\mathrm{cos}(5X)-\mathrm{sin}(5X)\right]-5\right]},\\\vspace{1.5ex}
\lambda_1(X)=\displaystyle{\frac{1}{15}\sqrt{676+650X(X+2)}+\frac{945+875X(X+2)}{50[468+450X(X+2)]}},\\
\lambda_2(X)=\displaystyle{\frac{1}{15}\sqrt{676+650X(X+2)}+\frac{11[27+25X(X+2)]}{60[26+25X(X+2)]}}.
\end{array}\right.
\end{aligned}
\label{eq:C4}
\end{equation}
\end{small}

\end{itemize}

The position fields and growth fields for the examples of shape-programming in Section \ref{sec:4.2} are listed below.
\begin{itemize}

\item \emph{Case\ 1}: an ellipsoid ($C_1/C_2=1$, $\beta_1/\beta_2=5$, $h_1=0.05$, $h_2=0.01$)
\begin{small}
\begin{equation}
\begin{aligned}
\left\{\begin{array}{l}\vspace{1.5ex}
x_1^{(0)}(R)=\displaystyle{\frac{3}{5}\mathrm{sin}\left(\pi R-\frac{\pi}{5}\right)+\frac{1}{50}},\\\vspace{1.5ex}
z_1^{(0)}(R)=\displaystyle{-\frac{3}{10}\left[\mathrm{cos}\left(\pi R-\frac{\pi}{5}\right)-1\right]},\\\vspace{1.5ex}
\lambda_{1R}(R)=\displaystyle{\frac{3\pi\sqrt{2C_{\lambda}^{(6)}(R)}}{20}}-\frac{9\pi}{125C_{\lambda}^{(6)}(R)},\\\vspace{1.5ex}
\lambda_{1\Theta}(R)=\displaystyle{\frac{1+30C_{\lambda}^{(7)}(R)}{50R}-\frac{9C_{\lambda}^{(7)}(R)}{250R\sqrt{2C_{\lambda}^{(6)}(R)}}},\\\vspace{1.5ex}
\lambda_{2R}(R)=\displaystyle{\frac{3\pi\sqrt{2C_{\lambda}^{(6)}(R) }}{20}}-\frac{9\pi}{25C_{\lambda}^{(6)}(R)},\\\vspace{1.5ex}
\lambda_{2\Theta}(R)=\displaystyle{\frac{1+30C_{\lambda}^{(7)}(R)}{50R}-\frac{9C_{\lambda}^{(7)}(R)}{50R \sqrt{2C_{\lambda}^{(6)}(R)}}},\\\vspace{1.5ex}
C_{\lambda}^{(6)}(R)=\displaystyle{3\mathrm{cos}\left(2\pi R-\frac{2\pi}{5}\right)+5},\\
C_{\lambda}^{(7)}(R)=\displaystyle{\mathrm{sin}\left(\pi R-\frac{\pi}{5}\right)}.
\end{array}\right.
\end{aligned}
\label{eq:C5}
\end{equation}
\end{small}

\item \emph{Case\ 2}: an oval surface ($C_1/C_2=1$, $\beta_1/\beta_2=5$, $h_1=0.05$, $h_2=0.01$)
\begin{small}
\begin{equation}
\begin{aligned}
\left\{\begin{array}{l}\vspace{1.5ex}
r_1^{(0)}(R)=\displaystyle{-\frac{1}{2}\mathrm{cos}\left(\pi R+\frac{3\pi}{10}\right)+\frac{1}{20}},\\\vspace{1.5ex}
z_1^{(0)}(R)=\displaystyle{\frac{1}{20}\left[\mathrm{sin}\left(\pi R+\frac{3\pi}{10}\right)-1\right]\left[3\mathrm{sin}\left(\pi R+\frac{3\pi}{10}\right)-10\right]},\\\vspace{1.5ex}
\lambda_{1R}(R)=\displaystyle{\frac{\pi\sqrt{2C_{\lambda}^{(8)}(R)}}{40}-\frac{9\pi C_{\lambda}^{(9)}(R)}{50C_{\lambda}^{(8)}(R)}},\\\vspace{1.5ex}
\lambda_{1\Theta}(R)=\displaystyle{\frac{C_{\lambda}^{(10)}(R)}{20R}+\frac{9C_{\lambda}^{(11)}(R)}{250R\sqrt{C_{\lambda}^{(8)}(R)}}},\\\vspace{1.5ex}
\lambda_{2R}(R)=\displaystyle{\frac{\pi\sqrt{2C_{\lambda}^{(8)}(R)}}{40}-\frac{9\pi C_{\lambda}^{(9)}(R)}{10C_{\lambda}^{(8)}(R)}},\\\vspace{1.5ex}
\lambda_{2\Theta}(R)=\displaystyle{\frac{C_{\lambda}^{(10)}(R)}{20R}+\frac{9C_{\lambda}^{(11)}(R)}{50R\sqrt{C_{\lambda}^{(8)}(R)}}},\\\vspace{1ex}
C_{\lambda}^{(8)}(R)=\displaystyle{278+69 \mathrm{cos}\left(2\pi R+\frac{3\pi}{5}\right)+9\mathrm{cos}\left(4\pi R+\frac{\pi}{5}\right)}\\\vspace{1.5ex}
\ \ \ \ \ \ \ \ \ \ \ \ \ \ \displaystyle{-78\mathrm{sin}\left(\pi R+\frac{3\pi}{10}\right)-78\mathrm{sin}\left(3\pi R+\frac{9\pi}{10}\right) },\\\vspace{1.5ex}
C_{\lambda}^{(9)}(R)=\displaystyle{26-9\mathrm{sin}\left(\pi R+\frac{3\pi}{10}\right)+3\mathrm{sin}\left(3\pi R+\frac{9\pi}{10}\right)},\\\vspace{1.5ex}
C_{\lambda}^{(10)}(R)=\displaystyle{1-10\mathrm{cos}\left(\pi R+\frac{3\pi}{10}\right)},\\
C_{\lambda}^{(11)}(R)=\displaystyle{\mathrm{cos}\left(\pi R+\frac{3\pi}{10}\right)\left[13-6\mathrm{sin}\left(\pi R+\frac{3\pi}{10}\right)\right]}.
\end{array}\right.
\end{aligned}
\label{eq:C6}
\end{equation}
\end{small}

\item \emph{Case\ 3}: an Ipomoea cairica ($C_1/C_2=5$, $\beta_1/\beta_2=1$, $h_1=0.03$, $h_2=0.03$)
\begin{small}
\begin{equation}
\begin{aligned}
\left\{\begin{array}{l}\vspace{1.5ex}
r_1^{(0)}(R)=\displaystyle{\frac{1}{50}\left[5\mathrm{tan}\left(\frac{4\pi R}{5}+\frac{12\pi}{25}\right)+5\mathrm{cot}\left(\frac{7\pi}{50}\right)+3\right]},\\\vspace{1.5ex}
z_1^{(0)}(R)=\displaystyle{-\frac{1}{2}\left[\mathrm{cos}\left(\pi R-\frac{\pi}{5}\right)-1\right]},\\\vspace{1.5ex}
\lambda_{1R}(R)=\displaystyle{\frac{\pi\sqrt{C_{\lambda}^{(12)}(R)}}{50}-\frac{3\pi C_{\lambda}^{(13)}(R)}{25C_{\lambda}^{(12)}(R)}},\\\vspace{1.5ex}
\lambda_{1\Theta}(R)=\displaystyle{\frac{C_{\lambda}^{(14)}(R)}{50R}-\frac{3}{10R\sqrt{C_{\lambda}^{(12)}(R)}}\mathrm{sin}\left(\pi R-\frac{\pi}{5}\right)},\\\vspace{1.5ex}
\lambda_{2R}(R)=\displaystyle{\frac{\pi\sqrt{C_{\lambda}^{(12)}(R)}}{50}-\frac{3\pi C_{\lambda}^{(13)}(R)}{5C_{\lambda}^{(12)}(R)}},\\\vspace{1.5ex}
\lambda_{2\Theta}(R)=\displaystyle{\frac{C_{\lambda}^{(14)}(R)}{50R}-\frac{3}{2R\sqrt{C_{\lambda}^{(12)}(R)}}\mathrm{sin}\left(\pi R-\frac{\pi}{5}\right)},\\\vspace{1.5ex}
C_{\lambda}^{(12)}(R)=\displaystyle{16\mathrm{sec}^4\left(\frac{4\pi R}{5}+\frac{12\pi}{25}\right)+625\mathrm{sin}^2\left(\pi R-\frac{\pi}{5}\right)},\\\vspace{1ex}
C_{\lambda}^{(13)}(R)=\displaystyle{\left[3\mathrm{cos}\left(\frac{4\pi R}{5}+\frac{8\pi}{25}\right)+13\mathrm{cos}\left(\frac{9\pi R}{5}+\frac{7\pi}{25}\right)\right]}\\\vspace{1.5ex}
\ \ \ \ \ \ \ \ \ \ \ \ \ \ \displaystyle{\times\mathrm{sec}^3\left(\frac{4\pi R}{5}+\frac{12\pi}{25}\right)},\\
C_{\lambda}^{(14)}(R)=\displaystyle{5\mathrm{tan}\left(\frac{4\pi R}{5}+\frac{12\pi}{25}\right) + 5\mathrm{cot}\left(\frac{7\pi}{50}\right)+3}.
\end{array}\right.
\end{aligned}
\label{eq:C7}
\end{equation}
\end{small}

\item \emph{Case\ 4}: a pot surface ($C_1/C_2=5$, $\beta_1/\beta_2=1$, $h_1=0.03$, $h_2=0.03$)
\begin{small}
\begin{equation}
\begin{aligned}
\left\{\begin{array}{l}\vspace{1.5ex}
r_1^{(0)}(R)=\displaystyle{\frac{1}{60}\left[\displaystyle{3-6\mathbf{e}^{\frac{3}{10}}+6\mathbf{e}^{\frac{3R}{2}}-20\mathrm{cos}\left(\frac{9\pi R}{5}\right) +20\mathrm{sin}\left(\frac{7\pi}{50}\right)}\right]},\\\vspace{1.5ex}
z_1^{(0)}(R)=\displaystyle{\frac{4}{5}\left(\mathbf{e}^{\frac{4R}{5}}-\mathbf{e}^{\frac{4}{25}}\right)},\\\vspace{1.5ex}
\lambda_{1R}(R)=\displaystyle{\frac{\sqrt{C_{\lambda}^{(15)}(R)}}{100}+\frac{C_{\lambda}^{(17)}(R)}{125C_{\lambda}^{(15)}(R)}},\\\vspace{1.5ex}
\lambda_{1\Theta}(R)=\displaystyle{\frac{C_{\lambda}^{(16)}(R)}{60R}-\frac{96\mathbf{e}^{\frac{4R}{5}}}{125R\sqrt{C_{\lambda}^{(15)}(R)}}},\\
\lambda_{2R}(R)=\displaystyle{\frac{\sqrt{C_{\lambda}^{(15)}(R)}}{100}+\frac{C_{\lambda}^{(17)}(R)}{25C_{\lambda}^{(15)}(R)}},\\\vspace{1.5ex}
\lambda_{2\Theta}(R)=\displaystyle{\frac{C_{\lambda}^{(16)}(R)}{60R}-\frac{96\mathbf{e}^{\frac{4R}{5}}}{25R\sqrt{C_{\lambda}^{(15)}(R)}}},\\\vspace{1.5ex}
C_{\lambda}^{(15)}(R)=\displaystyle{4096\mathbf{e}^{\frac{8R}{5}}+225\left[\mathbf{e}^{\frac{3R}{2}}+4\pi\mathrm{sin}\left(\frac{9\pi}{5}R\right)\right]^2},\\\vspace{1.5ex}
C_{\lambda}^{(16)}(R)=\displaystyle{3-6\mathbf{e}^{\frac{3}{10}}+6\mathbf{e}^{\frac{3R}{2}}-20\mathrm{cos}\left(\frac{9\pi R}{5}\right)+20\mathrm{sin}\left(\frac{7\pi}{50}\right)},\\
C_{\lambda}^{(17)}(R)=\displaystyle{144\mathbf{e}^{\frac{4R}{5}}\left[7\mathbf{e}^{\frac{3R}{2}}+8\pi\left[9\pi\mathrm{cos}\left(\frac{9\pi R}{5}\right)-4\mathrm{sin}\left(\frac{9\pi R}{5} \right)\right]\right]}.
\end{array}\right.
\end{aligned}
\label{eq:C8}
\end{equation}
\end{small}

\end{itemize}



\begin{thebibliography}{00}



\bibitem[Koch et al.(2008)]{koch2008} K. Koch, B. Bhushan, W. Barthlott, Diversity of structure, morphology and wetting of plant surfaces, Soft Matter 4 (2008) 1943-1963.

\bibitem[M{\"a}thger et al.(2009)]{math2009} L.M. M{\"a}thger, E.J. Denton, N.J. Marshall, R.T. Hanlon, Mechanisms and behavioural functions of structural coloration in cephalopods, J. R. Soc. Interface 6 (2009) S149-S163.

\bibitem[Li et al.(2010)]{li2010} B.W. Li, H.P. Zhao, X.Q. Feng, W.-W. Guo, S.C. Shan, Experimental study on the mechanical properties of the horn sheaths from cattle, J. Exp. Biol. 213 (2010) 479-486.

\bibitem[Fernandes and Gracias(2012)]{fern2012} R. Fernandes,  D.H. Gracias, Self-folding polymeric containers for encapsulation and delivery of drugs, Adv. Drug Deliver. Rev. 64 (2012) 1579-1589.

\bibitem[Li et al.(2017)]{li2017} T.F. Li, G.R. Li, Y.M. Liang, T.Y. Cheng, J. Dai, X.-X. Yang, B.Y. Liu, Z.D. Zeng, Z.L. Huang, Y.W. Luo, T. Xie, W. Yang, Fast-moving soft electronic fish, Sci. Adv. 5 (2017) e1602045.

\bibitem[Li et al.(2021)]{li2021} C. Li, Y. Xue, M. Han, L.C. Palmer, J.A. Rogers, Y.G. Huang, S.I. Stupp, Synergistic photoactuation of bilayered spiropyran hydrogels for predictable origami-like shape change, Matter 4 (2021) 1-14.

\bibitem[Stoychev et al.(2012)]{stoy2012} G. Stoychev, S. Zakharchenko, S. Turcaud, J.W.C. Dunlop, L. Ionov, Shape-Programmed Folding of Stimuli-Responsive Polymer Bilayers, ACS Nano 6 (2012) 3925-3934.

\bibitem[Egunov et al.(2016)]{egun2016} A.I. Egunov, J.G. Korvink, V.A. Luchnikov, Polydimethylsiloxane bilayer films with an embedded spontaneous curvature, Soft Matter 12 (2016) 45-52.

\bibitem[Ambrosi et al.(2011)]{ambr2011} D. Ambrosi, G.A. Ateshian, E.M. Arruda, S.C. Cowin, J. Dumais, A. Goriely, G.A. Holzapfel, J.D. Humphrey, R. Kemkemer, E. Kuhl, J.E. Olberding, L.A. Taber, K. Garikipati, Perspectives on biological growth and remodeling, J. Mech. Phys. Solids 59 (2011) 863-883.

\bibitem[Li et al.(2012)]{li2012} B. Li, Y.P. Cao, X.Q. Feng, H.J. Gao, Mechanics of morphological instabilities and surface wrinkling in soft materials: a review, Soft Matter 8 (2012) 5728-5745.

\bibitem[Liu et al.(2015)]{liu2015} Z.S. Liu, W. Toh, T.Y. Ng, Advances in mechanics of soft materials: a review of large deformation behavior of hydrogels, Inter. J. Appl. Mech. 07 (2015) 1530001.

\bibitem[Lubarda and Hoger(2002)]{luba2002} V.A. Lubarda, A. Hoger, On the mechanics of solids with a growing mass, Int. J. Solids Struct. 39 (2002) 4627-4664.

\bibitem[Ben Amar and Goriely(2005)]{ben2005} M. Ben Amar, A. Goriely, Growth and instability in elastic tissues, J. Mech. Phys. Solids 53 (2005) 2284-2319.

\bibitem[Goriely(2017)]{gori2017} A. Goriely, The Mathematics and Mechanics of Biological Growth, Springer New York, 2017.

\bibitem[Kondaurov and Nikitin(1987)]{kond1987} V.I. Kondaurov, L.V. Nikitin, Finite strains of viscoelastic muscle tissue, J. Appl. Math. Mech. 51 (1987) 346-353.

\bibitem[Rodriguez et al.(1994)]{rodr1994} E.K. Rodriguez, A. Hoger, A.D. McCulloch, Stress-dependent finite growth in soft elastic tissues, J. Biomech. 27 (1994) 455-467.

\bibitem[Skalak et al.(1996)]{skal1996} R. Skalak, S. Zargaryan, R.K. Jain, P.A. Netti, A. Hoger, Compatibility and the genesis of residual stress by volumetric growth, J. Math. Biol. 34 (1996) 889-914.

\bibitem[Taber and Humphrey(2001)]{tabe2001} L.A. Taber, J.D. Humphrey, Stress-Modulated Growth, Residual Stress, and Vascular Heterogeneity, J. Biomech. Eng. 123 (2001) 528-535.

\bibitem[Humphrey(2003)]{hump2003} J.D. Humphrey, Review Paper: Continuum biomechanics of soft biological tissues, P. Roy. Soc. A 459 (2003) 3-46.

\bibitem[Dai and Song(2014)]{dai2014} H.-H. Dai, Z.L. Song, On a consistent finite-strain plate theory based on three-dimensional energy principle, P. Roy. Soc. A 470 (2014) 20140494.

\bibitem[Song and Dai(2016)]{song2016} Z.L. Song, H.-H. Dai, On a consistent dynamic finite-strain plate theory and its linearization, J. Elast. 125 (2016) 149-183.

\bibitem[Wang et al.(2016)]{wang2016} J. Wang, Z.L. Song, H.-H. Dai, On a consistent finite-strain plate theory for incompressible hyperelastic materials, Int. J. Solids Struct. 78-79 (2016) 101-109.

\bibitem[Fu et al.(2021)]{fu2021} C.B. Fu, H.-H. Dai, F. Xu, Computing wrinkling and restabilization of stretched sheets based on a consistent finite-strain plate theory, Comput. Method. Appl. M. 384 (2021) 113986.

\bibitem[Wang et al.(2018)]{wang2018} J. Wang, D.J. Steigmann, F.-F. Wang, H.-H. Dai, On a consistent finite-strain plate theory of growth, J. Mech. Phys. Solids 111 (2018) 184-214.

\bibitem[Wang et al.(2019)]{wang2019} J. Wang, Q.Y. Wang, H.-H. Dai, P. Du, D.X. Chen, Shape-programming of hyperelastic plates through differential growth: an analytical approach, Soft Matter 15 (2019) 2391-2399.

\bibitem[Kadapa et al.(2021)]{kada2021} C. Kadapa, Z.F. Li, M. Hossain, J. Wang, On the advantages of mixed formulation and higher-order elements for computational morphoelasticity, J. Mech. Phys. Solids 148 (2021) 104289.

\bibitem[Mehta et al.(2021)]{meht2021} S. Mehta, G. Raju, P. Saxena, Growth induced instabilities in a circular hyperelastic plate, Int. J. Solids Struct. 226 (2021) 111026.

\bibitem[Li et al.(2022)]{li2022} Z.F. Li, Q.Y. Wang, P. Du, C. Kadapa, M. Hossain, J. Wang, Analytical study on growth-induced axisymmetric deformations and shape-control of circular hyperelastic plates, Int. J. Eng. Sci. 170 (2022) 103594.

\bibitem[Wang et al.(2022)]{wang2022} J. Wang, Z.F. Li, Z.L. Jin, A theoretical scheme for shape-programming of thin hyperelastic plates through differential growth. Math. Mech. Solids (2022), doi: 10.1177/10812865221089694.

\bibitem[Tsai et al.(2004)]{tsai2004} H. Tsai, T.J. Pence, E. Kirkinis, Swelling induced finite strain flexure in a rectangular block of an isotropic elastic material, J. Elast. 75 (2004) 69-89.

\bibitem[Schmidt(2007)]{schm2007} B. Schmidt, Plate theory for stressed heterogeneous multilayers of finite bending energy, J. Math. Pures Appl. 88 (2007) 107-122.

\bibitem[Delgado and Schmidt(2021)]{delg2021} M.D. Delgado, B. Schmidt,  A hierarchy of multilayered plate models, ESAIM-Contr. Optim. Cal. Var. 27 (2021) S16.

\bibitem[Dervaux and Ben Amar(2010)]{derv2010} J. Dervaux, M. Ben Amar, Localized growth of layered tissues, IMA J. Appl. Math. 75 (2010) 571-580.

\bibitem[Armon et al.(2011)]{armo2011} S. Armon, E. Efrati, R. Kupferman, E. Sharon,  Geometry and mechanics in the opening of chiral seed pods, Science 333 (2011) 1726-1730.

\bibitem[Budday et al.(2014)]{budd2014} S. Budday, P. Steinmann, E. Kuhl, The role of mechanics during brain development, J. Mech. Phys. Solids 72 (2014) 75-92.

\bibitem[Ben Amar and Bordner(2017)]{ben2017} M. Ben Amar, A. Bordner, Mimicking cortex convolutions through the wrinkling of growing soft bilayers, J. Elast. 129 (2017) 213-238.

\bibitem[Lucantonio et al.(2014)]{luca2014} A. Lucantonio, P. Nardinocchi, M. Pezzulla, Swelling-induced and controlled curving in layered gel beams, P. Roy. Soc. A 470 (2014) 20140467.

\bibitem[Nardinocchi et al.(2017)]{nard2017} P. Nardinocchi, E. Puntel, Swelling-induced wrinkling in layered gel beams, P. Roy. Soc. A 473 (2017) 20170454.

\bibitem[Pezzulla et al.(2016)]{pezz2016} M. Pezzulla, G.P. Smith, P. Nardinocchi, D.P. Holmes, Geometry and mechanics of thin growing bilayers, Soft Matter 12 (2016) 4435-4442.

\bibitem[van Rees et al.(2017)]{van2017} W.M. van Rees, E. Vouga, L. Mahadevan, Growth patterns for shape-shifting elastic bilayers, P. Natl. Acad. Sci. USA 114 (2017) 11597-11602.

\bibitem[Ackermann et al.(2022)]{acke2022} J. Ackermann, P.Q. Qu, L. LeGoff, M. Ben Amar, Modeling the mechanics of growing epithelia with a bilayer plate theory, Eur. Phys. J. Plus 137 (2022) 8.

\bibitem[Du et al.(2020)]{du2020} P. Du, H.-H. Dai, J. Wang, Q.Y. Wang, Analytical study on growth-induced bending deformations of multi-layered hyperelastic plates, Int J. Nonlin. Mech. 119 (2020) 103370.

\bibitem[Du et al.(2022)]{du2022} P. Du, J. Wang, Z.F. Li, W.C. Cai, On a finite-strain plate theory for growth-induced plane-strain deformations and instabilities of multi-layered hyperelastic plates, Int. J. Solids Struct. 236-237 (2022) 111348.

\bibitem[Yu et al.(2020)]{yu2020} X. Yu, Y. Fu, H.-H. Dai, A refined dynamic finite-strain shell theory for incompressible hyperelastic materials: equations and two-dimensional shell virtual work principle, Proc. R. Soc. A 476 (2020) 20200031.

\bibitem[Chen et al.(2021)]{chen2021} X.Y. Chen, H.-H. Dai, E. Pruchnicki, On a consistent rod theory for a linearized anisotropic elastic material: I. Asymptotic reduction method, Math. Mech. Solids 26 (2021) 217-229.

\bibitem[Wang F.F. et al.(2019)]{wangff2019} F.-F. Wang, D.J. Steigmann, H.-H. Dai, On a uniformly-valid asymptotic plate theory, Int. J. Non-lin. Mech. 112 (2019) 117-125.

\bibitem[Liu et al.(2016)]{liu2016} Y. Liu, J. Genzer, M.D. Dickey, "2D or not 2D": Shape-programming polymer sheets, Prog. Polym. Sci. 52 (2016) 79-106.

\bibitem[Sydney Gladman et al.(2016)]{glad2016} A. Sydney Gladman, E.A. Matsumoto, R.G. Nuzzo, L. Mahadevan, J.A. Lewis, Biomimetic 4D printing, Nat. Mater. 15 (2016) 413-418.

\bibitem[Yuk et al.(2017)]{yuk2017} H. Yuk, S.T. Lin, C. Ma, M. Takaffoli, N.X. Fang, X.H. Zhao, Hydraulic hydrogel actuators and robots optically and sonically camouflaged in water, Nat. Commun. 8 (2017) 14230.

\bibitem[Siefert et al.(2019)]{sief2019} E. Si$\mathrm{\acute{e}}$fert, E. Reyssat, J. Bico, B. Roman, Bio-inspired pneumatic shape-morphing elastomers, Nat. Mater. 18 (2019) 24-28.

\bibitem[Dias et al.(2011)]{dias2011} M.A. Dias, J.A. Hanna, C.D. Santangelo, Programmed buckling by controlled lateral swelling in a thin elastic sheet, Phys. Rev. E 84 (2011) 036603.

\bibitem[Jones and Mahadevan(2015)]{jone2015} G.W. Jones, L. Mahadevan, Optimal control of plates using incompatible strains, Nonlinearity 28 (2015) 3153-3174.

\bibitem[Acharya(2019)]{acha2019} A. Acharya, A Design Principle for Actuation of Nematic Glass Sheets, J. Elast. 136 (2019) 237-249.

\bibitem[Nojoomi(2021)]{nojo2021} A. Nojoomi, J. Jeon, K. Yum, 2D material programming for 3D shaping, Nat. Commun. 12 (2021) 603.

































\end{thebibliography}
\end{document}